\documentclass[rmp,aps,amsfonts,twocolumn,superscriptaddress,preprintnumbers]{revtex4}
\usepackage{subfig}
\usepackage{graphicx}
\usepackage{amsmath,latexsym}
\usepackage{epsfig}

\IfFileExists{srcltx.sty}{\usepackage[active]{srcltx}}

\newcommand{\met} {\mbox{${E\!\!\!\!/_{\rm T}}$}}

\begin{document}


\title{The origin of dark matter, matter-anti-matter asymmetry, and inflation}

\author{Anupam Mazumdar}
\affiliation{Lancaster University, Physics Department, Lancaster LA1 4YB, UK}
\affiliation{Niels Bohr Institute,  Blegdamsvej-17,  DK-2100, Denmark.}


\begin{abstract}
A rapid phase of accelerated expansion in the early universe, known as inflation, 
dilutes all matter except the vacuum induced quantum fluctuations. These are responsible for seeding 
the initial perturbations in the baryonic matter, the non-baryonic dark matter and the observed temperature anisotropy in the 
cosmic microwave background (CMB) radiation. To explain the universe observed today, the end of inflation must also 
excite a thermal bath filled with baryons, an amount of baryon asymmetry, and dark matter. We review the current
understanding of inflation, dark matter, mechanisms for generating matter-anti-matter asymmetry, and the prospects for testing them at 
ground and space based experiments.

\end{abstract}

\maketitle 



\tableofcontents


\section{Introduction}

This review aims at building a consistent picture of the early universe where the three pillars of modern cosmology: inflation, baryogenesis and the synthesis of dark matter can be understood in a testable framework of physics beyond the Standard Model (SM).

Inflation ~\cite{Guth:1980zm}, which is a rapid phase of accelerated expansion of space, is the leading model that explains the origin of matter; during this phase, primordial density perturbations are also stretched from sub-Hubble to super-Hubble length scales~\cite{Mukhanov:1990me}. A strong support for such an inflationary scenario comes from the precision measurement of these perturbations in the cosmic microwave background (CMB) radiation, e.g. by the Cosmic Background Explorer (COBE)~\cite{Smoot:1992td} and the Wilkinson Microwave Anisotropy Probe (WMAP)~\cite{Komatsu:2010fb} satellites. However, one of the most serious challenges faced by inflationary models is that only a few of them provide clear predictions for crucial questions regarding the nature of the matter created after inflation and the mode of exiting inflation in a vacuum that can excite the SM degrees of freedom ({\it d.o.f}) ~\cite{Mazumdar:2010sa}.

From observations we know that the current universe contains $4.6\%$ atoms, $23\%$ non-relativistic, non-luminous 
dark matter, and the rest in the form of dark energy. While some $13.7$~billion years ago it was $37\%$ atoms, photons and 
neutrinos, and $63\%$ non-relativistic dark matter~\cite{Komatsu:2010fb}. Therefore, it is mandatory that 
the inflationary vacuum must excite these SM baryons, and create the right abundance of dark matter. Since the success of Big Bang 
Nucleosynthesis (BBN)~\cite{Iocco:2008va} requires an asymmetry between the baryons and anti-baryons of order one part in $10^{10}$, 
it is necessary that the baryonic asymmetry must have been created dynamically in the early universe before the BBN~\cite{Sakharov:1967dj}.

\begin{figure}
\vspace*{-0.0cm}
\includegraphics[width=7.5cm]{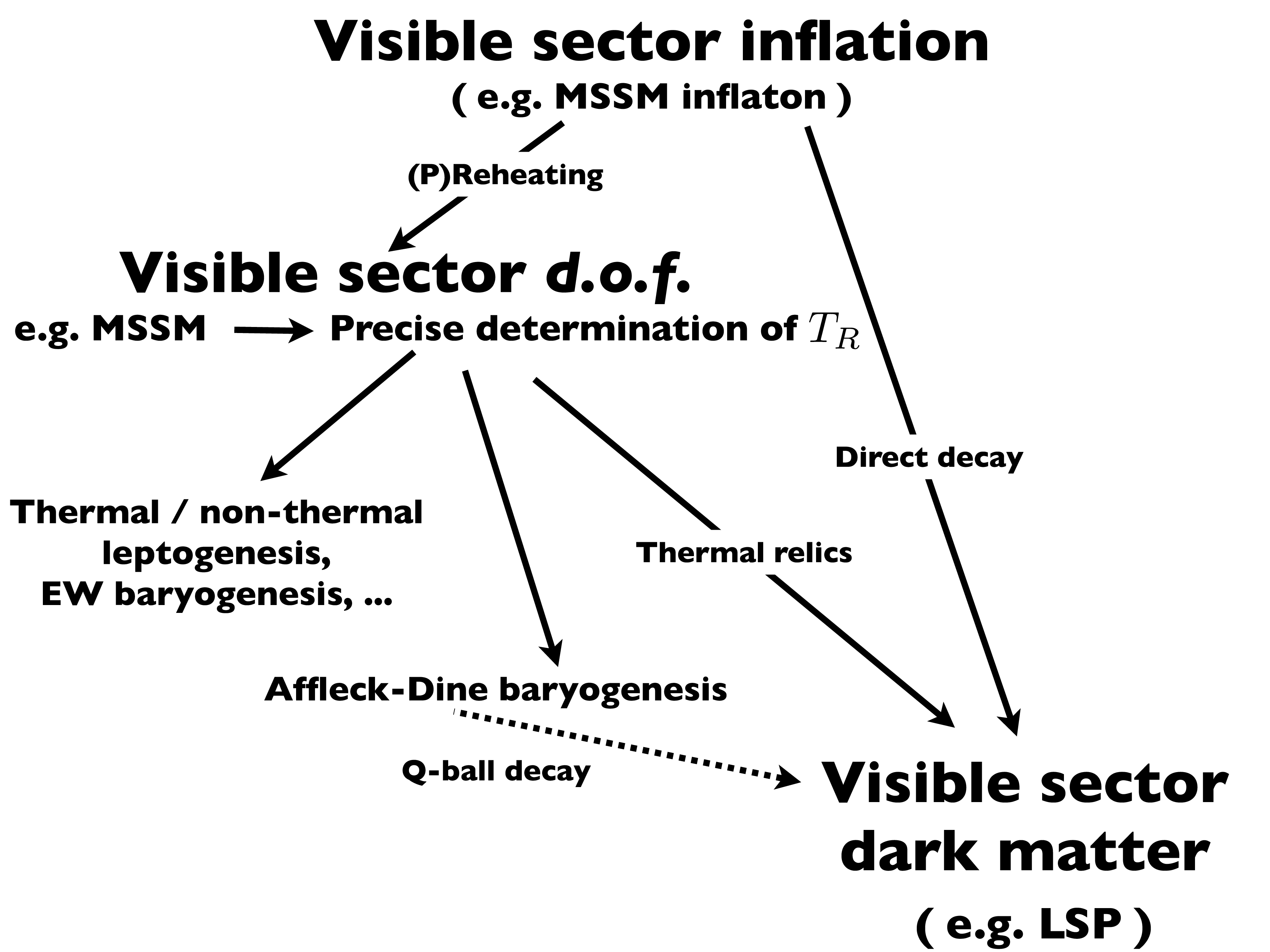}
\caption{An illustration of a visible sector model for the early universe.  EW stands for the electroweak.\label{Rev-plan}}
\end{figure}

The prime question is what sort of {\it visible sector} beyond the SM would accomplish all these goals -- inflation, matter creation, and 
seed perturbations for the CMB. Beyond the scale of electroweak SM (at energies above $\geq 100-1000$~GeV) there are plethora of 
candidates, e.g.~\cite{Bustamante:2009us}. However the low scale supersymmetry (SUSY) provides an excellent platform, 
which have been built on the success of the electroweak physics~\cite{Nilles:1983ge,Haber:1984rc, Martin:1997ns,Chung:2003fi}.
The minimal supersymmetric extension of the SM, known as MSSM,  or its minimal extensions, provides many testable imprints 
at the collider experiment~\cite{Nath:2010zj}. In particular, the lightest SUSY particle (LSP) 
can be electrically neutral, and will be an ideal candidate for the weakly interacting massive particle (WIMP) as a 
dark matter~\cite{Goldberg:1983nd,Ellis:1983ew}, whose abundance can now be calculated from the direct decay of the inflaton, or from the decay products 
of the inflaton, as shown in Fig.~\ref{Rev-plan}. 

If such a visible sector, with the known gauge interactions, can also provide us with an inflationary potential capable of matching the current CMB data,
then we would be able to identify the origin of the inflaton, its mass and couplings, and the vacuum energy density within a testable theory, such as 
the MSSM. The inflaton's {\it gauge invariant} 
couplings would enable us to ascertain the post-inflationary dynamics, and the exact mechanism for particle creation from the inflaton's 
coherent oscillations, known as (p)reheating~\cite{Allahverdi:2010xz}. We would be able to precisely determine the largest 
reheat temperature, $T_{R}$, of the post-inflationary universe, during which all the MSSM {\it d.o.f} come in {\it chemical} 
and in {\it kinetic} equilibrium for the first time ever.  Once the relevant {\it d.o.f} are created it would be possible to build a coherent picture where 
we will be able to understand the origin of baryogenesis and the dark matter in a consistent framework as illustrated in Fig.~\ref{Rev-plan}.


This review is divided into three parts. In the first part we will discuss the origin of inflation, and how
to connect the models of inflation to the current CMB observations. We will keep our discussions general and
provide some examples of non-SUSY models of inflation. We will mainly focus on SUSY based models and its generalization to
supergravity (SUGRA). We will discuss the epoch of reheating, preheating and thermalization for an MSSM based models of inflation.  
In the second part of the review, we will focus on baryogenesis. We will state the conditions for generating baryogenesis. We will 
discuss electroweak baryogenesis, baryogenesis induced by lepton asymmetry, known as leptogenesis, and the MSSM based 
Afflck-Dine baryogenesis which can create non-topological solitons, known as Q-balls. In the third part, we will consider general 
properties of dark matter, various mechanisms for creating them, some well motivated candidates, and link the origin of dark matter to 
the origin of inflation within SUSY. We will briefly discuss the ongoing searches of WIMP as a dark matter candidate.


\section{Particle physics origin of inflation}\label{PPOI}

There are two classes of models of inflation, which have been discussed extensively in the 
literature, e.g. see reviews~\cite{Lyth:1998xn,Mazumdar:2010sa}. In the first class, the inflaton field belongs to the 
hidden sector (not charged under the SM gauge group). The direction along which the 
inflaton field rolls belongs to an absolute gauge singlet, whose couplings to the {\it visible sector} -- such as that of the SM or MSSM fields are 
not determined a-priori. A singlet inflaton would couple to the visible and hidden sectors without any biased -- such as the 
case of gravity which is a true singlet, and a color and flavor blind. 
In the second class, the inflaton candidate distinctly belongs to the visible sector, where the inflaton is charged under the 
SM or its minimal extension beyond the SM gauge group. This has many advantages, which we will discuss in some details.

Any inflationary models are required to be tested by the amplitude of the density perturbations 
for the observed large scale structures~\cite{Mukhanov:1990me}. Therefore the predictions for the CMB 
fluctuations are the most important ones to judge the merits of the models, which would contain information about
the power spectrum, the tilt in the spectrum, running in the tilt, and tensor to scalar ratio. These observable quantities
can be recast in terms of the properties of the potential which we will discuss below.
From the particle origin point of view, one of the successful criteria is to end inflation in the right vacuum - where the 
SM baryons are excited naturally for a successful baryogenesis before BBN~\cite{Mazumdar:2010sa}.


\subsection{Properties of inflation}\label{POI}

The inflaton direction which leads to a graceful exit needs to be flat with a non-negligible slope provided  by 
a potential $V(\phi)$ which dominates the energy density of the universe. A completely flat 
potential, or a false vacuum with a very tiny tunneling rate to a lower vacuum, would render inflation future eternal, 
but not past~\cite{Borde:1993xh,Borde:2001nh,Linde:1986fd,Linde:1983gd,Linde:1996hg,Linde:1993xx}.
A past eternal inflation is possible only if the null geodesics are past complete~\cite{Biswas:2010zk,Biswas:2005qr}.
The slow-roll inflation assumes that the potential dominates over the 
kinetic energy of the inflaton $\dot\phi^2 \ll V(\phi)$, and $\ddot 
\phi \ll V^{\prime}(\phi)$, therefore the Friedmann and the
Klein-Gordon equations are approximated as:
\begin{eqnarray}
\label{slowr1}
H^2 &\approx & {V(\phi)}/{3M_{\rm P}^2}\,, \\
\label{slowr2}
3H\dot\phi &\approx &-V^{\prime}(\phi)\,,
\end{eqnarray}
where prime denotes derivative with respect to $\phi$. 
There exists slow-roll conditions, which constrain the shape of the potential,
are give by:
\begin{eqnarray}
\label{ep1}
\epsilon(\phi)&\equiv&\frac{M_{\rm P}^2}{2}\left({V^{\prime}}/{V}\right)^2
\ll 1\,,\\
\label{eta1}
|\eta(\phi)|&\equiv&{M_{\rm P}^2}\left |{V^{\prime\prime}}/{V}\right| \ll 1\,.
\end{eqnarray}
These conditions are necessary but not sufficient for inflation.  The slow-roll conditions are violated when
$\epsilon \sim 1$, and $\eta \sim 1$, which marks the end of inflation.

However, there are certain models
where this need not be true, for instance in hybrid inflation models
\cite{Linde:1993cn}, where inflation comes to an end via a phase transition,
or in oscillatory models of inflation where slow-roll conditions are
satisfied only on average \cite{Damour:1997cb,Liddle:1998pz}, or inflation happens in oscillations~\cite{Biswas:2009fv}, or in fast roll inflation where 
the slow-roll conditions are never met~\cite{Linde:2001ae}. The K-inflation where only the kinetic term 
dominates where there is no potential at all~\cite{ArmendarizPicon:1999rj}.

One of the salient features of the slow-roll inflation is that there exists a
late time attractor behavior, such that  the evolution of a scalar field after sufficient e-foldings become
independent of the initial conditions \cite{Salopek:1990jq}.

The number of e-foldings between, $t$, and the end of inflation, $t_{end}$, is defined by:
\begin{equation}
N\equiv \ln\frac{a(t_{end})}{a(t)}=\int_{t}^{t_{end}}Hdt\approx
\frac{1}{M_{\rm P}^2}\int_{\phi_{end}}^{\phi}\frac{V}{V^{\prime}}d\phi\,,
\end{equation}
where $\phi_{end}$ is defined by $\epsilon(\phi_{end})\sim 1$, provided
inflation comes to an end via a violation of the slow-roll conditions.
The number of e-foldings can be related to the Hubble crossing
mode $k=a_{k}H_{k}$ by comparing with the present Hubble length $a_{0}H_{0}$.
The final result is \cite{Liddle:2003as}
\begin{eqnarray}
\label{efoldsk}
N(k)= 62-\ln\frac{k}{a_0H_0}-\ln\frac{10^{16}{\rm GeV}}{V_{k}^{1/4}}+
\ln\frac{V_{k}^{1/4}}{V_{end}^{1/4}}
-\frac{1}{3}\ln\frac{V_{end}^{1/4}}
{\rho_{R}^{1/4}}\,,
\end{eqnarray}
where the subscripts $end$ ($\rm R$) refer to the end of inflation (end of
reheating). Today's Hubble length would correspond to $N_Q\equiv N(k=a_0H_0)$ number of 
e-foldings, whose actual value would depend on the equation of state, i.e. $\omega=p/\rho$ ($p$ denotes the pressure, 
$\rho$ denotes the energy density), from the end of 
inflation to radiation and matter dominated epochs.  A high scale inflation with a prompt reheating with relativistic species would yield
approximately, $N_Q\approx 50-60$. A significant modification can take place if the scale of inflation is low
\cite{Lyth:1995hj,Lyth:1995ka,Mazumdar:1999tk,Mazumdar:2001ya,ArkaniHamed:1999gq,Green:2002wk}.


\subsection{Density Perturbations}\label{DP}

\subsubsection{Scalar perturbations}\label{SP}

Small inhomogeneities in the scalar field, $\phi(\vec x,t) =\phi(t)+\delta\phi(\vec x,t)$, such that $\delta\phi\ll\phi$,
induce perturbations in the background metric, but the separation between the background metric
and a perturbed one is not unique. One needs to choose a gauge.  A simple choice would be to
fix a gauge where the non-relativistic limit of the full perturbed Einstein equation can be recast as
a Poisson equation with a Newtonian gravitational potential, $\Phi$. The induced metric
can be written as, e.g.~\cite{Mukhanov:1990me}:
\begin{equation}
\label{gauge}
ds^2=a^2(\tau)\left[(1+2\Phi)d\tau^2-(1-2\Psi)\delta_{ik}dx^{i}dx^{k}\right]\,,
\end{equation}
Only in the presence of Einstein gravity and when the spatial part of the energy momentum tensor is diagonal,
i.e. $\delta T^{i}_{j}=\delta^{i}_{j}$, it follows that $\Phi=\Psi$. 

During inflation the massless inflaton (with mass squared: $m^{2}\sim V^{''}\ll H^{2}$) perturbations, $\delta \phi$, are stretched outside the Hubble patch. 
One can track their perturbations from a sub-Hubble to that of a super-Hubble length scales. Right at the time when the wave numbers 
are crossing the Hubble patch, one finds a solution for $\delta\phi$ as
\begin{equation}
\label{pertphi}
\langle |\delta\phi_{k}|^2\rangle=({H(t_{\ast})^2}/{2k^3})\,,
\end{equation}
where $t_\ast$ denotes the instance of Hubble crossing. One can define a power spectrum for
the perturbed scalar field
\begin{equation}
\label{spect}
{\cal P}_{\phi}(k)=\frac{k^3}{2\pi^2}\langle |\delta\phi_{k}|^2\rangle=
\left[\frac{H(t_{\ast})}{2\pi}\right]^2 \equiv
\left. \left[\frac{H}{2\pi}\right]^2\right|_{k=aH}\,.
\end{equation}
Note that the phase of $\delta\phi_{k}$ can be arbitrary, and therefore,
inflation has generated a Gaussian perturbation. Now, one has to calculate the power spectrum for the metric perturbations.
For a critical density universe
\begin{equation}
\delta_{k}\equiv \left.\frac{\delta\rho}{\rho}\right|_{k}=-\frac{2}{3}
\left(\frac{k}{aH}\right)^2\Phi_{k}\,,
\end{equation}
where $\Phi_{k}(t)\approx (3/5)H(\delta \phi_{k}/\dot \phi)|_{k=aH}$.
Therefore, one obtains:
\begin{equation}
\label{pspect}
\delta_{k}^2\equiv \frac{4}{9}{\cal P}_{\Phi}(k)\, =\frac{4}{9}\frac{9}{25}
\left(\frac{H}{\dot\phi}\right)^2\left(\frac{H}{2\pi}\right)^2\,,
\end{equation}
where the right hand side can be evaluated at the time of horizon exit
$k=aH$. The temperature anisotropy seen by the observer in the matter
dominated epoch is proportional to the Newtonian potential, 
$\Delta T_{k}/T=-(1/3)\Phi_{k}$.

Besides tracking the perturbations in the longitudinal gauge with the help of Newtonian potential, there 
exists another useful gauge known as the comoving gauge. By definition, this choice of gauge requires a
comoving hypersurface on which the energy flux vanishes, and the relevant perturbation amplitude is known 
as the comoving curvature perturbation, $\zeta_{k}$~\cite{Lukash:1980iv,Mukhanov:1990me}. 
For the super-Hubble modes, $k\rightarrow 0$, the comoving curvature perturbation, 
$\zeta_{k}$ is a conserved quantity, and it is proportional
to the Newtonian potential, $\zeta_{k}=-(5/3)\Phi_{k}$. Therefore, 
$\delta_{k}$ can also be expressed in terms of
curvature perturbations \cite{Liddle:1993fq,Liddle:2000cg}
\begin{equation}
\delta_{k}=\frac{2}{5}\left(\frac{k}{aH}\right)^2{\zeta}_{k}\,,
\end{equation}
and the corresponding power spectrum
$\delta_{k}^2=({4}/{25}){\cal P}_{\zeta}(k)=({4}/{25})(H/\dot\phi)^2(H/2\pi)^2$.
With the help of the slow-roll equation $3H\dot\phi=-V^{\prime}$,
and the critical density formula $3H^2M^2_{\rm P}=V$, one obtains
\begin{eqnarray}\label{eq:powerspectrum1}
\delta_{k}^2 &\approx &\frac{1}{75\pi^2 M_{\rm P}^6}\frac{V^3}{V^{\prime 2}}\,
=\frac{1}{150\pi^2 M_{\rm P}^4}\frac{V}{\epsilon}\,\nonumber \\
{\cal P}_{\zeta}(k) &=& \frac{1}{24\pi^2M_{\rm P}^4}\frac{V}{\epsilon}\,,
\end{eqnarray}
where we have used the slow-roll parameter
$\epsilon\equiv (M_{\rm P}^2/2)(V^{\prime}/V)^2$.
The COBE satellite measured the CMB anisotropy and fixes the normalization
of ${\cal P}_{\zeta}(k)$ on very large scales. If we assume that the primordial 
spectrum can be approximated by a power law
(ignoring  the gravitational waves and the $k-$dependence of the power 
$n_s$)~\cite{Komatsu:2008hk}
\begin{equation}\label{eq:primordialpowerspectrum}
{\cal P}_{\zeta}(k)\simeq (2.445\pm 0.096)\times 10^{-9}\left(\frac{k}{k_0}\right)^{n_s-1}~,
\end{equation}
where $n_s$ is called the spectral index (or spectral tilt), the reference 
scale is: $k_0=7.5a_{0}H_{0}\sim 0.002~{\rm Mpc}^{-1}$, and the error bar on the 
normalization is given at $1\sigma$, and
\begin{equation}
 n_s(k_0)=0.960\pm 0.13
\end{equation}
It is important to stress that these central values and error bars vary 
significantly when other parameters are introduced to fit the data, in part 
because of degeneracies between parameters (in particular $n_s$ with 
$\Omega_b h^2$, the optical depth $\tau$, its running, the tensor-to-scalar ratio, $r$, and the 
fraction of cosmic strings). The spectral index $n(k)$ is defined as
\begin{equation}\label{spectind}
n(k)-1\equiv \frac{d\ln{\cal P}_{\zeta}}{d\ln k}\,.
\end{equation}
This definition is equivalent to the power law behavior if $n(k)$ is close to
a constant quantity over a range of $k$ of interest. One particular value of 
interest is $n_s\equiv n(k_0)$.  If $n_s=1$, the spectrum is flat and known as Harrison-Zeldovich 
spectrum~\cite{Harrison:1969fb,Zeldovich:1969sb}. For $n_s\neq 1$, the spectrum 
is tilted, and $n_s>1$ ($n_s<1$) is known as a blue (red) spectrum.
In the slow-roll approximation, this tilt can be expressed in terms of the 
slow-roll parameters and at first order:
\begin{equation}
\label{spectind4}
n_s-1=-6\epsilon +2\eta +\mathcal{O}(\epsilon^2,\eta^2,\epsilon\eta,\xi^2)~,
\end{equation}
where
\begin{equation}
\xi^2\equiv M_{\rm P}^4\frac{V^{\prime}(\mathrm{d}^3V/\mathrm{d}\phi^3)}{V^2}\,,
\quad\sigma^3\equiv M_{\rm P}^6\frac{V^{\prime 2}(\mathrm{d}^4V/\mathrm{d}\phi^4)}{V^3}\,.
\end{equation}
The running of these parameters are given by~\cite{Salopek:1990jq}. Since the
slow-roll inflation requires that $\epsilon \ll 1, |\eta|\ll 1$, therefore naturally predicts small
variation in the spectral index within $\Delta \ln k\approx 1$~\cite{Kosowsky:1995aa}
\begin{equation}
\frac{\mathrm{d} n(k)}{\mathrm{d}\ln k}=-16\epsilon\eta+24\epsilon^2+2\xi^2\,.
\end{equation}
It is possible to extend the calculation of metric perturbation beyond the
slow-roll approximations based on a formalism similar to that developed 
in Refs.~\cite{Mukhanov:1985rz,Sasaki:1986hm,Mukhanov:1989rq,Kolb:1995iv}.


\subsubsection{Multi-field perturbations}\label{Multi-field perturbations}

Inflation can proceed along many flat directions with many light fields. Their perturbations
can be tracked conveniently in a comoving gauge, on large scales 
${\zeta}=-H\delta\phi/\dot\phi$ remains a good conserved quantity, 
provided  each field follow slow-roll condition. The comoving curvature perturbations can be 
related to the number of e-foldings, $N$, given by
\cite{Salopek:1995vw,Sasaki:1995aw}
\begin{equation}
\label{multi1}
{\zeta} =\delta N=({\partial N}/{\partial \phi_{a}})\delta\phi_{a}\,,
\end{equation}
where $N$ is measured by a comoving observer while passing from flat
hypersurface (which defines $\delta\phi$) to the comoving hypersurface
(which determines $\zeta$). The repeated indices
are summed over and the subscript $a$ denotes a component of 
the inflaton~\cite{Lyth:1998xn,Lyth-new-book}.
If the random fluctuations along $\delta\phi_{a}$ have an amplitude $(H/2\pi)^2$, one obtains:
\begin{equation}
\label{multi5}
\delta_{k}^2=\frac{4}{25}{\cal P}_{\zeta}=\frac{V}{75\pi^2~M_{\rm P}^2}\frac{\partial N}{\partial\phi_{a}}
\frac{\partial N}{\partial\phi_{a}}\,.
\end{equation}
For a single component
$\partial N/\partial \phi\equiv (M_{\rm P}^{-2}V/ V^{\prime})$, and then
Eq.~(\ref{multi5}) reduces to Eq.~(\ref{eq:powerspectrum1}). By using 
slow-roll equations we can again define the spectral index
\begin{equation}
\label{multi6}
n-1=-\frac{M_{\rm P}^2V_{,a}V_{,a}}{V^2}-\frac{2}{M_{\rm P}^2 N_{,a}
N_{,a}}+2\frac{M_{\rm P}^2N_{,a}N_{,b}V_{,ab}}{V N_{,c}N_{,c}}\,,
\end{equation}
where $V_{,a}\equiv \partial V/\partial \phi_{a}$, and similarly
$N_{,a}\equiv\partial N/\partial \phi_{a}$. For a 
single component we recover Eq.~(\ref{spectind4}) from Eq.~(\ref{multi6}). 
In the case of multi-fields, one has to distinguish adiabatic from isocurvature
perturbations. Present CMB data rules out pure isocurvature perturbation 
spectrum~\cite{Beltran:2004uv,Komatsu:2008hk},
although a mixture of adiabatic and isocurvature perturbations remains
a possibility. 

\subsubsection{Gravitational waves}

During inflation stochastic gravitational 
waves are expected to be produced similar to the scalar perturbations
\cite{Grishchuk:1974ny,Grishchuk:1989ss,Allen:1987bk,Sahni:1990tx}. 
For reviews on gravitational waves, see
\cite{Mukhanov:1990me,Maggiore:1999vm}.
The gravitational wave perturbations are described by a line element
$ds^2+\delta ds^2$, where
\begin{equation}
{d}s^2=a^2(\tau)(\mathrm{d}\tau^2-\mathrm{d}x^{i}\mathrm{d}x_{i})\,,\quad \delta 
{d}s^2=-a^2(\tau)h_{ij} \mathrm{d}x^{i} \mathrm{d}x^{j}\,.
\end{equation}
The gauge invariant and conformally invariant $3$-tensor $h_{ij}$ is symmetric,
traceless $\delta^{ij}h_{ij}=0$, and divergenceless $\nabla_{i}h_{ij}=0$
($\nabla_{i}$ is a covariant derivative). Massless spin
$2$ gravitons have two transverse  degrees of freedom ({\it d.o.f})

For the Einstein gravity, the gravitational wave equation of motion follows
that of a massless Klein Gordon equation \cite{Mukhanov:1990me}. Especially,
for a flat universe
\begin{equation}
\ddot h^{i}_{j}+3H\dot h^{i}_{j}+\left({k^2}/{a^2}\right)h^{i}_{j}=0\,,
\end{equation}
As  any massless field, the gravitational waves also feel the quantum
fluctuations in an expanding background. The spectrum mimics that of
Eq.~(\ref{spect})
\begin{equation}
{\cal P}_{\rm grav}(k)=\left.\frac{2}{M_{\rm P}^2}\left(\frac{H}{2\pi}\right)^2
\right|_{k=aH}\,.
\end{equation}
Note that the spectrum has a Planck mass suppression, which suggests that the
amplitude of the gravitational waves is smaller compared to that of the
scalar perturbations. Therefore it is usually assumed that their
contribution to the CMB anisotropy is small. The corresponding
spectral index can be expanded in terms of the slow-roll parameters at first 
order as
\begin{equation}
r\equiv \frac{{\cal P}_{\rm grav}}
{ {\cal P}_{\zeta}}=16\epsilon~,\quad 
n_t=\frac{\mathrm{d}\ln{\cal P}_{\rm grav}(k)}{\mathrm{d}\ln k}\simeq -2\epsilon,\,.
\end{equation}
Note that the tensor spectral index is negative. It is expected that PLANCK could 
detect gravity waves if $r\gtrsim 0.1$, however the spectral index will be 
hard to measure in forthcoming experiments. The primordial gravity waves 
can be generated for large field value inflationary models.  Using the 
definition of the number of e-foldings it is possible to derive the range 
of $\Delta \phi$~\cite{Lyth-new-book,Lyth:1996im,Hotchkiss:2008sa})
\begin{equation}
16\epsilon = r<0.003\left({50}/{N}\right)^2\left({\Delta \phi}/
{M_{\rm P}}\right)~.
\end{equation}
Note that it is possible to get sizable, $r$, for $\Delta\phi \ll M_{\rm P}$ in 
assisted inflation (discussed below), and in inflection point inflation discussed in Ref.~\cite{BenDayan:2009kv}. 
If the tensor-to-scalar ratio $r$ and/or a running $\alpha_s$ are 
introduced, the best fit for $n_{s}$ and error bars~ (at $1\sigma$)
$n_s=1.017^{+0.042}_{-0.043}~,\quad \alpha_s=-0.028\pm 0.020$
$n_s=0.970\pm 0.015~,\quad r< 0.22~~(\mathrm{at}~2\sigma) $,
$n_s=1.089^{+0.070}_{-0.068}~,\quad r< 0.55~~(\mathrm{at}~2\sigma)~, \quad 
\alpha_s=-0.053 \pm 0.028$~\cite{Komatsu:2008hk}.
These data therefore suggest that a red spectrum is favored ($n_s=1$ 
excluded at $2.5\sigma$ from WMAP and at $3.1\sigma$ when other data sets are included) 
if there is no running.


\subsection{Generic models of inflation}

\subsubsection{High scale models of inflation}

The most general form for the potential of a gauge singlet scalar field $\phi$
contains an infinite number of terms,
\begin{equation}
\label{simplesinglepot}
V=V_0+\sum_{\alpha=2}^\infty \frac{\lambda_\alpha}{M_{\rm P}^{\alpha-4}}\phi^{\alpha}~.
\end{equation}
The renormalizable terms allows to prevent all terms
with $\alpha\geq 4$. By imposing the parity $Z_2$, under which $\phi\rightarrow -\phi$,
allows to prevent all terms with $\alpha$ odd. Most phenomenological models of
inflation proposed initially assume that one or two terms in
Eq.~(\ref{simplesinglepot}) dominate over the others, though some do contain an
infinite number of terms.


\vspace{-0.5cm}
\paragraph{\bf Power-law chaotic inflation:}
The simplest inflation model by the number of free parameters is perhaps the chaotic
inflation~\cite{Linde:1983gd} with the potential dominated by only one
of the terms in the above series
\begin{equation}\label{eq:potentialchaotic}
V=\frac{\lambda_\alpha}{M_{\rm P}^{\alpha-4}}\phi^{\alpha}~,
\end{equation}
with $\alpha$ a positive integer. The first two slow-roll parameters are
given by
\begin{equation}
\epsilon= \frac{\alpha^2}{2}\frac{M_{\rm P}^2}{\phi^2}~, \quad \quad
\eta =\alpha(\alpha-1)\frac{M_{\rm P}^2}{\phi^2}\,.
\end{equation}
Inflation ends when $\epsilon = 1$, reached for $\phi_e=\alpha M_{\rm P}/
\sqrt{2}$. The largest cosmological scale becomes super-Hubble when
$\phi_Q=\sqrt{2N_Q\alpha}M_{\rm P}$, which is super Planckian; this is
the first challenge for this class of models. The spectral index for
the scalar and tensor to scalar ratio read:
\begin{equation}
n_s=1-\frac{2+\alpha}{2N_Q+\alpha/2}~, \qquad
r=\frac{4\alpha}{N_Q+\alpha/4}\,.
\end{equation}
The amplitude of the density perturbations, if normalized at the COBE
scale, yields to extremely small coupling constants;
$\lambda_\alpha \ll 1$ (for i.e. $\lambda_4 \simeq 3.7\times 10^{-14}$).
The smallness of the coupling, $\lambda_\alpha / M_{\rm P}^{\alpha-4}$, is often
considered as an unnatural fine-tuning. Even when dimension full, for
example if $\alpha=2$, the generation (and the stability) of a mass scale,
$\sqrt{\lambda_2} M_{\rm P} \simeq 10^{13}$~GeV, is a challenge in theories
beyond the SM, as they require unnatural cancellations.  
These class of models have an interesting behavior for initial conditions
with a large phase space distribution where there exists a late attractor
trajectory leading to an end of inflation when the slow-roll conditions
are violated close to the Planck
scale~\cite{Linde:1983gd,Linde:1985ub,Brandenberger:1990wu,Kofman:2002cj}.

Note that the above mentioned monomial potential can be a good approximation
to describe in a certain field range  for various models of inflation
proposed and motivated from particle physics; natural inflation
when the inflaton is a pseudo-Goldstone boson \cite{Freese:1990rb},
or the Landau-Ginzburg potential when the inflaton is a Higgs-type
field~\cite{Bezrukov:2007ep}. The necessity of super Planckian
VEVs represents though a challenge to such embedding in particle physics and supergravity (SUGRA).

\vspace{-0.5cm}
\paragraph{\bf Exponential potential:}
An exponential potential also belongs to the large field models:
\begin{equation}
V(\phi)=V_{0}\exp\left(-\sqrt{\frac{2}{p}}\frac{\phi}{M_{\rm P}}\right)\,.
\label{exp-eq}
\end{equation}
It would give rise to a power law expansion $a(t)\propto t^p$,
so that inflation occurs when $p>1$. The case $p=2$ corresponds to the
exactly de Sitter evolution and a never ending accelerated expansion. Even for
$p\neq 2$, violation of slow-roll never takes place, since
$\epsilon(\phi)=1/p$ and inflation has to be ended by a phase transition
or gravitational production of particles~\cite{Lyth:1998xn,Copeland:2000hn}.

The confrontation to the CMB data yields: $n_s=1-2/p$ and $r=16/p$; the
model predicts a hight tensor to scalar ratio and it is within the one
sigma contour-plot of WMAP (with non-negligible $r$) for $p\in[73-133]$.


\subsubsection{Assisted inflation}

Many heavy fields could collectively assist inflation by increasing the effective Hubble friction term
for all the individual fields~\cite{Liddle:1998jc}. This idea can be illustrated with the help of $'m'$ identical 
scalar fields with an exponential potentials, see Eq.~(\ref{exp-eq}), where now $\phi \longrightarrow \phi_{i}$, where
$i=1,2,\cdots, m$. For a particular solution; where all the scalar fields are equal:
$\phi_1 = \phi_2 = \cdots = \phi_m$.
\begin{eqnarray}
H^2 & = & \frac{1}{3 M_{\rm P}^2}  m  
[ V(\phi_1) +   \dot{\phi}_1^2/2] \,, \\
\ddot{\phi}_1 & = & - 3 H \dot{\phi}_1 - {dV(\phi_1)}/{d\phi_1} \,.
\end{eqnarray}
These can be mapped to the equations of a model with a single scalar field
$\tilde{\phi}$ by the redefinitions
$\tilde{\phi}_1^2 = m \, \phi_1^2 \quad ; \quad  \tilde{V} = m\, V
    \quad ; \quad \tilde{p} = mp $,
so the expansion rate is $a \propto t^{\tilde{p}}$, provided that $\tilde{p} > 1/3$.
The expansion becomes quicker the more scalar fields there are. In particular,
potentials with $p < 1$, which for a single field are unable to support
inflation, can do so as long as there are enough scalar fields to make
$mp>1$.

In order to calculate the density perturbation produced in
multi-scalar field models, we recall the results from Eq.~(\ref{multi5}). Since $N = - \int H \, dt$, we have
$\sum_i \frac{\partial N}{\partial \phi_i} \dot{\phi}_i = -H$,  we yield:
${\cal P}_{\zeta} = ( {H}/{2\pi})^2 ({1}/{m})(H^2/\dot{\phi}_1^2)$.

Note that this last expression only contains one of the scalar fields, chosen
arbitrarily to be $\phi_1$. The estimation for the spectral tilt is given by :
$n -1 =-{2}/{mp}$, which matches that produced by a
single scalar field with $\tilde{p} = mp$. The more scalar fields there are,
the closer to scale-invariance is the spectrum that they produce. The above calculation can 
be repeated for arbitrary slopes, $p_{i}$. In which case the spectral tilt would have been given by
$n=1-2/\tilde p$, where $\tilde p=\sum p_i$. The above scenario has been generalized to
study arbitrary exponential potentials with couplings, 
$V=\sum^{n} z_s \exp(\sum^{m}\alpha_{sj}\phi_j)$~\cite{Copeland:1999cs,Green:1999vv}.


\vspace{-0.5cm}\paragraph{\bf Assisted chaotic inflation:}
Multi-scalar fields of chaotic type has interesting properties~\cite{Jokinen:2004bp}:
\begin{equation}
V\sim \sum_{i}f\left({\phi_{i}^{n}}/{M_{\rm P}^{n-4}}\right)
\end{equation}
(for $n\geq 4$). The chaotic inflation can now be driven at VEVs, $\phi_{i}\ll M_{\rm P}$, below the Planck
scale~\cite{Kanti:1999ie,Kanti:1999vt}. The {\it effective} slow-roll parameters
are given by: $\epsilon_{eff}=\epsilon/m \ll 1$ and $|\eta_{eff}|=|\eta|/m\ll 1$, where $\epsilon,~\eta$ are the slow-roll
parameters for the individual fields. Inflation can now occur for field VEVs~\cite{Jokinen:2004bp}:
\begin{equation}
\frac{\Delta \phi}{M_{\rm P}}\sim \left(\frac{600}{m}\right)\left(\frac{N_Q}{60}\right)\left(\frac{\epsilon_{eff}}{2}\right)^{1/2} \ll 1\,,
\end{equation}
where $N_Q$ is the number of e-foldings.
Obviously, all the properties of chaotic inflation can be preserved at VEVs $\ll M_{\rm P}$,
including the prediction for the tensor to scalar ration for the stochastic gravity waves, i.e. $r=16\epsilon_{eff}$.
For $\epsilon_{eff}\sim 0.01$ and $m\sim 100$, it is possible to realize a sub-Planckian inflation, the spectral
tilt close to the flatness:  $n_s-1=-6\epsilon_{eff}+2\eta_{eff}$, and large tensor to scalar ratio, i.e. $r=0.16$.


\vspace{-0.5cm}\paragraph{\bf N-flation:}
Amongst various realizations of assisted inflation, N-flation is perhaps the most interesting one.
The idea is to have $N\sim 300(M_{\rm P}/f)\sim 10^{4}$ number of axions, where $f$ is the axion decay constant, of order 
$f\sim 0.1 M_{\rm P}^{-1}$ drive inflation simultaneously with a leading order 
potential~\cite{Dimopoulos:2005ac}:
\begin{equation}
\label{axion}
V=V_0+\sum_{i}\Lambda_{i}^4\cos(\phi_{i}/f_{i})+...
\end{equation}
where $\phi_i$ are axion fields correspond to the partners of K\"ahler moduli. The ellipses contain higher order 
contributions. 
In a certain Type-IIB compactification, it is assumed 
that all the moduli are heavy and thus stabilized by prior dynamics, including that of the volume modulus. Only the 
axions of $T_i=\phi_i/f_i+iM_s^2R_i^2$ are light~\cite{Dimopoulos:2005ac}.  The assumption of 
decoupling the dynamics of K\"ahler modulus from the axions is still a debatable issue, see~\cite{Kallosh:2007ig}. 
After rearranging the potential for the axions, and expanding them around
their minima for a canonical choice of the kinetic terms, the Lagrangian simplifies to the lowest order in expansion:
\begin{equation}
{\cal L}=\frac{1}{2}\partial_{\mu}\phi_{i}\partial^{\mu}\phi_{j}-\sum_{i}\frac{1}{2}m_{i}^2\phi_{i}^2+\cdots \,.
\end{equation}
The exact calculation of $m_i$ is hard, assuming all of the mass terms to be the same $m_{i}\sim 10^{13}$~GeV, and
$N> (M_{\rm P}/f)^2$, it is possible to match the current observations with a tilt in the spectrum, $n\sim 0.97$,
and {\it large} tensor to scalar ratio: $r\sim 8/{ N}_Q\sim 0.13$ for ${N}_Q\sim 60$. There are also realizations of 
assisted inflation via branes~\cite{Mazumdar:2001mm,Cline:2005ty,Becker:2005sg}.


\subsubsection{Hybrid inflation}\label{sec:originalhybridinflation}

The end of inflation can happen via a waterfall triggered by a Higgs 
(not necessarily the SM Higgs) field coupled to the inflaton, first discussed 
in~\cite{Linde:1991km,Linde:1993cn,Copeland:1994vg}. The model is based on
the potential given by~\cite{Linde:1991km,Linde:1993cn}
\begin{equation} \label{eq:potenhyb2d}
V(\phi,\psi) = \frac{1}{2} m^2 \phi^2 + \frac{\lambda}{4} \left(\psi^2
- M^2 \right)^2 +\frac{\lambda'}{2} \phi^2 \psi^2~,
\end{equation}
where $\phi$ is the inflaton and $\psi$ is the Higgs-type field.
$\lambda$ and $\lambda'$ are two positive coupling constants, $m$
and $M$ are two mass parameters.  It is the most general form
(omitting a quartic term $\lambda'' \phi^4$) of renormalizable
potential satisfying the symmetries: $\psi \leftrightarrow -\psi$
and $\phi \leftrightarrow -\phi $. Inflation takes place along the
$\psi=0$ valley and ends with a tachyonic instability for the
Higgs-type field. The critical point of instability occurs at:
\begin{equation}
\phi_{\rm c} = M \sqrt{{\lambda}/{\lambda'}}\,.
\end{equation}
The system then evolves toward its true minimum at $V=0$,
$\langle\phi\rangle=0$, and $\langle\psi\rangle=\pm M$.

The inflationary valley, for $\langle\psi\rangle=0$, where the last $50-60$
e-foldings of inflation is assumed to take. This raises the issue of initial conditions
for $(\phi,\psi)$ fields and the fine tuning required to initiate inflation~\cite{Tetradis:1997kp,Lazarides:1997vv,Panagiotakopoulos:1997if,Mendes:2000sq,Clesse:2008pf}. In Ref.~\cite{Clesse:2008pf} it was found that when
the initial VEV of the inflaton, $\phi\ll M_{\rm P}$,  a subdominant
but non-negligible part of the initial conditions for the phase space leads to a successful
inflation, i.e. around less than $15\%$ depending on the model parameters.
Initial conditions with super-Planckian VEVs for $\phi\gg M_{\rm P}$ automatically leads to a 
successful inflation similarly to chaotic inflation.
In the inflationary valley, $\langle\psi\rangle=0$, the effective potential is given by:
\begin{equation}
V_{\rm eff}(\phi) \simeq \frac{\lambda M^4 }{4}+\frac{1}{2} m^2 \phi^2~,
\end{equation}
The model predicts a blue tilt in the spectrum, i.e. $n_s>1$, in the small field regime, 
$\phi_Q < M_{\rm P}$, which is slightly disfavored by the current data.

Two variations of the hybrid inflation idea were proposed assuming that the term $\phi^2$ is negligible. 
The two-field scalar potentials are of the form:
\begin{equation}\label{eq:models:shiftedvalleypotential}
V_{pq}(\phi,\psi)=M^4\left[1-\left(\frac{\phi_{\ast}}{\phi}\right)^p\right]^2+\lambda \phi^2\psi^q~.
\end{equation}
They share the common feature of having an inflationary trajectory during
which $\langle\psi\rangle$ is varying and not vanishing. For $(p,q)=(1,2)$, the model is known 
as {\bf Mutated hybrid}~\cite{Stewart:1994pt}, and $(p,q)=(4,6)$ corresponds to {\bf Smooth hybrid} inflation
~\cite{Lazarides:1995vr}.  The latter involves non-renormalizable terms of order $M_{\rm P}^{-2}$ to keep the potential
bounded from below. 

%
%
The potential is valid in the large field limit $\phi\gg \phi_*$, since in the
small field limit, the potential is not bounded from below and
should be completed. For mutated,  the model predicts a red spectral index
and negligible tensor to scalar ratio,
$n_s -1 \simeq -{3}/{(8N_Q)}\simeq 0.97$, and $
r\simeq {3m}/{(2\lambda N_Q^{3/2})}\ll {3}/{(8N_Q^2)}\sim 10^{-4}$,
if we assume $N_Q\simeq 60$. For smooth, the end of slow-roll inflation happens 
by a violation of the conditions; $\epsilon,\eta \ll 1$, since no waterfall transition
takes place. This allows the predictions for the spectral
index to be
$n_s -1 \simeq -{5}/{(3N_Q)}\simeq 0.97$~\cite{Lazarides:1995vr},
and the ratio for tensor to scalar is found to be negligible.

\subsubsection{Inflection point inflation}\label{SAISMI}

One of the challenges for inflation is to realize inflation at low scales, preferably below $M_{\rm P}$, with the right tilt and 
the amplitude of the power spectrum. Inflection point inflation admits a large amount of flexibility in the field space -- similar to the analogy of a ball 
rolling on an elastic surface following the least action principle. With the help of two independent parameters, $A$ and $B$, it is possible 
to obtain a large range of tilt in the spectrum, while keeping the amplitude of the perturbations intact. Let us consider a  simple realization of such a potential:
\begin{equation}
\label{genericpotential}
 V(\phi)=A\phi^2-C\phi^3+B\phi^4\,,
\end{equation}
where $C=f(A,~B)$ in order to obtain a {\it point of inflection} suitable for inflation. The VEV at which inflation occurs is intimately related to the two independent parameters and can happen at wide ranging scales below $M_{P}$, and for wide ranging values of $(A,~B)$.

Here we will generalize this potential to any generic potential $V$ which can be written in the following form (here ${\prime}$ denotes differentiation with respect to $\phi$)~\cite{Enqvist:2010vd,Hotchkiss:2011am}:
\begin{eqnarray}\label{steppot}
V= V_0 + a (\phi-\phi_0) + \frac{b}{2}(\phi-\phi_0)^2 + \frac{c}{6}
(\phi-\phi_0)^3 + \cdot\cdot\,  
\end{eqnarray}
where $V_0 \equiv V(\phi_0),~ a \equiv V^{\prime}(\phi_0),~ b \equiv V^{\prime \prime}(\phi_0),~ c \equiv V^{\prime \prime \prime}(\phi_0)$,
which is the Taylor expansion, truncated at $n=3$, around a reference point
$\phi_0$, which we choose to be the point of inflection where $V^{\prime \prime}(\phi_0) = 0$, or saddle point 
where  $V^{\prime\prime}(\phi_{0})=V^{\prime}(\phi_{0})=0$.
The higher order terms in Eq.~(\ref{steppot}) can be neglected during inflation, provided that
\begin{eqnarray}
\vert V^{\prime \prime \prime}_0 \vert \gg \left \vert \frac{d^m V}{d\phi^m}(\phi_0)\right \vert\, \vert \phi_e-\phi_0 \vert^{m-3},
\qquad m\geq 4\,, \label{constr2}
\end{eqnarray}
where $\phi_e$ corresponds to the field value at the end of inflation. Assuming that the slow-roll parameters are small in the 
vicinity of the inflection point $\phi_0$, and that the velocity ${\dot \phi}$ is negligible, the potential energy
$V_0$ gives rise to a period of inflation.

Inflation ends at the point $\phi_e$ where $\vert \eta \vert \sim 1$. By solving the equation of motion,
the number of e-foldings of inflation during the slow-roll motion of the inflaton from $\phi$ to $\phi_e$,
where $\phi_0 - (\phi_0 - \phi_e) < \phi < \phi_0 + (\phi_0 - \phi_e)$, is found to be~\cite{Enqvist:2010vd}
%
\begin{eqnarray}
\label{N2}
{N} &=& \frac{V_0}{M_{\rm P}^2} \sqrt{\frac{2}{{a}{c}}} \left[ F_0(\phi_e)-F_0(\phi)\right], \nonumber \\
F_0(z)&=&{\rm arccot} \left(\sqrt{\frac{{c}}{2{a}}} (z - \phi_0) \right) ~.
\end{eqnarray}
It useful to define the parameters  $X = \frac{{a} M_{\rm P}}{\sqrt{2}V_0}$ and 
$Y = \sqrt{\frac{{c}}{{a}}} {N} M_{\rm P} X$. Note that $X$ is the square root of the slow-roll parameter $\epsilon$ at the point of inflection.
The slow-roll parameters can then be recast in the following form:
%
\begin{eqnarray}
\epsilon &=& \frac{2V_0^2}{c^2 M_{\rm P}^6 {N}^4} \left( \frac{Y}{S} \right)^4 \, , \label{epsilon3} \\
\eta &=& -\frac{2}{{N}} \, \frac{Y}{S} \left( {\sqrt{1-X} \cos Y - \sqrt{X} \sin Y}\right)\,, \label{eta3} \\
\xi^2 &=& \frac{2}{{ N}^2}  \left( \frac{Y }{S} \right)^2\, \label{xi3}
\end{eqnarray}
where $S=\sqrt{1-X} \sin Y + \sqrt{X} \cos Y$.
One can solve Eqs.~(\ref{epsilon3}-\ref{xi3}), for $X$, $Y$ and ${N}$ in terms of the slow-roll. The equations
are non-linear and in general cannot be solved analytically. However, since $\epsilon \ll \vert \eta \vert$,~$\xi$,
one can find a closed form solution provided that $V_0^{1/4} \leq 10^{16}$~GeV and $X \leq \sqrt{\epsilon} \ll 1$~\cite{Allahverdi:2006we,BuenoSanchez:2006xk,Enqvist:2010vd,Hotchkiss:2011am}:
\begin{eqnarray}
{\cal P}^{1/2}_{\zeta} &\equiv& \frac{1}{\sqrt{24 \pi^2}} \frac{V^{1/2}_0}{\epsilon^{1/2} M^2_{\rm P}} =  \frac{V_0^{1/2}}{2\pi\sqrt{6} M_{\rm P}^2 X} \sin^2Y\, ,  \label{amplitude2} \\
n_s &\equiv& 1 + 2 \eta - 6 \epsilon = 1-\frac{4}{{N}_{Q}} \, Y\cot Y\, , \label{ns2} \\
\alpha &=& -\frac{4}{{N}^2_{Q}} \left( \frac{Y}{\sin Y} \right)^2\,. \label{dns2}
\end{eqnarray}
 One can derive the properties of a saddle point inflation provided $Y/\sin Y\rightarrow 1$, and $Y\cot Y\rightarrow 1$. The model favors the 
current observations by matching the COBE normalization and the spectral tilt ranging from $n_{s}\in [0.93,1.0]$. For instance, the lowest 
value corresponds to the saddle point inflation for $N_{Q}\sim 60$.


\subsection{Supersymmetric models}\label{SUSY-models}

One of the most compelling virtues of SUSY is that it can protect the quadratically divergent contributions to the scalar 
mass, which arise in one-loop computation from the fermion contribution and quartic self interaction of the scalar field. 
Such corrections generically spoil the flatness of the inflaton potential. 
The quadratic divergence is independent of the mass of the scalar field and cancel, exactly 
if $\lambda_s=\lambda_f^2$, where $\lambda_f$ is the fermion Yukawa and $\lambda_s$  is the 
quartic scalar coupling. However this procedure fails at 2-loops and one requires fine tuning of
the couplings order by order in perturbation theory. In the case of the SM Higgs, a precision of roughly one part in $10^{17}$ is 
required in couplings to maintain the Higgs potential,  often known as the {\it hierarchy problem} or 
the {\it naturalness problem}. The electroweak symmetry is still broken by the Higgs 
mechanism, but the quadratic divergences in the scalar sector are absent.  In the SUSY limit 
the fermion and scalar masses are degenerate, but the SUSY has to be broken {\it softly} at the TeV scale 
in such a way that it does not spoil the solution to the hierarchy problem, see~\cite{Nilles:1983ge,Haber:1984rc, Martin:1997ns,Chung:2003fi}.

The matter fields for $N=1$ SUSY are chiral superfields 
$\Phi=\phi+\sqrt{2}\theta\psi+\theta\theta F$, which
describe a scalar $\phi$, a fermion $\psi$ and a scalar auxiliary field $F$.  The 
SUSY scalar potential $V$ is the sum of the $F$- and $D$-terms:
\begin{eqnarray}
\label{fplusd}
V= \sum_i |F_i|^2+\frac 12 \sum_a g_a^2D^aD^a\,, \nonumber \\
F_i\equiv {\partial W\over \partial \phi_i},~~D^a=\phi^\dagger T^a \phi~\,,
\label{fddefs}
\end{eqnarray}
where $W$ is the superpotential, and $\phi_i$ transforms 
under a gauge group $G$ with the generators of the Lie algebra given by $T^{a}$. Note that all the
kinetic energy terms are included in the $D$-terms. For inflation, the effects of supergravity (SUGRA) becomes important.
At tree level, $N=1$ SUGRA potential is
given by the sum of $F$ and $D$-terms, see \cite{Nilles:1983ge}
\begin{eqnarray}
\label{eq:partphys:sugrapotential}
V=e^{\frac{K(\phi_i,\phi^{\ast i}) }{M_{\rm P}^2}}\left[\left(K^{-1}\right)^{j}_{i}F_{i}F^{j}-\frac{3
|W|^2}{M_{\rm P}^2}\right]+\nonumber \\
\frac{g^2}{2}{\rm Re}f^{-1}_{ab}{\hat D}^{a}{\hat D}^{b}\,, \\
F^{i}=W^{i}+K^{i}\frac{W}{M_{\rm P}^2}\,,~~
{\hat D}^{a}=-K^{i}(T^{a})^{j}_{i}\phi_{j}+\xi^{a}\,.
\end{eqnarray}
where we have added the Fayet-Iliopoulos contribution $\xi^{a}$ to the
$D$-term, and $\hat D^{a}=D^{a}/g^{a}$, where $g^a$ is gauge coupling.
Here $K(\phi_i,\phi^{\ast i})$ is the K\"ahler potential, which is a function of the
fields $\phi_i$, and $K^i\equiv \partial K/\partial \phi_i$. In the simplest case, at tree-level
$K=\phi^{\ast i}\phi_{i}$ (and $K^{j}_{i}=(K^{-1})^{j}_{i}=\delta^{j}_{i}$). In general  the 
K\"ahler potential can be expanded as: 
$K=\phi_{i}\phi^{\ast i}+ (k^{ij}_k \phi_i\phi_j\phi^{\ast k}+c.c.)/M_{\rm P}+
(k^{ij}_{kl}\phi_i \phi_j\phi^{\ast k}\phi^{\ast l}\phi^{\ast k}\phi^{\ast l}+k^{ijk}_{l}\phi_i\phi_j\phi_k\phi^{\ast l}
+c.c.)/M_{\rm P}^2+\cdots )$.  The kinetic terms for the scalars 
take the form: 
\begin{equation}
\frac{\partial^2K}{\partial \phi_{i}\partial \phi_{j}^{\ast}}D_{\mu}\phi_{i} D^{\mu}\phi^{\ast}_{j}\,.
\end{equation}
The real part of the gauge kinetic function matrix is given by ${\rm Re}f_{ab}$. 
In general, $f_{ab}=\delta_{ab}(1/g_a^2+f^{i}_{a}\phi_{i}/M_{\rm P}+\cdots)$. The gauginos masses are typically 
given by $m_{\lambda^{a}}={\rm Re}[f^{i}_{a}]\langle F_{i}\rangle/2M_{\rm P}$. For a universal gaugino
masses, $f^{i}_{a}$ are the same for all the three gauge groups of MSSM.
In the simplest case, it is just a constant, 
$f_{ab}=\delta_{ab}/g_a^2$, and the kinetic terms for the gauge potentials, $A^{a}_{\mu}$, are given by: 
\begin{equation}
\frac{1}{4}({\rm Re} f_{ab})F^{a}_{\mu\nu}F_{a}^{\mu\nu}\,.
\end{equation}
SUGRA will be broken if one or more of the $F_{i}$ obtain a VEV. The gravitino, spin $\pm 3/2$ 
component of the graviton, then absorb the Goldstino component to become massive. Requiring
classically $\langle V\rangle=0$, as a constraint to obtain the zero cosmological constant,
one obtains 
\begin{equation}
m_{3/2}^2=\frac{\langle K^{i}_{j}F_{i}F^{\ast j}\rangle }{3M_{\rm P}^2}=e^{\langle K\rangle/M_{\rm P}^2}
\frac{|\langle W\rangle|^2}{M_{\rm P}^4}\,.
\end{equation}
 %


\subsubsection{F-term inflation}\label{sec:ftermhybridinflation}


The most well-known model of SUSY inflation driven by $F$-terms is of the
hybrid type and based on the superpotential~\cite{Copeland:1994vg,Dvali:1994ms,Linde:1997sj}
\begin{equation}
\label{eq:superpotentialFterm}
W=\kappa S(\Phi \overline{\Phi}-M^2)~.
\end{equation}
where, $S$ is an absolute gauge singlet, while $\Phi$ and
$\overline{\Phi}$ are two distinct superfields belonging to complex
conjugate representation, and $\kappa$ is an arbitrary constant fixed by the
CMB observations. It is desirable to obtain an effective singlet $S$ superfield 
arising from a higher gauge theory such as GUT~\cite{Langacker:1980js}, however to our knowledge 
it has not been possible to implement this idea, see the discussion in~\cite{Mazumdar:2010sa}. Typically $S$ would 
have other (self)couplings which would effectively ruin the flatness required for hybrid inflation. 

This form of potential is protected from additional destabilizing
contributions with higher power of $S$, if $S$, $\Phi$ and $\overline{\Phi}$
carrying respectively the charges $+2$, $\alpha$ and $-\alpha$ under
R-parity. Then $W$ carries a charge $+2$ so that the action
$\mathcal{S}=\int \mathrm{d}^2\theta \; W +\dots$ is invariant.

The tree level scalar potential derived from Eq.~(\ref{eq:superpotentialFterm}) reads
\begin{equation}\label{eq:potentialFterm3field}
V_{\rm tree}(S,\phi,\overline{\phi})= \kappa^2|M^2-\overline{\phi}\phi|^2+\kappa^2|S|^2
(|\phi|^2+|\overline{\phi}|^2)^2,
\end{equation}
where we have denoted by $S,\phi,\overline{\phi}$ the scalar components
of $S,\Phi,\overline{\Phi}$. Note the similarity between Eq.~(\ref{eq:potenhyb2d}) and
Eq.~(\ref{eq:potentialFterm3field}), where $m=0$, and both $\lambda$ and $\lambda'$ are
replaced by only $\kappa^2$. We will also assume $\phi^*=\overline{\phi}$ along this direction, and 
the kinetic terms for the superfields are minimal, i.e. with a k\"ahler potential:
$K=|S|^2+|\Phi|^2+|\overline{\Phi}|^2$.

Let us define two effective
real scalar fields canonically normalized, $\sigma\equiv \sqrt{2}
\mathrm{Re}(S)$, and $\psi\equiv 2\mathrm{Re}(\Phi)=2\mathrm{Re}
(\overline{\Phi})$, the overall potential can then be recast as:
\begin{equation}\label{eq:potentialFterm2field}
V_{\rm tree}(\sigma,\psi)= \kappa^2\left(M^2-\frac{\psi^2}{4}\right)^2+\frac{\kappa^2}{4}
\sigma^2\psi^2~.
\end{equation}
The global minimum of the potential is
located at $S=0$, $\phi\overline{\phi}=M^2$. At large VEVs,
$S > S_c\equiv M$, the potential also possesses a local valley
of minima (at $\langle\psi\rangle = 0$) in which the field $\sigma$,
now rolls on with $V_{\rm tree}=V_0\equiv \kappa^2 M^4$. This non-vanishing value of
the potential both sustain the inflationary dynamics and induces a
SUSY breaking. 

 This induces a splitting in the mass of the
fermionic and bosonic components of the superfields $\Phi$ and
$\overline{\Phi}$, with $m^2_B(S)=\kappa^2 |S|^2 \pm \kappa^2 M^2$ and
$m^2_F=\kappa^2 |S|^2$. Note that this description is valid only as
long as $S$ is sufficiently slow-rolling such that $\kappa^2 |S|^2 |\Phi|^2$
can be considered as a mass term.
Therefore radiative corrections do not exactly cancel
out~\cite{Dvali:1994ms,Lazarides:2000ck}, and provide a one-loop potential
which can be calculated by using the Coleman-Weinberg formula~\cite{Coleman:1973jx}, 
\begin{eqnarray}\label{induced-one-loop}
V_{\rm 1-loop}(\phi) &=&V_{inf}(\phi)+\Delta V\,\nonumber \\
\Delta V &=& \frac{1}{64\pi^2}\sum_{i}(-)^{F_i}M_{i}(\phi)^4\ln\frac{M_{i}(\phi)^2}{\Lambda(\phi)^2}\,,
\end{eqnarray}
where $V_{inf}$ is now the renormalized potential, $\Lambda(\phi)$ is the
renormalization mass scale. The sum extends over all helicity states $i$, $F_{i}$ is the fermion number, and $M(\phi)$
is the mass of the i-th state. One obtains:
\begin{eqnarray}\label{eq:effectivepotentialFtermvalley}
V_{\rm 1-loop}(S)=\frac{\kappa^4\mathcal{N}M^4}{32\pi^2}\left[2\ln
\frac{s^2\kappa^2}{\Lambda^2}+(z+1)^2\ln(1+z^{-1})\right. \nonumber \\
\left. + (z-1)^2\ln(1-z^{-1})\right]~,
\end{eqnarray}
where $z={|S|^2}/{M^2}\equiv x^2$, $\Lambda$ represents a non-physical energy scale of renormalization and
$\mathcal{N}$ denotes the dimensionality.  Note that the perturbative approach of Coleman and Weinberg breaks down when
close to the inflection point at $z\simeq 1$. For small coupling $\kappa$, the slow-roll conditions
(for $\eta$) are violated infinitely close to the critical point, $z=1$, which
ends inflation. 

The normalization to COBE allows to fix the scale $M$ as a function of
$\kappa$. If the breaking of $G$ does not produce cosmic strings, the
contribution to the quadrupole anisotropy simply comes from the
inflationary contribution (see Eq.~(\ref{eq:powerspectrum1})) and the
observed value can be obtained even with a coupling $\kappa$ close to
unity \cite{Dvali:1994ms}. Small values of $\kappa$ can render
the scale of inflation very low, as low as the TeV
scale~\cite{Randall:1995dj,Randall:1994fr,BasteroGil:1997vn,BasteroGil:2002xs}.

However it has been shown that the formation of cosmic strings at the end of
$F$-term inflation is highly probable when the model is embedded in SUSY
GUTs~\cite{Jeannerot:2003qv}. In this case, the normalization to COBE
receives two contributions, one from inflation and other from cosmic strings~\cite{Jeannerot:1997is,Rocher:2004my}, 
which affects the relation $M(\kappa)$ at large $\kappa$, and imposes
new stringent bounds on $M\lesssim 2\times 10^{15}$ GeV,
and~\cite{Rocher:2004et,Jeannerot:2005mc}
\begin{equation}\label{eq:constraintFtermcoupling}
\kappa \lesssim 7\times 10^{-7}({126}/{N_{Q}})~,
\end{equation}
by demanding that the cosmic strings cam at best contribute less than $\lesssim 10\%$  of 
isocurvature fluctuations \cite{Bevis:2007gh}. Once $M$ is fixed, the spectral index $n_s$ can be computed 
as the range is found to be: $n_s\in [0.98,1]$ whether cosmic strings form or not~\cite{Senoguz:2003zw,Jeannerot:2005mc}, and 
by including the soft-SUSY breaking terms within minimal kinetic terms in the K\"ahler potential, 
the spectral index can be brought down to  $0.928\leq n_s \leq 1.008$~\cite{Rehman:2009nq}.


\subsubsection{SUGRA corrections and solutions}\label{SUGRA-eta}

For inflaton VEVs below the Planck scale, the SUGRA effects can become important 
and may ruin the flatness of the potential. The $N=1$ SUGRA potential is now given by
Eq.~(\ref{eq:partphys:sugrapotential}), where the $F$-terms containing an
additional exponential factor. Various cross terms between the K\"ahler and the superpotential
leads to the soft breaking mass term for the light scalar fields
\cite{Dine:1983ys,Bertolami:1987xb,Copeland:1994vg,Dine:1995kz,Dine:1995uk,Linde:1997sj}
\begin{equation}
\label{msoftH}
m^2_{SUGRA} \sim m^2_{susy}+\frac{V}{3M_{\rm P}^2} \sim {\cal O}(1)H^2\,,
\end{equation}
where $m_{susy}\sim {\cal O}(100)$~GeV  contains soft-SUSY breaking mass term for the low scale
SUSY breaking scenarios. Once the inflaton gets a mass $\sim H$, the contribution to the second
slow-roll parameter $\eta$ becomes order unity and the slow roll inflation ends, i.e.
$ |\eta| \equiv {M_{\rm P}^2}{V^{\prime \prime}}/{V} \sim {m^2_{SUGRA}}/{H^2} \sim {\cal O}(1) $.
This is known as the SUGRA-$\eta$ problem.

When there are more than one chiral superfields, as in the $F$-term hybrid
model, it can be possible to cancel the dominant ${\cal O}(1)H$ correction
to the inflaton mass by choosing an appropriate K\"ahler term \cite{Stewart:1994ts,Copeland:1994vg}. 
For non-minimal K\"ahler potentials, such as
\begin{equation}\label{eq:models:FtermKahlerNonMin}
K =|S|^2 +|\Phi|^2+|\overline{\Phi}|^2+\kappa_S |S|^4/M_{\rm P}^2+\dots~,
\end{equation}
the kinetic terms $K^{ij}\partial_\mu \Phi_i\partial^\mu \Phi_j^*$ are
non-minimal because $K^{ij}\neq \delta^{ij}$. One obtains:
$\left(\partial_{SS^{*}}K\right)^{-1}\sim 1-4\kappa_S |S|^2/M_{\rm P}^2+\dots $
One again obtains a problematic contribution to the inflaton mass, i.e. $\kappa_S\times {\cal O}(1)H$.
Several mechanisms have been proposed to tackle this $\eta$-problem. One can impose, $\kappa_S\sim 10^{-3}$, which is 
sufficient to keep the slow roll inflation safe, but without much physical motivation. For a generic inflationary model it is not
possible to compute $\kappa_s$ at all.


\vspace{-0.5cm}\paragraph{\bf Shift and Heisenberg symmetry:}
 Safe non-minimal K\"ahler potentials have also been
proposed~\cite{Brax:2005jv,Antusch:2009ef,Pallis:2009pq,BasteroGil:1998te} making use of
the shift symmetry. Under this symmetry, a superfield
$S\rightarrow S+iC$, where $C$ is a constant.~\cite{Kawasaki:2000yn,Kawasaki:2000ws} to
protect the K\"ahler potential of the form $K(\Phi,\bar{\Phi})
\rightarrow K(\Phi+\bar{\Phi})$. This symmetry generates an exactly flat direction for an inflaton field and a
non-invariance of the superpotential induces some slope to its potential
to allow slow-roll at the loop level. Another symmetry - the Heisenberg symmetry - has also been invoked to
protect the form of the K\"ahler potential~\cite{Antusch:2008pn}, where the effective K\"ahler is a no-scale type
of the form $K=\ln (\Phi_i)$. This solves the SUGRA-$\eta$-problem by
canceling the exponential term $\exp (K)$. However note that K\"ahler potentials generically obtains 
quantum corrections unlike the {\it non-renormalization theorem} which 
can only protect the superpotential terms~\cite{Grisaru:1979wc}. Such corrections are hard to compute 
without knowing the ultra-violet completion, and the exact matter sector for the inflationary model~\cite{Berg:2005yu,Berg:2004ek,Berg:2005ja}.
Note that none of these papers considered MSSM matter sector.

\vspace{-0.5cm}\paragraph{\bf  Inflection point inflation:}\label{SUGRA-inflection}
For any smooth potential, it is possible to drive inflation near the saddle point, $V'=0,~V''=0, V'''\neq 0$, or
near the point of inflection,~$V'\neq 0, V''=0, V'''\neq 0$. These are special points where the effective mass term
of the inflaton vanishes and the potential does not suffer through SUGRA-$\eta$ 
problem~\cite{Allahverdi:2006iq,Allahverdi:2006we,Mazumdar:2011ih}. In the saddle point case the potential can be made so flat 
that inflation can be driven eternally~\cite{Allahverdi:2006iq,Allahverdi:2007wh}. 

\begin{figure}[t]
\includegraphics[scale=0.3]{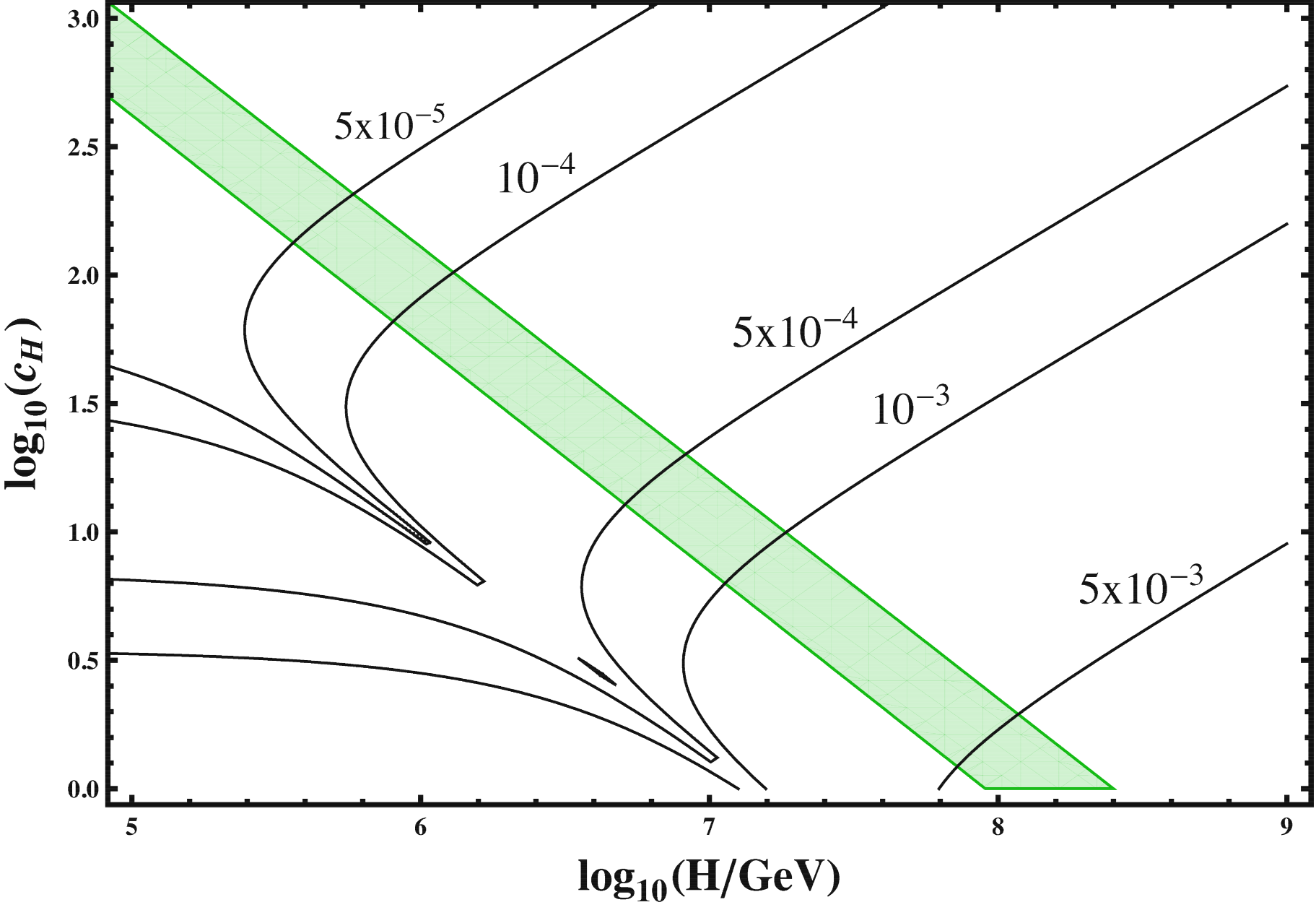}
\caption{Regions of parameter space (green) for the potential in Eq.~\eqref{superHubblepot} that satisfy the WMAP 7-year  constraints on 
$(\delta_{H},~n_{s})$.  Contour lines of $\vert\beta_\mathrm{CMB}\vert$ are shown in black, for the values of $\vert\beta_\mathrm{CMB}\vert$ indicated for $n=6$ case. For ${\cal O}(1)H$ correction to the inflaton mass the value of $\beta \sim10^{-2}$~\cite{Mazumdar:2011ih}.}
\label{figure:c_H-H}
\end{figure}

From the  low-energy perspective, the most generic and dangerous SUGRA corrections to the inflaton potential ( with minimal and non-minimal K\"ahler potentials for $\phi$ ) would have a large vacuum energy contribution. To complicate further, one may even assume that the flatness of $\phi$ is lifted by non-renormalizable contribution to the potential~\cite{Mazumdar:2011ih}:
\begin{eqnarray}
\label{superHubblepot}
V(\phi) = V_{c}+\frac{c_H H^{2}}{2}{\left\vert \phi \right\vert}^2 - \frac{a_H H}{nM_P^{n-3}}\phi^n\,
+ \frac{\left\vert \phi \right\vert^{2(n-1)}}{M_P^{2(n-3)}}\,,
\end{eqnarray}
where $V_{c}=3H^{2}M_{\rm P}^{2}$.
As mentioned above the interesting observation is that, in fact, there always exists a range of field values, $\Delta \phi$, for which a full potential
admits a {\it point of inflection} with all known sources of corrections taken into account. Now, {\it all} the uncertainties in the corrections to the 
K\"ahler potential can be absorbed in the full potential, such that the flat region admits a slow roll inflation with $\Delta \phi \gg \phi_{0}$~\cite{Mazumdar:2011ih}. The condition for this inflection point  is $a_H^2\approx8(n-1)c_H$, where we 
characterize the fine-tuning by $\beta$ defined as: 
\begin{equation}
\frac{a_H^2}{8(n-1)c_H} = 1-\frac{(n-2)^2}{4}\beta^2.
\end{equation}
When $\vert\beta\vert$ is small, a point of inflection $\phi_0$ exists such that $V^{\prime\prime}\left(\phi_0\right) =0$, with
\begin{equation}
\phi_0 = \left(\sqrt{\frac{c_H}{2(n-1)}} H M_{\rm P}^{n-3}\right)^{\frac{1}{(n-2)}}\,.
\end{equation}
We can Taylor expand the potential about $\phi_0$ as discussed in section~\ref{SAISMI}, and analyze 
the CMB constraints as shown in Fig.~\ref{figure:c_H-H}.

\subsubsection{D-term inflation}

In Refs.~\cite{Stewart:1994ts,Binetruy:1996xj,Halyo:1996pp}, it was noticed that a perfectly flat inflaton potentials can be constructed using
a constant contribution coming from the $D$-term. In addition, the  SUGRA-$\eta$
problem arising in $F$-term models does not appear for $D$-terms driven
inflation  because the $D$-sector of the potential does not receive exponential contributions
from non-minimal SUGRA. The model however
requires the presence of a Fayet-Iliopoulos (FI) term $\xi$, and 
therefore a $U(1)_\xi$ symmetry which generates it.
For a K\"ahler potential $K(\Phi_m,\Phi_n)$, the
$D$-terms $$D^{a}=-g_{a} [D_a=\phi_i {(T_a)^i}_j K^j+\xi_a]$$ (where
$K^m\equiv \partial K/\partial \Phi_m$) give rise to a scalar potential:
\begin{equation}
V(\phi,\phi^{\ast})=\frac{1}{2}[\mathrm{Re} f(\phi)]^{-1}
\sum D^{a}D_{a}+\mathrm{F-terms}\,
\end{equation}
where $g_{a}$ and $T^{a}$ are respectively the gauge coupling constants and
the generators of each factors of the symmetry of the action, $'a'$ running
over all factors of the symmetry, and $f(\phi)$ is the gauge kinetic
function. If this symmetry contains a factor $U(1)_\xi$, the most general action then allows for the
presence of a constant contribution $\xi$. 

The simplest realization of $D$-term inflation reproduces the hybrid
potential with three chiral superfields, $S$, $\phi_{+}$, and $\phi_{-}$
with non-anomalous $U(1)_\xi$ (an abelian theory is said to be anomalous if
the trace of the generator is non-vanishing $\sum q_n\neq 0$) charges
$q_n=0,+1,-1$~\cite{Binetruy:1996xj,Halyo:1996pp}. The
superpotential can be written as
\begin{equation}\label{eq:dtermsuperpotential}
W^D = \lambda S\phi_{+}\phi_{-}\,.
\end{equation}
In what follows, we assume the
minimal structure for $f(\Phi_i)$ (i.e., $f(\Phi_i)$=1) and take the
minimal K\"ahler potential, i.e. $K=|\phi_-|^2
+|\phi_+|^2+|S|^2$.. Then the scalar potential reads
\begin{equation}\label{DpotenSUGRAtot}
\begin{split}
&V^{\rm D-SUGRA}_{\rm tree}=
\lambda^2 \exp\left({\frac{|\phi_-|^2+|\phi_+|^2+|S|^2}{M^2_{\rm
P}}}\right)\, \\
&\left[|\phi_+\phi_-|^2\left(1+\frac{|S|^4}{M^4_{\rm
P}}\right)+|\phi_+S|^2 \left(1+\frac{|\phi_-|^4}{M^4_{\rm
P}}\right) \right.\\
&\left. +|\phi_-S|^2 \left(1+\frac{|\phi_+|^4}{M^4_{\rm P}}\right)
+3\frac{|\phi_-\phi_+S|^2}{M^2_{\rm P}}\right] +\\
&\frac{g_\xi^2}{2}\left(|\phi_+|^2-|\phi_-|^2+\xi\right)^2~,
\end{split}
\end{equation}
where $g_\xi$ is the gauge coupling of $U(1)_\xi$.
The global minimum of the potential is obtained
for $\langle S\rangle =0$ and $\langle\Phi_-\rangle =\sqrt {\xi}$,
which is SUSY preserving but induces the breaking of $U(1)_\xi$.
For $S>S_{inst}\equiv g_\xi\sqrt{\xi}/\lambda$ the potential is
minimized for $|\phi_+|=|\phi_-|=0$ and therefore, at the tree level,
the potential exhibits a flat inflationary valley, with vacuum energy
$V_0=g_\xi^2\xi^2/2$. 

The radiative corrections depend on the splitting
between the effective masses of the components of the superfields
$\Phi_+$ and $\Phi_-$, because of the transient $D$-term SUSY breaking.
The radiative corrections are given by the Coleman-Weinberg
expression~\cite{Coleman:1973jx} and the full potential inside the
inflationary valley reads
\begin{equation}\label{eq:VexactDsugrainsidevalley}
\begin{split}
&V^{\rm D-SUGRA}_{\rm eff}=
\frac{g_\xi^2\xi^2}{2} \Bigg\{1+\frac{g_\xi^2}{16\pi^2}
\Bigg[2\ln \frac{\lambda^2|S|^2}{\Lambda^2} \exp
\left(\frac{|S|^2}{M_{\rm P}^2}\right)+\\
&(z+1)^2\ln(1+z^{-1})+(z-1)^2\ln(1-z^{-1})\Bigg]\Bigg\}~,
\end{split}
\end{equation}
with $z=({\lambda^2 |S|^2 }/{g_\xi^2\xi}) e^{{|S|^2}/{M^2_{\rm
P}}}$.
Inflation ends when the slow-roll conditions break down, that is for
$z_{\rm end}\simeq 1$, and the predictions for the inflationary
parameters are very similar to the previous discussion on $F$-term
inflation.


\subsection{MSSM flat direction inflation}

So far we have discussed inflationary models where the inflaton sector belongs to the hidden sector
(not charged under the SM gauge group), such models will have at least one SM {\it gauge singlet} component, 
whose couplings to other fields and mass are chosen {\it just} to match the CMB observations. These models are simple
but lack proper embedding within MSSM or its minimal extensions.

In order to construct a predictable hidden sector model of inflation, one must know all the inflaton 
couplings to the {\it hidden} and {\it visible} matter. One such unique model has been constructed within type IIB string theory, where it was 
found that all the inflaton energy is transferred to exciting the {\it hidden} matter~\cite{Cicoli:2010ha,Cicoli:2010yj}, and the universe could be 
prematurely dominated by the hidden sector dark matter. Such obstacles do not arise if the last phase of inflation occurs within MSSM.


\subsubsection{Introducing MSSM and its flat directions}

In addition to the usual quark and lepton superfields,
MSSM has two Higgs fields, $H_u$ and $H_d$. Two Higgses are needed
because $H^\dagger$ is forbidden in the superpotential. 
The superpotential for the MSSM is given by, see~\cite{Nilles:1983ge,Haber:1984rc, Martin:1997ns,Chung:2003fi}
\begin{equation}
\label{mssm}
W_{MSSM}=\lambda_uQH_u u+\lambda_dQH_d d+\lambda_eLH_d e~
+\mu H_uH_d\,,
\end{equation}
where $H_{u}, H_{d}, Q, L, u, d, e$ in
Eq.~(\ref{mssm}) are chiral superfields, and the dimensionless Yukawa couplings
$\lambda_{u}, \lambda_{d}, \lambda_{e}$ are $3\times 3$ matrices in the family
space. We have suppressed the gauge and family indices. The $H_{u}, H_{d}, Q, L$ fields are
$SU(2)$ doublets, while $u, d, e$ are  $SU(2)$ singlets. The last term is the $\mu$
term, which is a SUSY version of the SM Higgs boson mass.
Terms proportional to $H_{u}^{\ast}H_{u}$ or $H^{\ast}_{d}H_{d}$ are
forbidden in the superpotential, since $W_{MSSM}$ must be analytic in the
chiral fields. $H_{u}$ and $H_{d}$ are required not only because they  give
masses to all the quarks and leptons, but also for the cancellation of
gauge anomalies. The Yukawa matrices determine the masses and CKM mixing
angles of the ordinary quarks and leptons through the neutral components of
$H_{u}=(H^{+}_{u},H^{0}_{u})$ and $H_{d}=(H^{0}_{d}H^{-}_{d})$.
Since the top quark, bottom quark and tau lepton are the heaviest fermions
in the SM, we assume that only the third family, $(3,3)$ element of the matrices
$\lambda_{u}, \lambda_{d}, \lambda_{e}$ are important.

The $\mu$ term provides masses to the Higgsinos
\begin{equation}
{\cal L} \supset -\mu(\tilde H_{u}^{+}\tilde H_{d}^{-}-\tilde H^{0}_{u}
\tilde H^{0}_{d})+{c.c} \,,
\end{equation}
and contributes to the Higgs $(mass)^2$ terms in the scalar potential  through
\begin{equation}
\label{higgsmass}
-{\cal L} \supset V \supset |\mu|^2(|H^{0}_{u}|^2+|H^{+}_{u}|^2+|H^{0}_{d}|^2+
|H^{-}_{d}|^2)\,.
\end{equation}
Note that Eq.~(\ref{higgsmass}) is positive definite. Therefore, it
cannot lead to electroweak symmetry breaking without including
SUSY breaking $(mass)^2$ soft terms for the Higgs fields,
which can be negative. Hence, $|\mu|^2$ should almost cancel the
negative soft $(mass)^2$ term in order to allow for a Higgs VEV
of order $\sim 174$~GeV. That the two different sources of masses
should be precisely of same order is a puzzle for which
many solutions has been suggested~\cite{Kim:1983dt,Giudice:1988yz,Casas:1992mk}.

Within MSSM one can construct gauge invariant $D$-and $F$-flat directions, for
the list of MSSM flat directions see~\cite{Dine:1995kz,Gherghetta:1995dv}.
A flat direction can be represented by a composite gauge invariant
operator, $X_m$,  formed from the product of $k$ chiral superfields
$\Phi_i$ making up the flat direction: $X_m=\Phi_1\Phi_2\cdots \Phi_m$.
The scalar component of the superfield $X_m$ is related to the
order parameter $\phi$  through $X_m=c\phi^m$~\cite{Dine:1995kz}.

An example of a $D$-and $F$-flat direction is provided by~\cite{Enqvist:2003gh,Dine:2003ax}
\begin{equation}
\label{example}
H_u=\frac1{\sqrt{2}}\left(\begin{array}{l}0\\ \phi\end{array}\right),~
L=\frac1{\sqrt{2}}\left(\begin{array}{l}\phi\\ 0\end{array}\right)~,
\end{equation}
where $\phi$ is a complex field parameterizing the flat direction,
or the order parameter, or the AD field. All the other fields are
set to zero. In terms of the composite gauge invariant operators,
we would write $X_m=LH_{u}~(m=2)$. Note that a flat direction necessarily carries a global $U(1)$ quantum
number, which corresponds to an invariance of the effective Lagrangian for the
order parameter $\phi$ under phase rotation $\phi\to e^{i\theta}\phi$.
In the MSSM the global $U(1)$ symmetry is $B-L$. For example, the
$LH_u$-direction  has $B-L=-1$.

From Eq.~(\ref{example}) one clearly obtains 
$F_{H_u}^*=\lambda_uQ u +\mu H_d=F_{L}^*=\lambda_dH_d e\equiv 0$
for all $\phi$. However there exists a non-zero F-component given
by $F^*_{H_d}=\mu H_u$. Since $\mu$ can not be much larger than the
electroweak scale $M_W\sim {\cal O}(1)$~TeV, this contribution is of
the same order as the soft SUSY breaking masses, which are
going to lift the degeneracy. Therefore, following \cite{Dine:1995kz}, one may
nevertheless consider $LH_u$ to correspond to a F-flat direction.
The relevant $D$-terms read
\begin{equation}
\label{Dterm0}
D^a_{SU(2)}=H_u^\dagger\tau_3H_u+L^\dagger\tau_3L=\frac12\vert\phi\vert^2
-\frac12\vert\phi\vert^2\equiv 0\,.
\end{equation}
Therefore the $LH_u$ direction is also $D$-flat.


\subsubsection{Gauge invariant inflatons of MSSM}~\label{GIIOM}
A simple observation was first made in \cite{Allahverdi:2006iq,Allahverdi:2006cx,Allahverdi:2006we},
where the inflaton properties are directly related to the soft SUSY breaking mass term and the A-term of the 
MSSM. Within MSSM, it is  possible to lift the flatness of the gauge invariant 
combinations of squarks and sleptons away from the point of {\it enhanced gauge symmetry} by the $F$-term, while maintaining the $D$-flatness. 

\vspace{-0.5cm}\paragraph{\bf Squarks and sleptons driven inflation:}
Let us consider a non-renormalizable superpotential term~\cite{Enqvist:2003gh,Dine:2003ax}:
\begin{equation} \label{supot}
W_{non} = \sum_{n>3}{\lambda_n \over n}{\Phi^n \over M_{\rm P}^{n-3}}\,,
\end{equation}
Where $\Phi=\phi\,\exp[i\theta]$, while $\theta$ is the phase term is a {\it gauge invariant} superfield which
contains the flat direction.  Within MSSM (with conserved $R$-parity) all the flat directions are
lifted by the  non-renormalizable operators with $4\le n\le 9$~\cite{Gherghetta:1995dv}. 
Two distinct directions are: $udd$ and $LLe$, up to an overall phase factor they are
parameterized by:
\begin{eqnarray}
\label{example}
u^{\alpha}_i=\frac1{\sqrt{3}}\phi\,,~
d^{\beta}_j=\frac1{\sqrt{3}}\phi\,,~
d^{\gamma}_k=\frac{1}{\sqrt{3}}\phi\,.\\
L^a_i=\frac1{\sqrt{3}}\left(\begin{array}{l}0\\ \phi\end{array}\right)\,,~
L^b_j=\frac1{\sqrt{3}}\left(\begin{array}{l}\phi\\ 0\end{array}\right)\,,~
e_k=\frac{1}{\sqrt{3}}\phi\,, \label{example-1}
\end{eqnarray}
where $1 \leq \alpha,\beta,\gamma \leq 3$ ($\alpha \neq \beta \neq \gamma$)
are color indices, and $1\leq i,j,k \leq 3$ ($j \neq k$)
denote the quark families for $udd$,  and $1 \leq a,b \leq 2$ ($a \neq b$) are the weak isospin 
indices and $1 \leq i,j,k \leq 3$ ($i \neq j \neq k$) denote the lepton families for $LLe$.
Both these directions are lifted by $n=6$ non-renormalizable
operators~\cite{Gherghetta:1995dv},
\begin{eqnarray}\label{suppot-1}
W_6\supset\frac{1}{M_{\rm P}^3}(LLe)(LLe)\,,~~
W_6\supset \frac{1}{M_{\rm P}^3}(udd)(udd)\,.
\end{eqnarray}
{\it Rest of the directions within MSSM are lifted by hybrid operators of type, 
$W \sim (1/M_{\rm P}^{n-3})\Psi\Phi^{n-1}$, which does not lead to cosmologically 
flat potential viable for slow-roll inflation}~\cite{Allahverdi:2006iq,Allahverdi:2006cx,Allahverdi:2008bt}.

The scalar potential along these directions includes softly broken SUSY
mass term for $\phi$ and an $A$-term gives rise to a specific potential
~\cite{Allahverdi:2006iq,Allahverdi:2006cx,Allahverdi:2008bt}
\begin{equation} 
\label{scalar-pot}
V(\phi) = {1\over2} m^2_\phi\, |\phi|^2 - A {\lambda\phi^6 \over 6\,M^{6}_{\rm P}} + \lambda^2
{{|\phi|}^{10} \over M^{6}_{\rm P}}\,,
\end{equation}
The
$A$-term is a positive quantity with dimension of mass. Note that the first and third terms in Eq.~(\ref{scalar-pot}) are positive
definite, while the $A$-term leads to a negative contribution along
the directions whenever $\cos(n \theta + \theta_A) < 0$. The  above potential is similar to Eq.(\ref{steppot}). It is 
possible to find a {\it point of inflection}, $\phi_{0}$, provided that ${A^2 / 40 m^2_{\phi}} \equiv 1 + 4 \alpha^2$,
where $\alpha^2 \ll 1$, and at the lowest orders in ${\cal O}(\alpha^{2})$, we obtain:
\begin{eqnarray}
\label{pot}
&&V(\phi_0) = \frac{4}{15}m_{\phi}^2\phi_0^2 + \cdots \, ,~~~
V'(\phi_0) = 4 \alpha^2 m^2_{\phi} \phi_0 \, + \cdots \, , \nonumber\\
&& V^{\prime \prime}(\phi_0) = 0,~~~~V^{\prime \prime \prime}(\phi_0) = 32{m_{\phi}^2}/{\phi_0} + \cdots \, . \nonumber \\
&&\phi_0 = \left({m_\phi M^{3}_{\rm P}/ \lambda \sqrt{10}}\right)^{1/4} + {\cal O}(\alpha^2)\,.\label{phi-0}
\end{eqnarray}
In the case of 
gravity-mediated SUSY breaking scenarios, $m_{\phi} \sim A \sim m_{3/2} \sim (100~{\rm GeV}-1~{\rm TeV})$. Therefore the
 condition $A^{2}\sim 40 m_{\phi}^{2}$ can indeed be satisfied. Inflation occurs within an interval
$\vert \phi - \phi_0 \vert \sim {\phi^3_0 / 60 M^2_{\rm P}} \ll M_{\rm P}$,
in the vicinity of the point of inflection, $\phi_0 \sim {\cal O}(10^{14}~{\rm GeV})$. Within which the slow-roll 
parameters, $\epsilon, \eta \ll 1$. The Hubble expansion rate during inflation is given by
\begin{equation} \label{hubble}
H_{\rm MSSM} \simeq \frac{1}{\sqrt{45}}\frac{m_{\phi}\phi_0}{M_{\rm P}}
\sim (100~{\rm MeV}-1~{\rm GeV})\,.
\end{equation}
The amplitude of density perturbations $\delta_H$ (see Eqs.~(\ref{eq:powerspectrum1}, \ref{amplitude2},\ref{ns2}) and the scalar spectral index $n_s$ are given 
by~\cite{Allahverdi:2006iq,Allahverdi:2006cx,BuenoSanchez:2006xk,Allahverdi:2006wt}:
\begin{eqnarray} \label{ampl}
\delta_H &=& {8 \over \sqrt{5} \pi} {m_{\phi} M_{\rm P} \over \phi^2_0}{1 \over \Delta^2}
~ {\rm sin}^2 [{N}_{Q}\sqrt{\Delta^2}]\, \\
 \label{tilt}
n_s &=& 1 - 4 \sqrt{\Delta^2} ~ {\rm cot} [{N}_{Q}\sqrt{\Delta^2}], 
\end{eqnarray}
where $2 \times 10^{-6} \leq \Delta^2 \equiv 900 \alpha^2 {N}^{-2}_{Q} ({M_{\rm P} / \phi_0})^4 \leq 5.2 \times 10^{-6}$, 
and $N_{Q}\sim 50$. Running in the tilt is very small. In this case the universe thermalizes in to MSSM radiation
instantly in less than one Hubble time after the end of inflation~\cite{Allahverdi:2011aj}, see the discussion in Sect.~\ref{TR-MSSM}.

For $\phi_{0}\sim 10^{14}$~GeV, there is an apparent fine-tuning in the parameters $A/m_{\phi}=\alpha\sim 10^{-10}$,
which may look unpleasant. However note that this fine tuning between the two MSSM parameters in the ratio is energy dependent and valid only at the 
scale of inflation at $10^{14}$~GeV, but at the TeV scale where the soft masses would be measured at the collider there is no apparent
fine tuning in the parameters~\cite{Allahverdi:2010zp}. 

As shown in Sect.~\ref{SUGRA-inflection}, see Fig.~\ref{figure:c_H-H}, the SUGRA corrections will ameliorate the tuning down to  
$\beta \equiv \alpha \sim 10^{-2}$, virtually addressing any fine tuning required for the success of MSSM inflation.  
It was shown in Refs.~\cite{Allahverdi:2007wh,Allahverdi:2008bt}, that the {\it inflection point}
for the MSSM inflaton is an attractor solution, provided there exists a phase of inflation prior of that of the MSSM with $N\geq 10^{10}$ e-foldings.
Such large e-foldings can be generated within string theory landscape~\cite{Allahverdi:2007wh}, or within MSSM~\cite{Allahverdi:2008bt}.


\begin{figure}
\includegraphics[width=5.0cm]{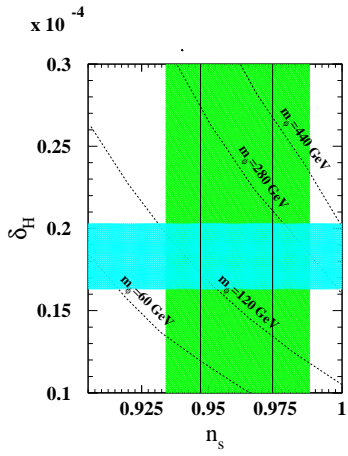}
\caption{$(\delta_{H},~n_s)$ is plotted for different
values of $m_{\phi}$ and $\lambda=1$. The $2\sigma$ region for $\delta_H$ is shown by the blue horizontal band and the $2\sigma$ 
allowed region of $n_s$ is shown by the vertical green band. The $1\sigma$ allowed region of $n_s$ is within the 
solid vertical lines~\cite{Allahverdi:2008bt}.} \label{nsdel0}
\end{figure}


\vspace{-0.5cm}\paragraph{\bf Renormalizable superpotential:}

The left handed neutrinos can be of Dirac type with an appropriate Yukawa 
coupling. The simplest way to obtain this would be  to augment
the SM symmetry by, $SU(3)_C\times SU(2)_L\times U(1)_Y\times U(1)_{B-L}$, where
$U(1)_{B-L}$ is gauged.  The relevant superpotential term is
\begin{equation}\label{supotN} 
W  \supset h {N} {H}_u { L} . 
\end{equation}
Here ${ N}$, ${ L}$ and ${H}_u$ are superfields containing
the RH neutrinos, left-handed (LH) leptons and the Higgs which gives
mass to the up-type quarks, respectively. Note that the ${N} {H}_u {L}$ monomial represents a
$D$-flat direction under the $U(1)_{B-L}$, as well as the SM gauge
group.

The value of $h$ needs to be small, i.e. $h \leq 10^{-12}$, in order
to explain the light neutrino mass, $\sim {\cal O}(0.1~{\rm eV})$
corresponding to the atmospheric neutrino oscillations detected by
Super-Kamiokande experiment.  The potential along this direction, after the minimization along the
angular direction, is found to be~\cite{Allahverdi:2006cx,Allahverdi:2007wt},
\begin{eqnarray} \label{flatpot}
V (\vert \phi \vert) = \frac{m^2_{\phi}}{2} \vert \phi \vert ^2 +
\frac{h^2}{12} \vert \phi \vert^4 \, - \frac{A h}{6\sqrt{3}}
\vert \phi \vert^3 \,.
\end{eqnarray}
For $A \approx 4 m_{\phi}$, there exists an {\it inflection point} for
which $V^{\prime}(\phi_0) \neq 0, V^{\prime \prime}(\phi_0) = 0$, where
inflation takes place
\begin{eqnarray} \label{sad} \phi_0 = \sqrt{3}\frac{m_{\phi}}{h}=
6 \times 10^{12} ~ m_{\phi} ~ \Big({0.05 ~
{\rm eV} \over m_\nu} \Big)\,,~~\nonumber \\
V(\phi_0) = \frac{m_{\phi}^4}{4h^2}=3 \times 10^{24} ~ m^4_{\phi} ~
\Big({0.05 ~ {\rm eV} \over m_\nu} \Big)^2 \,.
\label{sadpot}
\end{eqnarray}
The amplitude of density perturbations 
follows  from Eqs.~(\ref{eq:powerspectrum1},\ref{amplitude2},\ref{ns2})~\cite{Allahverdi:2006cx,Allahverdi:2007wt}.
\begin{equation} \label{amp} \delta_{H} \simeq
\frac{1}{5\pi}\frac{H^2_{inf}}{\dot\phi} \simeq 3.5 \times 10^{-27}
~ \Big( {m_\nu \over 0.05 ~ {\rm eV}} \Big)^2 ~  \Big({M_{\rm P}
\over m_{\phi}} \Big) ~ {N}_{Q}^2\,. \end{equation}
Here $m_{\phi}$ denotes the loop-corrected value of the inflaton
mass at the scale $\phi_0$ in Eqs.~(\ref{sad},\ref{amp}). The spectral tilt as 
usual has a range of values  $0.90 \leq n_s\leq 1.0$ ~\cite{Allahverdi:2006cx,Allahverdi:2007wt}. 


\vspace{-0.5cm}\paragraph{\bf MSSM Higgses as inflaton:}
The MSSM Higgses are another fine example of a visible sector inflaton provided some restrictions are met~\cite{Chatterjee:2011qr}.
The required superpotential is given by
\begin{equation}
W = \mu {H_{u}}. {H_{d}} +\sum_{k}\frac{\lambda_k}{k} \frac{\left({H_{u}}.
{H_{d}}\right)^k} { M^{2k-3}_{\rm P}},
\label{supot-Higgs}
\end{equation}
This is the $\mu$-term which were considered an ideal candidate to generate the density perturbations~\cite{Enqvist:2003qc,Enqvist:2004kg},
but now they can also provide the required vacuum energy to inflate the universe~\cite{Chatterjee:2011qr}. The scalar potential along the 
$H_{u} H_{d}$  $D$-flat direction is given by,
\begin{eqnarray}
\label{soft-pot-Higgs}
{V} (\varphi,\theta) & = & \frac{1}{2} m^{2}(\theta)\varphi^{2}
   +(-1)^{(k-1)} 2 \lambda_{k}^{'} \mu \cos ((2k-2)\theta) \varphi^{2k} \nonumber \\
  & + & 2 \lambda_{k}^{'2} \varphi^{2(2k-1)},
\end{eqnarray}
where $\varphi  =   \sqrt{2} |\phi| e^{i\theta}$, and $H_{u} = (1/\sqrt{2})(\phi, 0)^{T},~~
H_{d}  =  (1/\sqrt{2})^{-1}(0, \phi)^{T}$, and 
\begin{eqnarray}
m^{2}(\theta) & = & \frac{1}{2}(m_{H_u}^{2} + m_{H_d}^{2}  
               +  2 \mu^{2}-2 B \mu \cos 2\theta ), \\ 
\lambda^{'}_{k}& = & \frac{\lambda_k}{2^{(2k-1)}M^{2k-3}_{\rm P}}\,.
\label{lambda-k} 
\end{eqnarray}
For simplicity, we may assume $\mu$ and $ B$ to be real. This choice
is compatible with the experimental constraints, mainly from the Electron Dipole Moment 
measurements~\cite{Pospelov:2005pr}.  The inflection point can be obtained for $\theta=0, \pm \pi/2$,
for simplicity let us consider the case for $\theta=0$, and when the following condition is satisfied,  
$m^{2}_{0}= {k^{2} \mu^{2}}/{(2k-1)} + \tilde{\lambda}^{2}$,
where $\tilde{\lambda}$ is the tuning required to maintain the flatness of the potential. Although, this tuning could be 
harsh at the inflationary scale, $\varphi\sim 10^{14}$~GeV, but the ratio evolves to $m_{0}^{2}/\mu^{2} \sim {\cal O}(1)$
at the electroweak scale by virtue of running of the renormalization group equations~\cite{Chatterjee:2011qr}.
The amplitude of the CMB perturbations can be 
obtained for  $\lambda_{2}\sim 10^{-8}$ and  $\lambda_{3}\sim {\cal O}(1)$, it is possible to obtain a similar plot like Fig.~(\ref{nsdel0})
for Higgs mass $m(\theta=0)\sim 100-250$~GeV, which yields the spectral tilt in the range $0.93\leq n_{s}\leq 0.98$.


\subsection{Preheating, reheating, thermalization}

Reheating at the end of inflation is an important aspect of inflationary cosmology. Without reheating the
universe would be empty of matter, for a review see~\cite{Allahverdi:2010xz}. Reheating occurs through coupling of the inflaton
field $\phi$, to the SM matter. Such couplings must be present at least via
gravitational interactions.  In particular, if the inflaton is a SM
gauge singlet, the relevant couplings to SM 
are: $\frac{\lambda}{M}\phi(H\bar q_l)q_R\,,~~\frac{\lambda}{M}\phi F_{\mu\nu}F^{\mu\nu}\,, ~~g^{2}\phi^{2}\bar H H$,
where $M$ is the scale below which {\it all}  these effective operators are valid, $\lambda,~g \sim {\cal O}(1)$, $H$ is the SM 
Higgs doublet, and $q_l, q_R$ are the left  and the right handed SM fermions~\cite{Allahverdi:2007zz}. 

Similar couplings would arise 
if $\phi$ is replaced by right handed sneutrinos, axions, moduli, or any other hidden sector field. Being a SM singlet,
$\phi$ can as well couple to other hidden sectors, moduli, axions, etc. Since the hidden sectors are largely unknown, it becomes a 
challenge for a singlet inflaton to decay solely into the SM {\it d.o.f}~\cite{Cicoli:2010ha,Cicoli:2010yj}. 

After the end of inflation, the inflaton starts coherent oscillations around its minimum.
The frequency of oscillations are determined by the frequency of oscillations,  $\omega \sim m_{eff}\geq H_{inf}$. 
During this epoch the inflaton can 
decay perturbatively~\cite{Albrecht:1982mp,Turner:1983he,Dolgov:1989us,Kolb:1988aj}.
Averaging over many oscillations results in a pressureless equation of state where
$\langle p\rangle =\langle \dot\phi^2/2 -V(\phi)\rangle$
vanishes, so that the energy density starts evolving like a matter domination (in a quadratic potential) with
$\rho_{\phi}=\rho_{i}(a_{i}/a)^{3}$ (subscript $i$ denotes the
quantities right after the end of inflation). For $\lambda\phi^4$ potential the coherent oscillations yield an 
effective equation of state similar to that of a radiation epoch. If $\Gamma_{\phi}$ represents
the {\it total} decay width of the inflaton to pairs of fermions. This releases the energy 
into the thermal bath of relativistic particles
when $H(a)=\sqrt{(1/3M_{\rm P}^2)\rho_{i}}(a_{i}/a)^{3/2}\approx \Gamma_{\phi}$.
The energy density of the thermal bath is determined by the reheat
temperature $T_{R}$, given by:
\begin{equation}
T_{R}=\left(\frac{90}{\pi^2 g_{\ast}}\right)^{1/4}\sqrt{\Gamma_{\phi}
M_{\rm P}}=0.3\left(\frac{200}{g_{\ast}}\right)^{1/4}\sqrt{\Gamma_{\phi}
M_{\rm P}}\,,
\end{equation}
where $g_{\ast}$ denotes the effective relativistic {\it d.o.f} in the
plasma. However the inflaton decay products need to thermalize, which
requires acquiring {\it kinetic} and {\it chemical} equilibrium.  


\vspace{-0.5cm}
\subsubsection{Non-perturbative particle creation}\label{PPR}

If the inflaton coupling to the matter field is large, a completely new channel of reheating opens up due to the coherent nature 
of the inflaton field, proposed by \cite{Traschen:1990sw,Shtanov:1994ce,Kofman:1994rk,Kofman:1997yn}, known as {\it preheating}.
Let us first consider a simple toy model with interaction Lagrangian
\begin{equation} \label{intLag}
{\cal L}_{\rm int} \, = \, - \frac{1}{2} g^2 \chi^2  \phi^2 \, ,
\end{equation}
where $\chi$ is another scalar field, in a realistic set-up $\chi$ could take the role of the SM Higgs.
We can neglect the effect of expansion provided that the time period of preheating is small compared
to the Hubble expansion time $H^{-1}$, this is reasonable in many cases. 

The quantum theory of $\chi$ particle production in the external classical
inflaton background begins by expanding the quantum field $\hat{\chi}$ into
creation and annihilation operators $\hat{a}_{k}$ and $\hat{a}^{\dagger}_{k}$ as:
\begin{equation}
\hat{\chi}(t,\mathbf{x})=\frac{1}{(2\pi)^{3/2}}\int d^3k \left(\chi_k^*(t)\hat{a}_{k} e^{i\mathbf{k}\mathbf{x}}+\chi_k(t)\hat{a}^{\dagger}_{k} e^{-i\mathbf{k}\mathbf{x}}\right),
\end{equation}
where $k$ is the momentum. Since the equation of motion for $\chi$ is linear it
can be studied simply mode by mode in Fourier space. The mode functions then satisfy:
\begin{equation}\label{math3}
\ddot{\chi}_k+\left(k^2+m_{\chi}^2+g^2\Phi^2\sin^2{(m_{\phi}t)}\right)\chi_k \, = \, 0 \, ,
\end{equation}
where $\Phi$ is the amplitude of oscillation in $\phi$.
This is the Mathieu equation which is written in
the form
\begin{equation}\label{mat}
\chi''_k+(A_k-2q\cos{2z})\chi_k \, = \, 0 \, ,
\end{equation}
where the dimensionless time variable is $z = m_{\phi} t$
and a prime now denotes the derivative with respect to $z$.
Comparing the coefficients, we find
\begin{equation}\label{param1}
A_k=\frac{k^2+m_{\chi}^2}{m^2}+2q\qquad q=\frac{g^2\Phi^2}{4m_{\phi}^2}
\end{equation}
The growth of the mode function
corresponds to particle production~\cite{Birrell:1982ix}. 
It is well known that the above Mathieu equation~Eq.~(\ref{mat}) has instabilities for certain ranges of $k$:
\begin{equation}
\chi_k \, \propto \, {\rm exp}(\mu_k z) \, ,
\end{equation}
where $\mu_k$ is called the Floquet exponent. For small values of $q$, i.e. $q \ll 1$,
resonance occurs in a narrow instability band about $k = m$, known as 
a ``narrow resonance" band~\cite{Traschen:1990sw}. The resonance is much more efficient if $q \gg 1$ \cite{Kofman:1994rk,Kofman:1997yn}.
In this case, resonance occurs in broad bands, i.e. the bands include all
long wavelength modes $k \rightarrow 0$, known as {\it broad resonance}. This can be understood by studying 
the condition for particle production in the WKB approximation for
the evolution of $\chi$ field which is violated. In the WKB approximation:
$\chi_k \, \propto \, e^{\pm i\int\omega_kdt}$,
which is valid as long as the adiabaticity condition
\begin{equation} \label{adiab}
{d\omega_k^2}/{dt} \, \leq \, 2 \omega^3_k
\end{equation}
is satisfied. In the above, the effective frequency $\omega_k$ is given by
\begin{equation} \label{efffreq}
\omega_k \, = \, \sqrt{k^2 + m_{\chi}^2 + g^2 \Phi(t)^2 sin^2(m_{\phi}t)} \, ,
\end{equation}
By inserting the effective frequency Eq.~(\ref{efffreq}) into the condition Eq.~(\ref{adiab}) and
following some algebra, the adiabaticity condition is violated for momenta
\begin{equation}
k^2 \, \leq \, \frac{2}{3\sqrt{3}}gm_{\phi}\Phi-m_{\chi}^2.
\end{equation}
For modes with these values of $k$, the adiabaticity condition breaks
down in each oscillation period when $\phi$ is close to zero. The
particle number does not increase smoothly, but
rather in ``bursts"~\cite{Kofman:1994rk,Kofman:1997yn}.

The above analysis of neglecting the expansion of
the universe is self-consistent. However, as discussed in detail
in \cite{Kofman:1997yn}, the expansion of space can be included. The equation of motion for $\chi$ becomes
\begin{equation}\label{eom_chi_frw}
\ddot{\chi}_k+3H\dot{\chi}_k+\left(\frac{k^2}{a^2}+m_{\chi}^2+
g^2\Phi(t)^2\sin^2{(m_{\phi}t)}\right)\chi_k \, = \, 0.
\end{equation}
The adiabaticity condition is now violated for momenta satisfying:
\begin{equation}\label{ineq_adia2}
{k^2}/{a^2} \, \leq \, ({2}/{3\sqrt{3}})gm\Phi(t)-m_{\chi}^2.
\end{equation}
Note that the expansion of space makes broad resonance more
effective since more $k$ modes are red-shifted into
the instability band as time proceeds. The detailed
analysis yields the same expression for the resonance
band except for the exact value of the numerical coefficient of the
first term on the r.h.s..
Broad parametric resonance ends when $q \leq 1/4$.

In principle,  it is also possible to excite the fermions
non-perturbatively,  in spite of the fact that the occupation number of any fixed state
cannot be greater than one (because of the
Pauli exclusion principle)~\cite{Greene:1998nh,Baacke:1998di,Giudice:1999fb,Greene:2000ew,Peloso:2000hy}, and 
higher spin $\pm 3/2$ gravitinos~\cite{Maroto:1999ch,Kallosh:1999jj,Giudice:1999am,Nilles:2001ry,Nilles:2001fg,Kallosh:2000ve}.

\vspace{-0.5cm}
\paragraph{\bf Tachyonic prehetaing:} It is possible that effective frequency of certain mode can be
negative. For example in a symmetry breaking potential:
$V(\phi) \, = \, \frac{1}{4} \lambda (\phi^2 - \eta^2)^2$,
for small field values,
the effective mass of the fluctuations of $\phi$ is negative
and hence a ``tachyonic" resonance will occur, as studied
in \cite{Felder:2000hj}. For small field values, the
equation for the fluctuations $\phi_k$ of $\phi$ is
\begin{equation}
\ddot{\phi_k} + \bigl( k^2 - m_{\phi}^2 \bigr) \phi_k \, = 0 \, .
\end{equation}
The modes with $k < m$ grow with an exponent 
which approaches $\mu_k = 1$ in the limit $k \rightarrow 0$.
Given initial vacuum amplitudes for the modes $\phi_k$
at the intial time $t = 0$ of the resonance, the field
dispersion at a later time $t$ will be given by
\begin{equation}
\langle \delta \phi^2 \rangle \, = \, \int_0^m \frac{k dk}{4 \pi} e^{2 t \sqrt{m_{\phi}^2 - k^2}} \, .
\end{equation}
The growth of the fluctuations modes terminates once the
dispersion becomes comparable to the symmetry breaking
scale.

Tachyonic preheating also occurs in hybrid inflation models, see Eq.~(\ref{eq:potenhyb2d}). In this case, it is the fluctuations of
$\psi$ which have tachyonic form and which grow exponentially
\cite{Felder:2000hj}. Preheating in hybrid inflation was
first studied in \cite{GarciaBellido:1997wm} using the tools of broad
parametric resonance. 


\vspace{-0.5cm}
\paragraph{\bf End of preheating:} In the above analysis we have neglected the back-reaction of the
produced $\chi$ and $\phi$ particles on the dynamics. The presence of $\chi$ particles changes the effective
mass of the inflaton oscillations. This back-reaction effect is negligible as long as the change $\Delta m_{\phi}^2$ in the
square mass of the inflaton is smaller than $m^2_{\phi}$. In the Hartree approximation, the change 
in the inflaton mass due to $\chi$ particles is given by
$\Delta m_{\phi}^2 \, = \, g^2 \langle \chi^2 \rangle $~\cite{Kofman:1997yn}.
Another important condition is that the energy in the $\chi$ particles should be sub-dominant. Therefore,
$\rho_{\chi} \, \sim \, \langle (\nabla \chi)^2 \rangle \, \simeq \, k^2 \langle \chi^2 \rangle \ll m_{\phi}^{2}\langle \phi^{2}\rangle$,
It was found that $\rho_{\chi}$ is smaller than the potential energy of the inflaton
field at the time $t_1$ as long as the value $q$ at the time $t_1$ is larger
than $1$, i.e. $q(t_1) > 1$. This is roughly speaking the same as the
condition for the effectiveness of broad resonance \cite{Kofman:1997yn}.


\subsubsection{Thermalization}

Neither the perturbative decay of the inflaton nor the preheating mechanism
produce a thermal spectrum of decay products.  In a full thermal equilibrium the energy density
$\rho$ and the number density $n$ of relativistic particles
scale as: $\rho \sim T^4$ and $n \sim T^3$,
where $T$ is the temperature of the thermal bath. Thus, in
full equilibrium the average particle energy is given by:
$\langle E \rangle_{\rm eq} = \left( \rho/n \right)$, which obeys the scaling,
$\langle E \rangle_{\rm eq} \, \sim \, \rho^{1/4} \, \sim \, T $.

\vspace{-0.5cm}\paragraph{\bf Perturbative reheating and thermalization:}
If the inflaton decays perturbatively,
then right after the inflaton has decayed completely, the energy density
of the universe is given by: $\rho \, \approx \, 3 \left(\Gamma_{\phi} M_{\rm P}\right)^2\,,~~~
\langle E \rangle \, \approx \, m_{\phi} \, \gg \, \rho^{1/4} $. From
conservation of energy, the number density of decayed particles is:
$n \, \approx \, \left({\rho / m_{\phi}} \right) \, \ll \, \rho^{3/4} $. Hence, perturbative 
decay results in a dilute plasma that contains a
small number of very energetic particles. A local thermal equilibrium requires re-distribution of the energy among
different particles, {\it kinetic equilibrium}, as well as increasing
the total number of particles, {\it chemical equilibrium}. 
Therefore both number-conserving and number-violating reactions must be involved.

The most important processes for kinetic equilibration
are $2 \rightarrow 2$ scatterings with gauge boson exchange
in the $t$-channel. Due to an infrared singularity, these scatterings are very efficient even in a dilute
plasma~\cite{Davidson:2000er,Allahverdi:2005mz}.
Chemical equilibrium is achieved by changing
the number of particles in the reheat plasma.
In order to reach full equilibrium the total number of particles
must {\it increase} by a factor of $n_{\rm eq}/n$, where $n \approx
\rho/m$ and the equilibrium value is: $n_{\rm eq} \sim
\rho^{3/4}$. This can be a very large number, i.e.
$n_{\rm eq}/n\sim {\cal O}(10^3)$.
It was recognized in~\cite{Davidson:2000er,Allahverdi:2002pu}, see
also~\cite{Jaikumar:2002iq,Allahverdi:2000ss} that the most
relevant processes are $2 \rightarrow 3$ scatterings with gauge-boson
exchange in the $t-$channel. 
When these scattering become efficient, the number of particles
increases very rapidly, and full thermal equilibrium is established shortly after that~\cite{Enqvist:1993fm}.

\vspace{-0.5cm}\paragraph{\bf Non-perturbative preheating and thermalization:}
In this case the occupation numbers of the excited
quanta are typically very high after the initial stages of preheating,
Once the occupation numbers of the resonant modes become
sufficiently large, re-scattering of the fluctuations begins~\cite{Khlebnikov:1996mc,Khlebnikov:1996wr,Khlebnikov:1996zt,Micha:2002ey,Micha:2004bv} 
which terminates the phase of exponential growth of the occupation numbers. 
The evolution of the field fluctuations evolves to a regime of turbulent scaling  which is characterized by the
spectrum $n(k) \, \sim \, k^{-3/2} $~\cite{Micha:2002ey,Micha:2004bv},
which is non-thermal (for a thermal distribution we would expect $n(k ) \sim k^{-1}$).
The phase of turbulence ends once most of the energy has
been drained from the inflaton field. At this time quantum
processes take over and lead to the thermalization of the
spectrum.


\subsubsection{Calculation of $T_{R}$ within MSSM}\label{TR-MSSM}

In the case of MSSM inflation, the inflaton couplings to MSSM {\it d.o.f} are known~\cite{Allahverdi:2011aj}.  It is therefore possible
to track the thermal history of the universe from the end of inflation.  When the MSSM inflaton passes through minimum, i.e. $\phi=0$, the entire gauge symmetry gets restored and all the {\it d.o.f} associated with the MSSM gauge group become massless, which is known as the {\it point of enhanced gauge symmetry}. 

These are the massless modes which couple to the inflaton directly, for instance the {\it d.o.f} corresponding  to $SU(2)_W\times U(1)_Y$, or  that of $SU(3)_c\times U(1)_Y$. At VEVs away from the minimum, the same modes become heavy and therefore it is kinematically  unfavorable to excite them. The actual process of excitation depends on how strongly the adiabaticity condition for the time dependent vacuum is violated for the inflaton zero mode. 


\vspace{-0.5cm}\paragraph{\bf Couplings for LLe inflaton:}
Let us illustrate this with $L_{1}L_{2}e_{3}$ flat direction as an inflaton. The inflaton non-zero VEV completely breaks the 
$SU(2)_W \times U(1)_Y$ symmetry. This results in four massive real scalars, whose masses are obtained from the $D$-terms~\cite{Allahverdi:2011aj}
\begin{eqnarray}\label{scalcoupl}
V  \supset  {1 \over 12} g_W^2 \phi^2 (\chi^2_1 + \chi^2_2  +  \chi^2_3) + {1 \over 4} g_Y^{ 2} \phi^2 \chi^2_4 \, .
\end{eqnarray}
Here $g_W,~g_Y$ are the $SU(2)_W$ and $U(1)_Y$ gauge couplings respectively, and $\phi$ denote the inflaton, see Eq.~(\ref{example-1}).
The  $\chi$'s are Goldstone bosons from breakdown of $SU(2)_W \times U(1)_Y$. They are eaten by the Higgs mechanism and give rise to longitudinal components of the electroweak gauge fields. In the unitary gauge, they are completely removed from the spectrum.
The $\chi$ particles decay to squarks, the Higgs particles, and the ${\tilde L}_3,~{\tilde e}_1,~{\tilde e}_2$ sleptons with 
the decay rates given by: $\Gamma_{\chi_1} = \Gamma_{\chi_2} = \Gamma_{\chi_3} = {3~g_W^3 \phi /8 \pi \sqrt{6}}  \,,~
\Gamma_{\chi_4} = (9~g_Y^{3} \phi / 16 \pi \sqrt{2}) $.
Note that the decay rate is proportional to the VEV of the inflaton, which sets the mass of $\chi$ fields.
Couplings of the inflaton to the gauge fields are obtained from the flat direction kinetic terms~\cite{Allahverdi:2011aj}
\begin{equation} \label{gaugcoupl}
{\cal L} \supset \frac{g_W^2}{12} \phi^2 (2 W^{+, \mu} W^{-}_{\mu} + W^{\mu}_3 W_{3,\mu}) + \frac{g_Y^{ 2}}{4} \phi^2 B^{\mu} B_{\mu} \, , 
\end{equation}
where $W^{+} = {(W_1 - i W_2 )/ \sqrt{2}} ~ ~ , ~~W^{-} = {(W_1 + i W_2 )/ \sqrt{2}}$,
where $W_{i,\mu}$ and $B_\mu$ are the $SU(2)_W$ and $U(1)_Y$ gauge fields respectively.
The gauge fields decay to (s)quarks, Higgs and Higgsino particles, and $L_3,~e_1,~e_2$ (s)leptons
with the total decay widths: $ \Gamma_{W^+} = \Gamma_{W^-} = \Gamma_{W_3} = {(3~g_W^3 \phi/ 8 \pi \sqrt{6})},~
\Gamma_B = {(9~g_Y^{ 3} \phi/ 16 \pi \sqrt{2})} $. Couplings of the inflaton to fermions can also be found in a similar way~\cite{Allahverdi:2011aj}. 

\vspace{-0.5cm}\paragraph{\bf Instant preheating and thermalization:}
The fields that are coupled to the inflaton acquire a VEV-dependent mass that varies in time due to the inflaton oscillations. For illustration, we first 
focus on the $\chi_1$ scalar, which are produced every time the
inflaton goes through zero.
The Fourier eigenmodes of $\chi_1$ have the corresponding energy
\begin{equation} \label{eigenmodes}
\omega_k = \sqrt{k^2 + m_\chi^2 + g^2_W {\phi (t)}^2/6} \end{equation}
where $m_\chi$ is the bare mass of the $\chi$ field. The growth of the occupation number of mode $k$ can be computed 
exactly for the first zero-crossing, $n_{k,\chi}= \exp\!\left[{-{\pi\sqrt6(k^2+m_\chi^2)}/{(g_W \dot\phi_0)}}\!\right]$,
where the inflaton near the zero-crossing is given by ${\dot \phi}_0 = (2 V({\hat \phi}))^{1/2}$, from the conservation of energy, where ${\hat \phi}$ is the amplitude of the inflaton oscillations, $\hat\phi\simeq\phi_{0}/\sqrt{3}\sim 10^{13}$~GeV, where $\phi_{0}$ is the inflection point for inflation, 
Eq.~(\ref{phi-0}). Note that after a few oscillations, 
$\dot\phi_0 \simeq m_\phi\hat\phi$, since the expansion
rate during the inflaton oscillations is negligible by virtue of $m_{\phi}\sim 100$~GeV and $H(t)\leq H_{inf}\sim 1$~GeV.
The total number density of particles thus produced follows
\begin{equation} \label{ndensity}
n_{\chi_1} = \int_{0}^\infty\frac{d^3 k}{(2 \pi)^3} n_{k,\chi}
= {m_\phi^3 q^{3/4}\over 2\sqrt2 \pi^3} \,\exp\left({-\frac{\pi m_\chi^2}{2m_\phi^2\sqrt q }}\right) \,.
\end{equation}
where $q \equiv ({g^2_W \dot\phi_0^2}/{24m_\phi^4})\gg 1$.
This expression corresponds to the asymptotic value and assumes there is no perturbative decay of the 
produced $\chi$ particles. However, immediately after adiabaticity is restored,
$ \tau > \tau_* = \frac{1}{\sqrt2}\,q^{-1/4}$, $\chi_1$ particles decay into lighter particles (i.e. those particles that have no gauge coupling to the inflaton). In the case of $L_1 L_2 e_3$ inflaton these are the (s)quarks, $H_u$ Higgs(ino), and $L_3,~e_1,~e_2$ (s)leptons. 

Thus the fraction that is transferred from the inflaton to $\chi_1$'s, and through their prompt decay into relativistic squarks, at every inflaton zero-crossing, 
can be computed analytically, and they are given by,
\begin{equation} \label{transfer}
{\rho_{\chi_{rel}} \over \rho_\phi} \sim 0.0067\,g_W^2\,e^{{-\frac{\pi m_{\chi_1}^2}{2m_\phi^2\sqrt q_{W}}}}
+0.0166 \, g_Y^2e^{{-\frac{\pi m_{\chi_4}^2}{2m_\phi^2\sqrt q_Y}}}\,.
\end{equation}
The total number of {\it d.o.f} coupled to the ${\bf L_1L_2e_3}$ inflaton is $32$ ($4$ from scalars, $4\times 3=12$ from gauge fields, and 
$4 \times 4=16$ from fermions). Therefore the fraction of the inflaton energy density that is transferred to relativistic quarks and squarks, see Eq.~(\ref{transfer}), for $g_Y \sim g_W \sim 0.6$, has to be multiplied by (1+3+4)=8~\cite{Allahverdi:2011aj}:
\begin{equation} \label{ttransfer}
{\rho_{\rm rel} / \rho_\phi} \sim 10.6 \%~~~({\rm per~zero-crossing}).
\end{equation}
Note that this fraction is independent from the amplitude of oscillations. The draining the inflaton energy is quite efficient,
nearly $10\%$ of the inflaton energy density gets transferred to the relativistic species -- but not all the SM {\it d.o.f} are in thermal 
equilibrium after one oscillation. It takes near about 120 oscillations to reach the full {\it chemical} and {\it kinetic} equilibrium 
via processes requiring $2\leftrightarrow 2$ and $2\leftrightarrow 3$ interactions. However 
due to the hierarchy between $H_{inf}\sim 10^{-3}m_{\phi}$, this happens within a single Hubble time after the end of inflation. One can 
estimate the final reheat temperature~\cite{Allahverdi:2011aj}
\begin{equation} \label{Trh}
T_{R} = \left({30}/{\pi^2 g_*}\right)^{1/4} \rho_0^{1/4} \simeq 2 \times 10^8 ~ {\rm GeV} \, ,
\end{equation}
where $g_* = 228.75$ and $\rho_{0} = (4/15)m_{\phi}^{2}\phi_{0}^{2}$, see Eq.~(\ref{phi-0}).


\section{Matter-anti-matter asymmetry}

If (p)reheating can provide a thermal bath where {\it all} the SM 
quarks and leptons are excited, it is then an important question to ask -- why the present day galaxies and intergalactic
medium is primarily made up of baryons rather than anti-baryons? 

The baryon abundance in the universe is denoted by $\Omega_{b}\equiv \rho_{b}/\rho_{c}$, which defines the fractional baryon
density $\rho_{b}$ with respect to the critical energy density of the
universe: $\rho_{c}=1.88~h^2\times 10^{-29}~{\rm g~cm^{-3}}$. The
observational uncertainties in the present value of the Hubble constant;
$H_{0}=100~h~{\rm km \cdot s^{-1}\cdot Mpc^{-1}}\approx (h/3000){\rm Mpc^{-1}}$
are encoded in $h=0.73$~\cite{Kessler:2009ys}. It is useful to write in terms of the 
baryon and photon number densities 
\begin{equation}
\eta\equiv \frac{n_{b}-n_{\bar b}}{n_{\gamma}}=2.68\times 10^{-8}
\Omega_{b}h^2\,,
\end{equation}
where $n_{b}$ is the baryon number density and $n_{\bar b}$ is for
anti-baryons. The photon number density is given by
$n_{\gamma}\equiv (2\zeta(3)/\pi^2)T^3$.
The best present estimation of the baryon density comes from BBN, 
which is based on SM physics with 3 neutrino species \cite{Fields:2006ga,Cyburt:2008kw}
\begin{eqnarray}
\label{omegab}
0.019 \leq \Omega_{b}h^2\leq 0.024~~(95 \% CL)\,,\\
5.1\times 10^{-10} \leq \eta \leq 6.5 \times 10^{-10}~~(95\% CL)\,.
\end{eqnarray}
The observational data on $D$ and $^{4}He$ are consistent with each other and 
the expectations from the BBN analysis, but both prefer slightly higher value compared to the 
$^{7}Li$ abundance $Li/H|_{P}=(1.7\pm0.06\pm 0.44)\times 10^{-10}$, which is smaller than
$D$ and $^{4}He$ by at least $\sim 4.2\sigma$. The $^{7}Li$ abundance is primarily measured
in the stellar systems such as globular clusters.

From the acoustic peaks of the CMB the baryon fraction can be deduced. The 
WMAP data imply $\Omega_{b}h^{2}=0.02273\pm 0.00062$ or 
$\eta=6.23\pm0.17\times 10^{-10}$~\cite{Komatsu:2010fb}. The WMAP data relies on 
priors and the choice of number of  parameters, it is possible to yield baryon abundance
as low as $\Omega_{b}h^2=0.0175\pm 0.0007$~\cite{Hunt:2008wp}. In spite of systematic 
uncertainties the WMAP data is consistent with that of theoretical predictions from BBN.
The major unresolved problem is the $Li$ abundance, stellar $Li/H$ measurements 
are inconsistent with both WMAP and BBN data, and this could be an useful probe of new physics at
BBN, see for a review~\cite{Jedamzik:2009uy}.

Often in the literature the baryon asymmetry is given in relation
to the entropy density $s=1.8g_{\ast}n_{\gamma}$, where $g_{\ast}$
measures the effective number of relativistic species which itself
a function of temperature. At the present time $g_{\ast} \approx 3.36$,
while during BBN $g_{\ast} \approx 10.11$, rising up to $106.75$ at
$T\gg 100$~GeV. In the presence of supersymmetry at $T \gg 100$~GeV,
the number of effective relativistic species are nearly doubled to $228.75$.
The baryon asymmetry defined as the difference of baryon and
anti-baryon number densities relative to the entropy density, is
bounded by
\begin{equation}
\label{range2}
58.4(7.2)\times 10^{-11} \leq\frac{n_{b}-n_{\bar b}}{s}\leq
9.2(9.2)\times 10^{-11}\,,
\end{equation}
at $(95\% CL)$, where the numbers are CMB~\cite{Komatsu:2010fb}, and BBN bounds~\cite{Fields:2006ga,Cyburt:2008kw}, respectively.

If at the beginning $\eta =0$, then the origin of this small number can
not be understood in a CPT invariant universe by a mere thermal decoupling
of nucleons and anti-nucleons at  $T\sim 20$~MeV. The resulting asymmetry
would be too small by at least nine orders of magnitude, see
\cite{Kolb:1988aj}. Therefore it is important to seek mechanisms for generating
baryon asymmetry, for reviews see~\cite{Rubakov:1996vz,Riotto:1998bt,Dine:2003ax}.


\subsection{Requirements for baryogenesis}

As pointed out first by Sakharov \cite{Sakharov:1967dj}, baryogenesis requires three ingredients:
$(1)$ baryon number non-conservation, $(2)$ $C$ and $CP$ violation,
and $(3)$ out-of-equilibrium condition. All these conditions
are believed to be met in the very early universe.


\vspace{-0.5cm}\paragraph{\bf Baryon number non-conservation:}
In the SM, baryon number $B$ is violated by non-perturbative instanton
processes \cite{'tHooft:1976up,'tHooft:1976fv}. Due to chiral anomalies both baryon number
$J^{\mu}_{B}$ and lepton number $J^{\mu}_{L}$ currents are
not conserved \cite{Adler:1969gk,Bell:1969ts}. However the
anomalous divergences come with an equal
amplitude and an opposite sign. Therefore $B-L$ remains conserved,
while $B+L$ may change via processes which interpolate between the
multiple non-Abelian vacua of $SU(2)$. The probability for the $B+L$
violating transition is however exponentially suppressed
\cite{'tHooft:1976up,'tHooft:1976fv}, but at finite temperatures when 
$T\gg M_W$, baryon violating transitions are in fact copious~\cite{Manton:1983nd}.

The $B$ violation also leads to proton decay
in GUTs~\cite{Langacker:1980js}.  The dimension $6$ operator $(QQQL)/\Lambda$
generates observable proton decay unless $\Lambda \geq 10^{15}$~GeV. In the
MSSM the bound is $\Lambda \geq 10^{26}$~GeV
because the decay can take place via a dimension $5$ operator. In the MSSM
superpotential there are also terms which can lead to $\Delta L =1$ and
$\Delta B=1$. Similarly there are other processes such as neutron-anti-neutron
oscillations in SM and in SUSY theories which lead
to $\Delta B =2$ and $\Delta B=1$ transitions. These
operators are constrained by the measurements of the proton lifetime,
which yield the bound  $\tau_{p} \geq 10^{33}$~years~\cite{Nakamura:2010zzi}.


\vspace{-0.5cm}\paragraph{\bf $C$ and $CP$ violation:}
The maximum $C$ violation occurs in weak interactions while neutral Kaon is an
example of $CP$ violation in the quark sector which has a relative
strength $\sim 10^{-3}$ \cite{Nakamura:2010zzi}. $CP$ violation is also expected
to be found in the neutrino sector. Beyond the SM there are many sources
for $CP$ violation. An example is the axion proposed for solving the
strong $CP$ problem in $QCD$ \cite{Peccei:1977hh,Peccei:1977ur}.

\vspace{-0.5cm}\paragraph{\bf Departure from thermal equilibrium:}
If $B$-violating processes are in thermal equilibrium, the inverse processes
will wash out the pre-existing asymmetry $(\Delta n_{b})_0$
\cite{Weinberg:1979bt}. This is a consequence of $S$-matrix unitarity
and $CPT$-theorem. However there are several
ways of obtaining an out-of-equilibrium process in the early
universe. Departure from a thermal equilibrium cannot be achieved
by mere particle physics considerations but is coupled to
the dynamical evolution of the universe.

\vspace{-0.5cm}
\subparagraph{\bf Out-of-equilibrium decay or scattering:}
The condition for out-of-equilibrium decay or scattering is that
the rate of interaction must be smaller than the expansion rate 
of the universe, i.e. $\Gamma < H$. The universe in a thermal equilibrium 
can not produce any asymmetry,
rather it tries to equilibrate any pre-existing asymmetry. 


\vspace{-0.5cm}
\subparagraph{\bf Phase transitions:}
They are ubiquitous in the early universe. The
transition could be of {\it first}, or of {\it second} (or of still higher)
order. First order transitions proceed by barrier penetration and
subsequent bubble nucleation resulting in a temporary departure from
equilibrium. The QCD and possibly electroweak phase transitions are examples
of first order phase transitions. The nature and details
of QCD phase transition is still an open
debate~\cite{Rajagopal:1992qz,Karsch:2000kv}.
Second order phase transitions have no barrier between
the symmetric and the broken phase. They are continuous and equilibrium
is maintained throughout the transition.


\vspace{-0.5cm}\subparagraph{\bf Non-adiabatic motion of a scalar field:}
Any complex scalar field carries $C$ and $CP$, but the symmetries
can be broken by terms in the Lagrangian. This can
lead to a non-trivial trajectory of a complex scalar field in the
phase space.
If a coherent scalar field is trapped in a local minimum
of the potential and if the shape of the potential changes to become a
maximum, then the field may not have enough time to readjust with
the potential and may experience completely non-adiabatic motion. This is
similar to a second order phase transition but it is the non-adiabatic
classical motion which prevails over the quantum fluctuations, and therefore,
departure from equilibrium can be achieved. If the field condensate carries
a global charge such as the baryon number, the motion can charge up the
condensate. This is the basis for the Affleck-Dine
baryogenesis~\cite{Affleck:1984fy}.

\subsection{Sphalerons}

At finite temperatures   $B+L$  violation in the SM can be large
due to sphaleron transitions  between degenerate gauge vacua  with different 
Chern-Simons numbers~\cite{Manton:1983nd,Klinkhamer:1984di}.
Thermal scattering produces sphalerons which in
effect decay in $B+L$ non-conserving ways below $10^{12}$~GeV
\cite{Bochkarev:1987wf}, and thus can exponentially wash away $B+L$ asymmetry.
The three important ingredients which play important role are following.


\vspace{-0.5cm}
\paragraph{\bf Chiral anomalies:}
In the SM there is classical conservation of the baryon and
lepton number currents $J^{\mu}_{B}$ and $J^{\mu}_{L}$, but
because of chiral anomaly (at the quantum level) the currents are not 
conserved~\cite{Adler:1969gk,Bell:1969ts}. Instead \cite{'tHooft:1976up},
\begin{eqnarray}
\partial_{\mu}J^{\mu}_{B}&=&-\frac{\alpha_{2}}{8\pi}N_{g}
W^{\mu\nu}_{i}\tilde W_{i\mu\nu}+
\frac{\alpha_{1}}{16\pi}N_{g}F^{\alpha\beta}\tilde F_{\alpha\beta}\,, \nonumber \\
\partial_{\mu}J^{\mu}_{L}&=&-\frac{\alpha_{2}}{8\pi}N_{g}
W^{\mu\nu}_{i}\tilde W_{i\mu\nu}+
\frac{\alpha_{1}}{16\pi}N_{g}F^{\alpha\beta}
\tilde F_{\alpha\beta}\,,
\end{eqnarray}
where $N_{g}$ is the number of generations, $\alpha_2$ and $\alpha_1$
($W_{i\mu\nu}$ and $F_{\mu\nu}$) are respectively the $SU(2)$ and $U(1)$
gauge couplings (field strengths). Note that at the quantum level $B+L\neq 0$ is violated,
but $B-L=0$ is still conserved.


\vspace{-0.5cm}
\paragraph{\bf Gauge theory vacua:}
in the $SU(2)$ gauge group, the vacua are
classified by their homotopy class $\{\Omega_{n}({\rm r})\}$,
characterized by the winding number $n$ which labels the so called
$\theta$-vacua \cite{'tHooft:1976fv,Polyakov:1976fu}. A gauge invariant quantity
is the difference in the winding number (Chern-Simons number)
\begin{equation}
\label{cs1}
N_{CS}\equiv n_{+}-n_{-}=\frac{\alpha_{2}}{8\pi}\int d^4x W^{\mu\nu}_{a}
\tilde W_{a\mu\nu}\,.
\end{equation}
In the electroweak sector the field density $W\tilde W$ is related
to the divergence of $B+L$ current. Therefore a change in $B+L$ reflects
a change in the vacuum configuration determined by the difference in winding 
number
\begin{equation}
\Delta{(B+L)}=-\frac{\alpha_{2}}{4\pi}
N_{g}\int d^4x W_{a}^{\mu\nu}\tilde W_{a\mu\nu}=-2N_{g}N_{CS}\,.
\end{equation}
For three generations of SM leptons and quarks the minimal violation is
$\Delta(B+L)=6$. Note that the proton decay $p\rightarrow e^{+}\pi^{0}$
requires $\Delta(B+L)=2$, so that despite $B$-violation, proton decay is
completely forbidden in the SM. The probability amplitude for tunneling  from an $n$ vacuum at
$t\rightarrow -\infty$ to an $n+N_{CS}$ vacuum at $t\rightarrow +\infty$
can be estimated by the WKB method~\cite{'tHooft:1976fv}
\begin{equation}
P(N_{CS})_{B+L}\sim \exp\left(\frac{-4\pi N_{CS}}{\alpha_{2}(M_{Z})}
\right)\sim 10^{-162N_{CS}}\,.
\end{equation}
The baryon number violation rate is  negligible at zero temperature, but as argued at
finite temperatures the situation is completely different~\cite{Manton:1983nd,Kuzmin:1985mm}.


\vspace{-0.5cm}
\paragraph{\bf Thermal tunneling:}
below the critical temperature of the electroweak phase transition,
the sphaleron rate is exponentially suppressed
\cite{Carson:1990jm}:
\begin{equation}
\label{gammalow}
\Gamma \sim 2.8\times 10^{5}\kappa T^4\left(\frac{\alpha_2}{4\pi}\right)^4
\left(\frac{E_{sph}(T)}{B(\lambda^2/g)}\right)^7e^{-E_{sph}/T}\,.
\end{equation}
where $\kappa$ is the functional determinant which can take the values
$10^{-4}\leq \kappa \leq 10^{-1}$ \cite{Dine:1992vs}.
Above the critical temperature the rate is however unsuppressed.
Since the Chern-Simons number changes at most by
$\Delta N_{CS} \sim 1$, one can estimate from Eq.~(\ref{cs1}) that
$\Delta N_{CS}\sim g_{2}^{2}l_{sph}^{2}W_{i}^{2}\sim 1\rightarrow W_{i}\sim ({1}/{g_{2}l_{sph}})$.
Therefore a typical energy of the sphaleron configuration is given by
$E_{sph}\sim l_{sph}^3(\partial W_{i})^2 \sim (1/g_{2}^2l_{sph})$.
At temperatures greater than the critical temperature there is no
Boltzmann suppression, so that the thermal energy $\propto T\geq E_{sph}$.
This determines the size of the sphaleron: $l_{sph}\geq {1}/{g_2^2T}$
Based on this coherence length scale one
can estimate the baryon number violation per volume $\sim l_{sph}^3$,
and per unit time $\sim l_{sph}$. On dimensional grounds the transition
probability would then be given by
\begin{equation}
\label{sphr}
\Gamma_{sph}\sim ({1}/{l_{sph}^3 t}) \sim \kappa(\alpha_{2}T)^4\,.
\end{equation}
where $\kappa$ is a constant which incorporates various uncertainties.
However, the process is inherently non-perturbative, and it has been
argued that damping of the magnetic field in a plasma suppresses the
sphaleron rate by an extra power of $\alpha_2$~\cite{Arnold:1996dy}, with
the consequence that $\Gamma_{sph}\sim \alpha_{2}^5T^4$. Lattice simulations
with hard thermal loops also give $\Gamma_{sph}\sim {\cal O}(10)\alpha_{2}^5T^4$~\cite{Moore:1998sw}.

\vspace{-0.5cm}
\paragraph{\bf Washing out $B+L$:}

Assuming that in the early universe the SM {\it d.o.f} are in equilibrium, the
transitions $\Delta N_{CS}=+1$ and $\Delta N_{CS}=-1$ are equally probable. The 
ratio of rates for the two transitions is given by ${\Gamma_{sph~+}}/{\Gamma_{sph~-}}=\exp(-\Delta F/T)$,
where $\Delta F$ is the free energy difference between the two vacua.
Because of a finite $B+L$ density, there is a net chemical potential
$\mu_{B+L}$. Therefore
one obtains \cite{Bochkarev:1987wf}
\begin{equation}
\label{old1}
\frac{dn_{B+L}}{dt}=\Gamma_{sph~+}-\Gamma_{sph~-}\sim N_{g}\frac{
\Gamma_{sph}}{T^3}n_{B+L}\,.
\end{equation}
It then follows that an exponential depletion of $n_{B+L}$ due to sphaleron
transitions remains active as long as
\begin{equation}
\label{active}
\frac{\Gamma_{sph}}{T^3} \geq H ~~\Rightarrow ~~T \leq \alpha_{2}^{4}
\frac{M_{\rm P}}{g_{\ast}^{1/2}}\sim 10^{12}~{\rm GeV}\,.
\end{equation}
This result imply that below $T =10^{12}$~GeV,
the sphaleron transitions can wash out any $B+L$ asymmetry being produced
earlier in a time scale $\tau \sim (T^3/N_{g}\Gamma_{sph})$. This seems to
wreck GUT baryogenesis based on $B-L$ conserving groups such as the
minimal $SU(5)$.


\subsection{Mechanisms for baryogenesis}

There are several scenarios for baryogenesis, the 
main contenders being GUT baryogenesis, electroweak baryogenesis, leptogenesis,
and baryogenesis through the decay of a field condensate, or
Affleck-Dine baryogenesis. Here we give a brief description of
these various alternatives.

\subsubsection{GUT-baryogenesis}

This model relied on out-of-equilibrium decays of heavy GUT gauge bosons
$X,Y\rightarrow qq$, and $X,Y\rightarrow \bar q\bar l$, for reviews see
\cite{Dolgov:1991fr,Kolb:1988aj}. The decay rate of the gauge boson 
goes as $\Gamma_{X}\sim \alpha_{X}M_{X}$,
where $M_{X}$ is the mass of the gauge boson and $\alpha^{1/2}_{X}$ is the
GUT gauge coupling. Assuming that the universe was in thermal equilibrium
at the GUT scale, the decay temperature is given by
\begin{equation}
T_{D}\approx g_{\ast}^{-1/4}\alpha_{X}^{1/2}(M_{X}M_{\rm P})^{1/2}\,,
\end{equation}
which is smaller than the gauge boson mass. Thus, at $T\approx T_{D}$,
one expects $n_{X}\approx n_{\bar X}\approx n_{\gamma}$, and hence the
net baryon density is proportional to the photon number density
$n_{B}=\Delta B n_{\gamma}$. However below $T_{D}$ the gauge boson abundances
decrease and eventually they go out-of-equilibrium. The net
entropy generated due to their decay heats up the universe with a temperature
which we denote here by $T_{R}$. Let us naively assume that the energy
density of the universe at $T_{D}$ is dominated by the $X$ bosons with
$\rho_{X} \approx M_{X}n_{X}$, and their decay products lead to radiation with
an energy density $\rho =(\pi^2/30)g_{\ast}T_{R}^4$, where
$g_{\ast}\sim {\cal O}(100)$ for $T\geq M_{GUT}$. Equating
the expressions for the two energy densities one obtains
\begin{equation}
n_{X}\approx \frac{\pi^2}{30}g_{\ast}\frac{T_{R}^4}{M_{X}}\,.
\end{equation}
Therefore the net baryon number comes out to be
\begin{equation}
\label{gutb}
B\equiv \frac{n_{B}}{s}\approx \frac{\Delta B n_{X}}{g_{\ast}n_{\gamma}}
\approx \frac{3}{4}\frac{T_{R}}{M_{X}}\Delta B\,.
\end{equation}
$T_{rh}$ is determined from the relation
$\Gamma^2_{X}\approx H^2(T_{D})\sim (\pi^2/90)g_{\ast}T^4_{R}/M_{\rm P}^2$.
Thus,
\begin{equation}
B\approx  \left(\frac{g_{\ast}^{-1/2}\alpha_{X}M_{\rm P}}{M_{X}}
\right)^{1/2}\Delta B\,.
\end{equation}
Uncertainties in $C$ and $CP$ violation are now hidden in $\Delta B$,
but can be tuned to yield total $B\sim 10^{-10}$ in many models.

Above we have assumed that the universe is in thermal equilibrium
when $T\geq M_{X}$. This might not be true, since for $2\leftrightarrow 2$
processes the scattering rate is given by $\Gamma\sim \alpha^2 T$, which
becomes smaller than $H$ at sufficiently high temperatures.
Elastic $2\rightarrow 2$ processes maintain thermal contact typically
only up to a maximum temperature $\sim 10^{14}$~GeV, while chemical
equilibrium is lost already at $T\sim 10^{12}$~GeV~\cite{Enqvist:1990dp,Elmfors:1993pz}.


\subsubsection{Electroweak baryogenesis}

A popular baryogenesis candidate is based on the
electroweak phase transition, during which one can in principle  meet all
the Sakharov conditions. There is the sphaleron-induced baryon number
violation above the critical temperature, various sources of $CP$
violation, and an out-of-equilibrium environment if the phase transition
is of the first order. In that case bubbles of broken $SU(2)\times U(1)_Y$
are nucleated into a symmetric background with a Higgs field profile
that changes through the bubble wall~\cite{Kuzmin:1985mm,Kuzmin:1987wn}.

There are two possible mechanisms which work in different regimes:
{\it local} and {\it non-local} baryogenesis. In the local case both $CP$ violation
and baryon number violation takes place near the bubble wall. This requires
the velocity of the bubble wall to be greater than the speed of the sound
in the plasma \cite{Ambjorn:1990wn,Turok:1990in,Turok:1990zg}, and the 
electroweak phase transition to be strongly first order with thin bubble walls. 

In the non-local case the bubble wall velocity speed is small
compared to the sound speed in the plasma.
In this mechanism the fermions, mainly the top quark and the tau-lepton,
undergo $CP$ violating interactions with the bubble wall, which results
in a difference in the reflection and the transmission probabilities for the
left and right chiral fermions. The net outcome is an overall chiral flux
into the unbroken phase from the broken phase. The flux is then converted
into baryons via sphaleron transitions inside the unbroken phase.
The interactions are taking place in a thermal equilibrium except for
the sphaleron transitions, the rate of which is slower than the rate
at which the bubble sweeps the space~\cite{Nelson:1991ab,Cohen:1993nk}.

For a constant velocity profile of the bubble, $v_{w}$, the net baryon  asymmetry is generated by:
\begin{equation}
n_{B}\simeq -\frac{\Gamma_{sph}}{T}\int dt~\mu_{B}\,,
\end{equation}
where $\mu_{B}$ is the chemical potential, which determines the tilt in the free energy of the 
sphaleron transitions, and numerically it is equivalent to: $\mu_{B}\equiv \rho(z-v_{w}t)/[(2N+5/3)T^{2}]$.
Here $\rho$ determines the profile of the bubble, and $N$ denotes the number of Higgs doublets.
The net baryon asymmetry can be calculated by following Eq.~(\ref{sphr}):
\begin{equation}
\frac{n_{B}}{s}\simeq \kappa \alpha_{2}^{4}\left(\frac{100}{\pi^{2}g_{\ast}}\right) \left(\frac{F_{z}}{v_{w}T^{3}}\right)\tau T\,,
\end{equation}
where $\tau$ is the transport time of the scattered fermions off the bubble wall, and $F_{z}\equiv \int _{0}^{\infty} dz \rho(z)$.
For the maximum wall velocity $v_{w}\sim 1/\sqrt{3}$ and a typical: $\tau T\sim 10-1000$ for top quarks, the 
maximum baryon asymmetry is given by:  ${n_{B}}/{s}\simeq 10^{-3}{F_{z}}/({v_{w}T^{3}})$. The function $F_{z}$ also
takes into account the $CP$ phase, in the favorable scenario one would expect $F_{z}/(v_{w}T^{3})\sim 10^{-6}$.
The details of the transport equations can be found in Refs.~\cite{Nelson:1991ab,Kainulainen:2001cn}.

One great challenge for the electroweak baryogenesis is the smallness
of $CP$ violation in the SM at finite temperatures. It has been pointed out that
an additional Higgs doublet \cite{McLerran:1990zh,Turok:1990zg}
would provide an extra source for $CP$ violation in the Higgs sector.
However, the situation is much improved in the MSSM where there are
two Higgs doublets $H_{u}$ and $H_{d}$, and two important sources of
$CP$ violation~\cite{Ellis:1982tk}. The Higgses couple to the charginos
and neutralinos at one loop level leading to a $CP$ violating contribution.
There is also a new source of $CP$ violation in the mass matrix of the
top squarks which can give rise to considerable $CP$ violation \cite{Huet:1995sh}.

Bubble nucleation depends on the thermal tunneling rate, and the expansion
rate of the universe. The tunneling rate has to overcome the expansion rate
in order to have a successful phase transition via bubble
nucleation at a given critical temperature $T_{c}>T_{t}>T_0$.
The effective potential for the Higgs at finite temperatures can be computed,
which takes the form:
\begin{equation}
V_{eff}(\phi,T)=(-\mu^2+\alpha T^2)\phi^{2} -\gamma T\phi^{3}+(\lambda/4)\phi^{4}
\end{equation}
%
The order parameter is given by the ratio of $\langle \phi(T_{c})/T_{c}\rangle \sim \gamma/\lambda$, which has to
be larger than one for first order phase transition.
For $T_{c}\sim 100$~GeV, 
one obtains the condition for the sphaleron energy~\cite{Shaposhnikov:1987tw,Rubakov:1996vz}
\begin{equation}
\frac{E_{sph}(T_c)}{T_c}\geq 7\log\left[\frac{E_{sph}(T_c)}{T_c}\right]+
9\log(10)+\log(\kappa)\,.
\end{equation}
which implies~\cite{Bochkarev:1990gb}
\begin{eqnarray}
\label{b1}
\frac{E_{sph}(T_{c})}{T_{c}}&\geq& 45\,~{\rm for}~\kappa =10^{-1}\,,
\end{eqnarray}
In terms of
the Higgs field value at $T_c$,
\begin{equation}
\frac{\phi(T_{c})}{T_{c}}=\frac{g}{4\pi B(\lambda/g_2)}
\frac{E_{sph}(T_c)}{T_c}\sim \frac{1}{36}\frac{E_{sph}(T_c)}{T_c}\,,
\end{equation}
where $g$ is gauge coupling of  $SU(2)_{L}$, and  $B\sim 1.87$. Then the bounds in
Eqs.~(\ref{b1}) translate to
\begin{eqnarray}
\label{b3}
{\phi(T_{c})}/{T_{c}}\geq 1.3,
\end{eqnarray}
which implies that the phase transition should be strongly first
order in order that sphalerons do not wash away all the produced baryon
asymmetry. This result is the main constraint on electroweak baryogenesis.

Lattice studies suggest that in the SM the phase transition
is strongly first order only below Higgs mass $m_{H}\sim 72$~GeV
\cite{Kajantie:1996mn,Rummukainen:1998as}. Above this scale
the transition is just a cross-over. Such a Higgs mass is clearly
excluded by the LEP measurements~\cite{Nakamura:2010zzi}, thus excluding electroweak
baryogenesis within the SM.  However, this opens up a possibility to include new physics 
beyond the SM.

\vspace{-0.5cm}
\paragraph{\bf Electroweak baryogenesis induced by new physics:}

it was pointed out that by modifying the SM Higgs self-interactions, especially the cubic term, it is possible to enhance the first order phase 
transition~\cite{Anderson:1991zb}. One such example has been considered in~\cite{Mohapatra:1992pk,Grojean:2004xa}
where non-renormalizable contribution to the Higgs potential has been considered of type:
\begin{equation}
V(\Phi)=\lambda \left( \Phi^\dagger\Phi-\frac{v^2}{2}\right)^2
 +\frac{1}{\Lambda^2}\left( \Phi^\dagger\Phi-\frac{v^2}{2}\right)^3
\label{eq:Vphi}
\end{equation}
where $\Phi$ is the SM electroweak Higgs doublet and $\Lambda$ is the scale of new physics which induces the corrections
below the energy scale of $\Lambda\sim {\cal O}(1)$~TeV. At zero temperature the CP-even
scalar state can be expanded in terms of its zero-temperature VEV, $\langle \phi\rangle =v_0\simeq 246$~GeV, 
and the physical Higgs boson $H$: $\Phi=\phi/\sqrt{2}=(H+v_0)/\sqrt{2}$.

The finite temperature effects are taken into account by adding a thermal mass to the potential: 
\begin{equation}
V(\phi,T)=  c T^2 \phi^{2} /2+V(\phi,0)
\end{equation}
where $c$ is given in the high-temperature expansion of the one-loop thermal potential:
\begin{equation}
c=\frac{1}{16}\left(4 y_t^2+ 3 g^2 + g'^2 + 4 \frac{m_H^2}{v_0^2} 
- 12 \frac{v_0^2}{\Lambda^2}\right),
\end{equation}
where $g$ and $g'$ are the $SU(2)_L$ and $U(1)_Y$ gauge couplings, and $y_t$ is the top Yukawa coupling.
Note that there is no trilinear term in the effective potential.

The critical temperature $T_c$ at which the minima $\phi \neq 0$
and $\phi=0$ are degenerate is given by
\begin{equation}
\label{T_c}
T_c^2= \frac{\Lambda^4 m_H^4+2\Lambda^2 m_H^2 v_0^4-3v_0^8}{16c\Lambda^2 v_0^4}.
\end{equation}
The VEV of the Higgs field at the critical 
temperature in terms of $m_H$, $\Lambda$ and $v_0$ is
\begin{equation}
\label{v_c}
\langle\phi^2 (T_c)\rangle =v^2_c = \frac{3}{2}v_0^2
-\frac{m_H^2\Lambda^2}{2v_0^2}.
\end{equation}
From Eqs.~(\ref{T_c}) and~(\ref{v_c}), one finds that for any given $m_H$
there is an upper bound on $\Lambda$ to make sure that the phase transition is always
first order ($v^2_c>0$), and there
is a lower bound on $\Lambda$ to make sure that the $T=0$ minimum 
at $\phi\neq 0$ is a global minimum ($T_c^2>0$).  These two combinations chart out a 
region where the phase transition is indeed first order:
\begin{equation}
\label{eq:Lambda-bounds}
\mathrm{max} \left( \frac{v_0^2}{m_H} , \frac{\sqrt{3} v_0^2}{\sqrt{m_H^2+2m_c^2}} \right)
<\Lambda < \sqrt{3}\frac{v_0^2}{m_H}
\end{equation}
where $m_c=v_0 \sqrt{(4 y_t^2 + 3 g^2 + g'^2)/8}\approx 200$~GeV. In order to ensure that the 
thermal mass correction is positive: $ c> 0 \rightarrow \Lambda> {\sqrt{3} v_0^2}/{\sqrt{m_H^2+2m_c^2}}$. 
For these ranges of $\Lambda$ the ratio $v_{c}/T_{c}>1$, ensuring a successful sphaleron transition for 
the Higgs mass $m_{H}\geq 115$~GeV. One nice aspect of this model is that the non-renormalizable 
scale $\Lambda$ can also be constrained from the precision electroweak observable, which can be 
tested in near future by the LHC~\cite{Grojean:2004xa}.


\subsubsection{Electroweak baryogenesis in MSSM}

In the MSSM the ratio $\Phi(T_{c})/T_{c}$ can be increased by virtue of the scalar loops
which can make the cubic term in the temperature dependent Higgs potential large;
$V_{eff}(\varphi,T)=(-\mu^2+\alpha T^2)\varphi^{2} -\gamma T\varphi^{3}+(\lambda/4)\varphi^{4}$.
In particular the right handed stop $\widetilde t_{R}$
coupling to the Higgs with a large Yukawa coupling. This leads to
a strong first order phase transition -- as the ratio of $\Phi(T_{c})/T_{c}\sim \gamma/\lambda\geq 1$, 
where $\gamma$ determines the order parameter~\cite{Carena:1996wj,Laine:1996ms,Cline:1997vk,Laine:1998qk,Cline:2000nw}.

The finite temperature cubic term is given by: $\gamma T\varphi^{3}\simeq (T/4\pi)[m^{2}_{\widetilde t_{R}}(\varphi, T)]^{3/2}$,
where the lightest right handed stop mass
\begin{equation}
m^2_{\widetilde t_{R}}\approx m^2_{U}+\xi T^{2}+0.15M^2_{Z}\cos(2\beta)+m^2_{t}\left(1-
\frac{{\widetilde A}^2_{t}}{m^2_{Q}}\right)\,,
\end{equation}
where ${\widetilde A}_{t}=A_{t}-\mu/\tan(\beta)$ is the stop mixing parameter, $A_{t}$
is the trilinear term in the MSSM superpotential, and $\mu$ is the soft-SUSY breaking 
mass parameter for the right-handed stop. The coefficient $\gamma $ of the cubic term 
$\gamma T\varphi^3$ in the effective potential reads
\begin{equation}
\gamma_{MSSM} \approx \gamma_{SM}+\frac{h_{t}^3\sin^3(\beta)}{4\sqrt{2}\pi}
\left(1-\frac{{\tilde A}^2_{t}}{m^2_{Q}}\right)^{3/2}\,,
\end{equation}
and can be at least one order of magnitude larger than
$\gamma_{SM}$. The implications for the particle spectrum are:

\begin{itemize}
\item A light right-handed stop: 
$120~{\rm GeV} \leq m_{\widetilde t_{1}}\leq 170~{\rm GeV}\leq m_{t}$.

\item A heavy left-handed stop: 
$m_{Q_{3}}\geq 2$~TeV.

\item A light SM-like Higgs: $m_{H}\leq 120$~GeV, for $5<\tan\beta < 10$.
\end{itemize}
The present LEP constraint on the lightest $CP$-even Higgs mass is
$m_{H}\geq 115$~GeV~\cite{Nakamura:2010zzi}. Note that within MSSM, the 
lightest Higgs mass is bounded by: $m^{2}_{H}\leq M_{Z}^{2}\cos^{2}2\beta$.
Hence, even an MSSM-based
electroweak baryogenesis may be at the verge of being ruled out.

MSSM also provides new $CP$ violating complex phases 
in the Higgsino sector, i.e. $arg(\mu M_{1,2})\geq 10^{-2}$, with 
$\mu,~M_{1,2}\leq 400$~GeV.  The $CP$-violating phases are
also constrained by the electric dipole moments. To match the
observational limit on $|d_{e}|< 1.6\times 10^{-27}$ e~cm~\cite{Regan:2002ta}, one requires 
first and second generation sfermion masses greater than $10$~TeV.
while the 2-loop electron dipole moment contribution comes out to be:
$|d_{e}|\geq 2\times 10^{-28}$~e~cm.

The  definitive test of the MSSM based electroweak baryogenesis
will obviously come from the Higgs and the stop searches at the 
LHC~\cite{Carena:2002ss,Chung:2008aya}.


\subsubsection{Electroweak baryogenesis beyond MSSM}
Some of these problems of MSSM can be resolved in nMSSM (next-to minimal SUSY SM), with the help 
of introducing an  extra singlet in the MSSM superpotential: 
$W=m^{2}S +\lambda SH_{u}H_{d}+W_{MSSM}$. The $S$ field gets a VEV to 
explain the $\mu\equiv \lambda \langle S\rangle $-term, but it also generates a singlet tadpole -- its contribution to 
the vacuum energy, $\delta V=t_{s}S\sim (1/16\pi^{2})^{n} (S/M_{\rm P})F_{s}^{2}$, can be suppressed with the help of discrete 
symmetries, ${\mathbb Z}^{R}_{5}$ or ${\mathbb Z}^{R}_{7}$, where $F_{s}\sim ~m_{soft}M_{\rm P}$~\cite{Abel:1995wk,Panagiotakopoulos:1998yw}.
As a result the soft-SUSY breaking Higgs potential becomes: 
\begin{equation}
V_{soft}=t_{s}(S+h.c.)+m^{2}_{s}|S|^{2}+a_{\lambda}(SH_uH_d+h.c.)+V_{MSSM}\,,
\end{equation}
Note that the trilinear, $a_{\lambda}SH_uH_d$ now contributes to the $\gamma $-term at the tree level, indicating 
potentially stronger first order phase transition even without a light stop and for $m_{H}> 120$~GeV. The 
$CP$ phases are distributed in gaugino  masses  as well as in the singlet, but not in the tree level of $a_{\lambda}$.

One can similarly proceed with 4 SM singlets, and the Higgs doublet as in the case of $U(1)^{\prime}$ electroweak
baryogenesis discussed in Ref.~\cite{Kang:2004pp}, for a review see~\cite{Kang:2009rd}, where the superpotential contains: 
\begin{equation}
W=h SH_{u}H_{d}+\lambda S_{1}S_{2}S_{3}+W_{MSSM}\,.
\end{equation}
 It is assumed that the $U(1)^{\prime}$ is broken at 
higher VEVs, such as $1-2$~TeV, and then the electroweak symmetry is broken at lower scales. The singlets 
$S_{1},~S_{2},~S_{3}$ have VEVs greater than those of $S$ and, $H_{u}$ and $H_{d}$. The mass of 
$Z'$ bosons are $M_{Z'}\sim {\cal O}(1)$~TeV. The tree level Higgs potential can now contain $CP$ violating 
 contributions from the phases $\beta_{1},~\beta_{2}$~\cite{Kang:2004pp,Kang:2009rd}:
\begin{eqnarray}
&&V_{soft}=V_{MSSM}+m^{2}_{s}|S|^{2}+\sum_{i=1}^{3}m_{S_{i}}^{2}|S_{i}|^{2}\, \nonumber \\
&&-2A_{h}h|S||H_{u}^{0}||H_{d}^{0}|\cos\beta_{3}
-2A_{\lambda}\lambda|S_{1}||S_{2}||S_{3}|\cos\beta_{4}\, \nonumber \\
&&-2m_{SS_{1}}^{2}|S||S_{1}|\cos\beta_{1}-2m^{2}_{SS_{2}}|S||S_{2}|\cos\beta_{2}\, \nonumber \\
&&-2|m^{2}_{S_{1}S_{2}}||S_{1}||S_{2}|\cos(-\beta_{1}+\beta_{2}+\gamma)
\end{eqnarray}
The potential can yield strong first order phase transition without large stop masses, and the new 
contributions to electron dipole moments can be tamed by tuning the Yukawa sector~\cite{Kang:2004pp,Kang:2009rd}.


\subsubsection{Thermal Leptogenesis}

At temperatures $10^{12}~{\rm GeV}\geq T\geq 100$~GeV, the $B+L$ is completely erased by the sphaleron transitions, a net
baryon asymmetry in the universe can still be generated from a non-vanishing
$B-L$ \cite{Harvey:1981cu,Fukugita:1986hr,Luty:1992un}, even if there were no baryon number violating
interactions. The lepton number violating interactions can produce baryon asymmetry, a process which is known as leptogenesis,
for recent reviews, see \cite{Buchmuller:2004nz,Buchmuller:2005eh,Davidson:2008bu}.

The lepton number violation requires physics beyond the SM. 
The most attractive mechanism arises in $SO(10)$ which is
left-right symmetric (for details, see~\cite{Langacker:1980js}),
and has a natural foundation for the see-saw mechanism 
\cite{Minkowski:1977sc,Mohapatra:1979ia,Yanagida,Gell-Mann} as it incorporates a singlet right-handed Majorana
neutrino $N_{R}$ with a mass $M_{R}$. A lepton number violation appears
when the Majorana right handed neutrino decays into the SM lepton doublet
and Higgs doublet, and their $CP$ conjugate state through
\begin{equation}
\label{ndecay}
N_{R}\rightarrow H+ l\,, \quad \quad N_{R}\rightarrow \bar H+\bar l\,,
\end{equation}
where ($H$) $l$ is the SM (Higgs) lepton. The relevant
$L$ violating interaction is then given by
\begin{equation}
\label{Linteract}
\mathcal{L} \supset \frac{1}{2} (M_N)_{ii} N_i N_i + y_{ij} N_i \bar{\ell}_j i \tau_2 H^* + {\rm h.c.}\;,
\end{equation}
where $i,~j=1,2,3$.
The above interaction is also responsible for generating the observed neutrino
masses via the canonical seesaw mechanism~\cite{Buchmuller:1997yu,Buchmuller:2000as,Buchmuller:2002rq,Akhmedov:2003dg}, as required by the neutrino
oscillation data \cite{GonzalezGarcia:2010er}. This mass turns out to
be $m_\nu \approx |y|^2 v^2/M_N$ with $v=174$~GeV, what implies
right-handed neutrino mass scale of $M_{N}\sim {\cal O}(10^{14})$ GeV for
$|y|\sim 1$ and $m_\nu\sim0.1$~eV.

Assuming a normal hierarchy in the heavy right handed neutrino sector, $M_{1}\ll M_{2},~M_{3}$
(corresponding to $N_{1},~N_{2},~N_{3}$).
The $CP$ asymmetry can be estimated from the $N_1$ decay, the asymmetry
is generated through the interference between tree level and one-loop
diagrams, which is given by \cite{Fukugita:1986hr,Luty:1992un,Flanz:1994yx,Covi:1996wh,Plumacher:1997ru}
\begin{eqnarray}\label{CP-phase}
\epsilon &= & \frac{\Gamma(N_{1}\rightarrow lH)-\Gamma(N_{1}\rightarrow \bar l\bar H)}
{\Gamma(N_{1}\rightarrow lH)+\Gamma(N_{1}\rightarrow lH)}\,,\\
&=& \frac{1}{8\pi}\frac{1}{yy^{\dagger}}\sum_{i=1,2,3} {\rm Im}[(yy^{\dagger})_{1i}]^{2}
f( {M_{i}^{2}}/M_{1}^2)\,,
\end{eqnarray}
where $f$ is a function which represents radiative corrections. In the case of SM, 
$f(x)=\sqrt{x}[(x-2)/(x-1)+(x-1)\ln(1+{1}/{x})]$, and in the case of MSSM,
$f(x)=\sqrt{x}[2/(x-1)+\ln(1+{1}/{x})]$. 

Let us  take an example of the SM where
the $CP$ phase can be labeled by, $|\epsilon | = 3 M_{1}/ (16 \pi v^2)
  \sqrt{\Delta m^2_{\rm atm}} ~\sin \delta $,
where $\Delta m^2_{\rm atm}$ is the atmospheric mass scale of light
neutrinos \cite{GonzalezGarcia:2010er} and $\delta$ is the effective
$CP$ violating phase. The total lepton asymmetry is then given by
\begin{equation}
\label{unfavored}
\eta_{L}=|\epsilon| Y_{N_{1}}\kappa
\end{equation}
where $Y_{N_{1}}$ is the abundance of the right handed Majorana neutrino $N_{1}$ and
$\kappa$ is a thermal wash-out factor, which takes into account that the scatterings such 
as $\bar \ell H\leftrightarrow \ell \bar H$ tend to wash out any lepton asymmetry being created.

In order to process the total lepton asymmetry into baryons,  we need to know 
the chemical potentials~\cite{Khlebnikov:1988sr}
\begin{equation}
\label{mu5}
B=\sum_{i}(2\mu_{qi}+\mu_{u_{R}i}+\mu_{d_{R}i})\,, \quad \quad
L=\sum_{i}(2\mu_{li}+\mu_{e_{R}i})\,,
\end{equation}
where $i$ denotes three leptonic generations. The Yukawa interactions
establish an equilibrium between the different generations
($\mu_{li}=\mu_{l}$ and $\mu_{qi}=\mu_{q}$, etc.), and one obtains
expressions for $B$ and $L$ in terms of the number of colors $N=3$, and the
number of charged Higgs fields $N_H$
\begin{equation}
B=-\frac{4N}{3}\mu_{l}\,,~~L=\frac{14N^2+9NN_{H}}{6N+3N_{H}}\mu_{l}\,,
\end{equation}
together with a relationship between $B$ and $B-L$ \cite{Khlebnikov:1988sr}
\begin{equation}
B=\left(\frac{8N+4N_{H}}{22N+13N_{H}}\right)(B-L)\,.
\end{equation}
The final asymmetry is then given by $B=(28/79)(B-L)$
in the case of SM and $B=(8/23)(B-L)$ for the MSSM~\cite{Khlebnikov:1988sr}

The baryon asymmetry based on the decays of right handed neutrinos in a thermal bath has been computed
within MSSM~\cite{Buchmuller:2002rq,Buchmuller:2004nz,Giudice:2003jh},
where besides the right handed neutrinos the right handed (s)neutrinos also participate in the interactions.
The decay of a RH (s)neutrino with mass $M_i$ results in a lepton
asymmetry via one-loop self-energy and vertex corrections, see Eq.~(\ref{CP-phase}).  If the asymmetry is mainly produced from the decay
of the lightest right handed states, and assuming hierarchical right handed (s)neutrinos
$M_1 \ll M_2,M_3$, we will have~\cite{Davidson:2002qv}
\begin{eqnarray} \label{baryontherm}
\eta & \simeq & 3 \times 10^{-10} 
\kappa  \left({m_3 - m_1 \over 0.05~{\rm eV}}\right) 
\left({M_1 \over 10^9~{\rm GeV}}\right)\,, 
\end{eqnarray}
for ${\cal O}(1)$ $CP$-violating phases ($m_{\nu_1} < m_{\nu_2} < m_{\nu_3}$ are the
masses of light mostly light handed neutrinos). Here $\kappa$ is the efficiency
factor accounting for the decay, inverse decay and scattering
processes involving the right handed states~\cite{Buchmuller:2002rq,Buchmuller:2003gz,Buchmuller:2004nz,Giudice:2003jh}.

A decay parameter $K$ can be defined as
\begin{equation} \label{decpar}
K \equiv {\Gamma_1 \over H(T = M_1)}\,,
\end{equation}
where $\Gamma_1$ is the decay width of the lightest right handed (s)neutrino.
If $K < 1$, the decay of right handed states will be out of equilibrium at all times. In this case
the right handed states, which are mainly produced via scatterings of the left handed
(s)leptons off the top (s)quarks and electroweak gauge/gaugino
fields, never reach thermal equilibrium.  The
cross-section for producing the right handed (s)neutrinos is $\propto
T^{-2}~(M^2_1)$, when $T > M_1~(< M_1)$, and hence most of them are
produced when $T \sim M_1$.  The efficiency factor reaches its maximum
value for $\kappa \simeq 0.1$ when ${m}_{\nu_1} = 10^{-3}$~eV. For
larger values of ${m}_{\nu_{1}}$ it drops again, because the
inverse decays become important and suppress the generated
asymmetry. Producing sufficient asymmetry then sets a lower bound,
$M_1 \geq 10^9$ GeV~\cite{Buchmuller:2002rq,Buchmuller:2003gz,Buchmuller:2004nz,Giudice:2003jh}.  Successful thermal
leptogenesis therefore requires that $T_{R} \geq 10^9$~GeV. 


\vspace{-0.5cm}
\paragraph{\bf Resonant leptogenesis:}
If the mass splitting between, say $M_{1},~M_{2}$,  is comparable to their
decay widths, the $CP$ asymmetry resonantly gets enhanced, see Eq.~(\ref{CP-phase}). For
example, let us consider $N_1$ and $N_2$ . The dominant contribution to 
the $CP$ asymmetry arises in the mixing of $N_1$ and $N_2$, 
and it is given by~\cite{Plumacher:1997ru,Pilaftsis:2003gt,Pilaftsis:2005rv}
\begin{equation}
\epsilon_1 = \frac{{\rm Im} (y^\dagger y)_{12}^2}{8\pi (y^\dagger y)_{11}}
\frac{(M_1^2-M_2^2)M_1 M_2}{(M_1^2-M_2^2)^2+ (M_2 \Gamma_2 - M_1\Gamma_1)^2}~.
\end{equation}
Now, assuming $M_1\sim M_2$ and $M_1-M_2 \sim \Gamma_1-\Gamma_2$  Therefore, a large $L$
asymmetry can be produced even if the initial abundance of $N_1$ and
$N_2$ is small. 


\vspace{-0.5cm}
\paragraph{\bf Flavored leptogensis:} 
So far we have assumed that all the leptonic flavors, i.e. $\tau,~\mu, e$,
behave alike in a thermal bath. Especially in a non-SUSY case, 
where we can imagine a thermal bath of SM {\it d.o.f} with a temperature $\leq 10^{12}$~GeV,
the $\tau$-Yukawa interactions are in thermal equilibrium, while temperatures
below $10^{9}$~GeV the muon Yukawa interactions are faster than the expansion
rate of the universe and the leptogenesis rate. Since the wash out factor $\kappa$ is inversely proportional
 to the lepton violating interaction rates, so each of the flavored symmetries is subject 
 to its own wash out effect~\cite{Abada:2006fw,Abada:2006ea}.  The above Eq.~(\ref{unfavored}) gets modified to:
 \begin{equation}
 \eta=Y_{N_{1}}\sum_{i=\tau,...}^{n_{f}}\kappa_{i}\epsilon^{i}, 
 \end{equation}
 where $n_{f}$ corresponds 
 to the number of active flavors participating in thermal interactions. For temperatures
 ranging $10^{9}~{\rm GeV}\leq T\leq 10^{12}$~GeV the flavored leptogenesis can enhance
 the net baryon asymmetry by a factor 2 or 3~\cite{Davidson:2008bu}.


\vspace{-0.5cm}
\paragraph{\bf Dirac leptogenesis:}
It is possible that the decay of a heavy particle accompanied by the $CP$ distributed the lepton number  
equally between left handed and right handed particles with the net lepton number zero. A specific example will
be when the decay gives rise to a negative lepton number in left-handed neutrinos, and a positive lepton number of 
equal magnitude in right-handed neutrinos. If the observed neutrinos are Dirac in nature with a small 
Yukawa couplings $h \sim 10^{-12}$, then the left and right handed neutrinos will not come to thermal 
equilibrium before the electroweak scale, $H\sim \Gamma \Rightarrow T^{2}/M_{\rm P}\sim h^{2}T$.  Since 
the sphalerons interact with the left-handed neutrinos,  violating $B + L$ and conserving $B-L$, part of the
lepton number in left handed neutrinos get converted into baryon number. Ar lower temperatures,  the 
universe contains a total positive baryon number, total positive lepton number, and $B- L = 0$~\cite{Dick:1999je}.


\vspace{-0.5cm}
\paragraph{\bf Leptogenesis via scattering:} If there exists a shadow world 
similar to the SM sector, but hidden, and the only mediator is the 
heavy singlet neutrinos $N$, then it is possible to realize leaking the lepton number
from the hidden to the visible sector~\cite{Bento:2001rc,BasteroGil:2002hs}. Let us consider a simple interaction between
hidden (lepton doublet $l'$ and Higgs $\phi'$) and visible sector fields (SM lepton doublet $l$,
and the SM Higgs $\phi$) via
\begin{equation}
\label{Yuk} 
h_{ia}l_i N_a \phi + h'_{ka}\ell'_k N_a \phi' + 
\frac{1}{2} M_{ab} N_a N_b  
+  {\rm H.C.}  
\end{equation} 
where the Yukawa interactions are given by $h_{ia}$ and $h_{ka}$.
After  integrating out the heavy neutrinos $N$ with a mass 
$M_a = g_a M$, where
$M$ being the overall mass scale and $g_a$ are order one 
real constants, we obtain an effective dimensional $5$ operators
\begin{equation}
\label{op} 
\frac{A_{ij}}{2 M} l_i l_j \phi \phi + 
\frac{D_{ik}}{M} l_i \l'_k \phi \phi' + 
\frac{A'_{kn}}{2 M} \ell'_k \ell'_n \phi' \phi'  + {\rm H.C.} \;,
\end{equation}
with coupling constant matrices of the form 
$A = h g^{-1} h^T$, $A' = h' g^{-1} h^{\prime T}$  
and $D = h g^{-1} h^{\prime T}$.  Let us suppose 
that the reheat temperatures in both hidden, $T_{R}'$, 
and visible sector, $T_{R}$,  are below $M$. The only
way the two sectors can interact via the 
lepton number violating scatterings 
mediated by the heavy neutrinos $N$ which stay out 
of equilibrium, since $T_R \ll M$.  The $CP$ phase
can be obtained in $l\phi \leftrightarrow \ell'\phi'$
and $\ell \phi \leftrightarrow \bar \ell'\bar\phi'$.
the net asymmetry is given by $\Delta\sigma = {3J\, S}/{32\pi^2 M^4}$,
where 
$J= {\rm Im\, Tr} [ (h^{'\dagger}h') g^{-2}(h^\dagger h) g^{-1}
(h^\dagger h)^\ast g^{-1}]$  is the $CP$-violation parameter 
and $S$ is the c.m. of energy square. The final $B-L$ asymmetry of the universe 
is given by~\cite{Bento:2001rc}
\begin{eqnarray}
B-L & = & \frac{ n_{B-L} }{s} =
\left[\frac{\Delta\sigma\, n_{\rm eq}^2 }{4 H s} \right]_{R} \,, \nonumber \\
&\approx & 10^{-8}J
\left(\frac{10^{12}~{\rm GeV}}{M}\right)^{4}\left(\frac{T_{R}}{10^{9}~{\rm GeV}}\right)^{3}
\end{eqnarray} 
where $s$ is the entropy density, and for Yukawa constants spreading in the range $0.1-1$
can achieve the right lepton asymmetry.

\vspace{-0.5cm}
\paragraph{\bf Non-thermal leptogenesis:}
There exist various scenarios of non-thermal leptogenesis~\cite{Lazarides:1991wu,Murayama:1992ua,Giudice:1999fb,Asaka:1999yd,Asaka:1999jb,Allahverdi:2002gz} 
which can work for $T_{R} \leq M_N$.  One classic example is when the right handed 
sneutrino, a scalar field, with mass $M_{N}$, dominates the energy density of the universe 
and decays into the SM leptons and Higgs to reheat the universe and simultaneously
creating the lepton asymmetry. The $CP$ asymmetry can be created again from the 
interference between a tree level and one-loop quantum corrections, which yields the net
asymmetry: 
\begin{equation}
\eta \sim   \frac{n_{L}}{s}\sim \epsilon \frac{\rho_{N}}{sM_{N}}
\sim \frac{3}{4}\epsilon \frac{T_{R}}{M_{N}}\,.
\end{equation}
A similar expression can be used if any sneutrino condensate decays after inflation~\cite{Berezhiani:2001xx,Postma:2003gc,Mazumdar:2004qv,Mazumdar:2003bs,Mazumdar:2003va,Mazumdar:2003wm}, in which case $T_{R}$ is replaced by the decay temperature of the sneutrino condensate, i.e. $T_{D}$.
The right handed neutrinos and sneutrinos could also be excited non-thermally during preheating if they couple to the inflaton, which would generate non-thermal leptogenesis~\cite{Giudice:1999fb}. 


\vspace{-0.5cm}
\paragraph{\bf Soft leptogenesis:} In a perfect SUSY preserving limit
the mass and the width of the right-handed neutrino and sneutrino
would be the same. Let us consider a single generation, where the mass is 
$M_N$, and their width is given by $ \Gamma = {Y^2 M_N}{4 \pi} = {m M_N^2}/{4 \pi v^2}\ ,
\qquad m \equiv {Y^2 v^2 / M_N}$, where $v \sim 174$~GeV is the Higgs VEV,
and $Y_{N}$ is the Yukawa coupling. However,  in a realistic scenario  we would expect 
soft SUSY breaking terms which would be relevant for soft-leptogenesis~\cite{Allahverdi:2003tu,Grossman:2003jv,Grossman:2005yi,D'Ambrosio:2003wy}:
\begin{equation}
\label{lsoft} 
{\cal L}_{\rm soft}=
\frac{B M_N}{2} \widetilde N \widetilde N
+A Y \widetilde L \widetilde N H + h.c.
\end{equation}
This model has one physical CP violating phase given by: $\phi=\arg(A B^*)$.
The soft SUSY breaking terms introduce mixing between the sneutrino
$\widetilde N$ and the anti-sneutrino $\widetilde{N}^\dagger$ in a
similar fashion to the $B^0 -\bar{B}^0$ and $K^0 - \bar{K}^0$
systems. The mass and width difference of the two sneutrino mass
eigenstates are given by
\begin{equation} \Delta m = |B|, \qquad \Delta \Gamma=
{2 |A| \Gamma \over M_N}.  
\end{equation}
 The CP violation in the mixing is responsible for generating the lepton-number
asymmetry in the final states of the $\widetilde N$ decay. This lepton
asymmetry is converted into the baryon asymmetry through the sphaleron
process.
The baryon to entropy ratio is given
by~\cite{D'Ambrosio:2003wy}:
\begin{eqnarray}
 \frac{n_B}{s} = - 10^{-3}\  \alpha
\left[ \frac{4 \Gamma |B|}{4 |B|^2 + \Gamma^2} \right]
\frac{|A|}{M_N} \sin \phi\, ,
\label{eq:asym}
\end{eqnarray}
where the efficiency parameter $\alpha$ depends on the mechanism that
produces the right-handed sneutrinos. In a thermal production, the largest 
conceivable value could be of order $\alpha\sim 0.1$ for the light neutrino mass
$m_{\nu}\sim 10^{-3}$~eV~\cite{D'Ambrosio:2003wy}. It may be slightly challenging to
fix the parameters to obtain the right lepton asymmetry either making $|B|/\Gamma$ or
$\Gamma/|B|$ small. The above requirement gives a non-trivial constraint on the parameters
\cite{Grossman:2003jv,D'Ambrosio:2003wy}:
\begin{equation}
 A \sim 10^2~{\rm GeV},~
M_N \leq 10^{8}~{\rm GeV},~B \leq 1 {\rm GeV},~ \phi \sim 1 .
\end{equation}
Small value of $|B|$ cannot be obtained in gravity mediated SUSY breaking scenarios, but it might be possible
to arrange within gauge mediated SUSY breaking~\cite{Grossman:2004dz}.

\subsubsection{Affleck-Dine Baryogenesis }

As we discussed already, within MSSM there exists {\it cosmologically flat} directions~\cite{Dine:1995kz,Gherghetta:1995dv}. 
Field fluctuations along such flat directions are smoothed out by inflation~\cite{Enqvist:2003gh},
which effectively stretches out any gradients, and only the zero mode of the scalar condensate remains. 
Baryogenesis can then be achieved by the perturbative decay of a condensate~\cite{Allahverdi:2006xh,Allahverdi:2008pf} that
carries baryonic charge, as was first pointed out by Affleck and Dine (AD)
\cite{Affleck:1984fy}. As we will discuss, the flat direction condensate
can get dynamically charged with a large $B$ and/or $L$ by virtue of $CP$-violating self-couplings.

In the original version \cite{Affleck:1984fy} baryons were
produced by  a direct decay of the condensate. It was however pointed out 
that  in the case of gauge mediated SUSY
breaking \cite{Kusenko:1997si,Kusenko:1997zq}, and in the
case of gravity mediated SUSY breaking~\cite{Enqvist:1997si,Enqvist:1998en,Enqvist:1999mv,Enqvist:2000gq}, that the AD flat direction
condensate in most cases is not stable but fragments and eventually forms
non-topological solitons called $Q$-balls \cite{Coleman:1985ki}. In gauge mediated SUSY breaking scenarios
these Q-balls can be made a long lived dark matter candidate~\cite{Kusenko:1997vp,Kusenko:2009iz}. 
For reviews see~\cite{Enqvist:2003gh,Dine:2003ax}.

Since, SUSY is broken by the finite energy density of the inflaton, the AD condensate
receives corrections  in the case of F-term inflation. Let us consider a generic superpotential
for the AD field given by Eq.~(\ref{supot}), then the effective potential for the AD field will be given by
\cite{Dine:1995uk,Dine:1995kz}
\begin{eqnarray}
\label{adpot0}
V(\phi)&=&-C_{I} H_{I}^2 {|\phi |}^2 + \left(a {\lambda}_d H {{\phi}^d \over d
M_{\rm P}^{d-3}} + {\rm h.c.}\right) + m^{2}_{\phi}{|\phi|}^2\,,\nonumber \\
&&+ \left(A_{\phi} {\lambda}_d \frac{{\phi}^d}{dM_{\rm P}^{d-3}} + {\rm h.c.}\right)
+|\lambda|^2\frac{|\phi|^{2d-2}}{M_{\rm P}^{2d-6}} \,.
\end{eqnarray}
The first and the third terms are the Hubble-induced and low-energy soft
mass terms, respectively, while the second and the fourth terms are the
Hubble-induced and low-energy $A$ terms. The last term is
the contribution from the non-renormalizable superpotential. The coefficients
$|C_{I}|,~a,~\lambda_{d}\sim {\cal O}(1)$, and the coupling 
$\lambda \approx 1/(d-1)!$. Note that low-energy $A_{\phi}$ term has a mass dimension.
The $a, ~A$-terms in Eq.~(\ref{adpot0}) violate the global $U(1)$ symmetry
carried by $\phi$. If $|a|$ is ${\cal O}(1)$, the phase $\theta$ of
$\langle \phi \rangle$ is related to the phase of $a$ through $n
\theta + {\theta}_a = \pi$; otherwise $\theta$ will take some random
value, which will generally be of ${\cal O}(1)$. This is the initial
$CP$-violation which is required for baryogenesis/leptogenesis. The AD baryogenesis 
is quite robust and can occur even in presence of positively large Hubble-induced corrections~\cite{Kasuya:2006wf}.

\begin{figure}
\includegraphics[width=6cm]{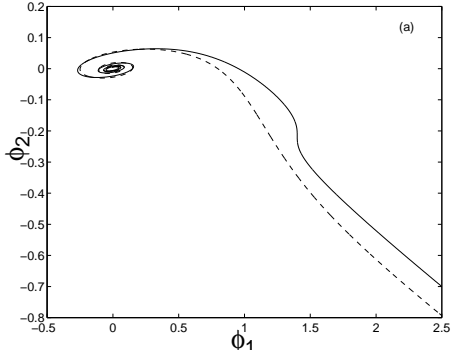}
\caption{The dynamical motion of AD condensate $\phi=\phi_{1}+i\phi_{2}$ for a gravity mediated case 
with $d=4$ (solid) and $d=6$ (dashed) with the initial condition $\theta_{i}=-\pi/10$.} \label{AD-pic}
\end{figure}

At large VEVs the first term dictates the
dynamics of the AD field. If $C_{I}<0$ , the absolute minimum
of the potential is  $\phi =0$ and during inflation the condensate will
evolve to its global minimum in one Hubble time. On the other hand 
if $C_{I} > 0$, the absolute value of the AD field settles during inflation to
the minimum given by $|\phi| \simeq (H_{I}M_{\rm P}^{d-3})^{1/d-2}$. After the end of inflation
the minimum of the condensate evolves from its initial large VEV to its global minimum
$\phi=0$, note that the dynamics is  non-trivial when the condensate starts oscillating
when $H(t)\sim m_{\phi}\sim {\cal O}(100)$~GeV. The dynamics of the AD condensate is 
non-trivial as shown in Fig.~(\ref{AD-pic}). 


If inflation is driven by  D-term, one does not get the Hubble induced mass
correction to the flat direction so that $C_{I},~a=0$.
Also the Hubble induced $a$-term is absent. However the Hubble induced mass
correction eventually dominates once D-term induced inflation comes to an
end. 

The baryon/lepton  number density is related to the dynamics of the AD field by
\begin{equation}
n_{B,L}=\beta i(\dot{\phi}^{\dagger}\phi - \phi^{\dagger}\dot{\phi})\,,
\end{equation}
where $\beta$ is corresponding baryon and/or lepton charge of the
AD field.The equation of motion for the AD field is given by
\begin{equation}
\ddot\phi + 3H\dot\phi +{\partial V(\phi)}/{\partial \phi^{\ast}}=0\,.
\end{equation}
The above two equations give rise to
\begin{eqnarray}
\label{bleq}
\dot n_{B,L}+3Hn_{B,L}&=&2\beta {\rm {\cal I}m}\left[\frac{\partial V(\phi)}
{\partial \phi^{\ast}}\phi\right]\,, \nonumber \\
&=&2\beta\frac{m_{\phi}}{d M_{\rm P}^{d-3}}{\rm {\cal I}m}(a\phi^{d})\,.
\end{eqnarray}
The net baryon and/or lepton number can be obtained by integrating the above 
equation
\begin{equation}
a^3(t)n_{B,L}(t)=2\beta |a|\frac{m_{\phi}}{M_{\rm P}^{d-3}}
\int^{t}a^3(t^{\prime})|\phi(t^{\prime})|^{d}\sin(\theta)~dt^{\prime}\,,
\end{equation}
Note that $''a''$ introduces an extra $CP$ phase which can be
parameterized by $\sin(\delta)$. Note that the asymmetry is not
governed by the Hubble induced $A$ term,  the amplitude
of the oscillations will be damped and so the $A$-term, which is proportional 
to a large power of $\phi$ will become gradually negligible. The net baryon and/or lepton
asymmetry is  given by~\cite{Dine:1995kz}
\begin{eqnarray}
\label{ap3}
n_{B,L}(t_{osc}) &=&  \beta \frac{2(d-2)}{3(d-3)}m_{\phi} \phi_{0}^{2}
\sin2 \theta \; \sin \delta\,, \nonumber \\
&\approx &  \beta \frac{2(d-2)}{3(d-3)}m_{\phi}\left(m_{\phi} M_{\rm P}^{d-3}\right)
^{2/(d-2)} \,,
\end{eqnarray}
where $\sin\delta \sim \sin 2\theta \approx {\cal O}(1)$.
When the inflaton decay products have completely thermalized with a
reheat temperature $T_{R}$, the baryon and/or lepton asymmetry
is given by
\begin{eqnarray}
\label{bleq1}
\frac{n_{B,L}}{s} &=&\frac{1}{4}\frac{T_{R}}{M_{\rm P}^2 H(t_{osc})^2}
n_{B,L}(t_{osc})\,,\nonumber \\
&=&\frac{d-2}{6(d-3)}\beta \frac{T_{R}}{M_{\rm P}^2 m_{\phi}}
\left(m_{\phi} M_{\rm P}^{d-3}\right)^{2/(d-2)} \,,
\end{eqnarray}
where we have used $H(t_{osc})\approx m_{\phi}$, and $s$ is the entropy
density of the universe at the time of reheating. For $d=4$, the
baryon-to-entropy ratio is
\begin{equation}
\label{assym1}
\frac{n_{B,L}}{s}\approx 1\times 10^{-10}\times \beta \left(\frac{1~{\rm TeV}}
{m_{\phi}}\right)\left(\frac{T_{R}}
{10^{9}~{\rm GeV}}\right)\,,
\end{equation}
and for $d=6$
\begin{eqnarray}
\label{assym2}
\frac{n_{B,L}}{s}&\approx &10^{-10}\times\beta\left(\frac{1~{\rm TeV}}{m_{\phi}}\right)^{1/2} 
\left(\frac{T_{R}}{100~{\rm GeV}}
\right)\,,
\end{eqnarray}
where we have taken the net $CP$ phase to be $\sim {\cal O}(1)$.
The asymmetry remains frozen unless there is additional entropy
production afterwards.  

The lepton asymmetry calculated above in
Eqs.~(\ref{assym1},\ref{assym2}) can be transformed into baryon number
asymmetry via sphalerons $n_{B}/s =(8/23)n_{L}/s$. AD leptogenesis
has important implications in neutrino physics also, because in the
MSSM, the $LH_{u}$ direction is lifted by the $d=4$ non-renormalizable operator
which also gives rise to neutrino masses~\cite{Dine:1995kz,Asaka:2000nb}:
\begin{equation}
W=\frac{1}{2M_{i}}\left(L_{i}H_{u}\right)^2=\frac{m_{\nu~i}}{2\langle H_{u}
\rangle^2}\left(L_{i}H_{u}\right)^2\,,
\end{equation}
where we have assumed the see-saw relation
$m_{\nu~i}=\langle H_{u}\rangle^2/M_{i}$ with diagonal entries for
the neutrinos $\nu_{i}$, $i=1,2,3$. The final $n_{B}/s$ can be
related to the lightest neutrino mass since the flat direction moves
furthest along the eigenvector of $L_{i}L_{j}$ which corresponds to the
smallest eigenvalue of the neutrino mass matrix~\cite{Dine:1995kz,Asaka:2000nb}.
\begin{equation}
\label{lnb}
\frac{n_{L}}{s}\approx 1\times 10^{-10}\times \beta\left(\frac{m_{3/2}}
{m_{\phi}}\right)\left(\frac{T_{R}}{10^{8}~{\rm GeV}}\right)\left(
\frac{10^{-6}~{\rm eV}}{m_{\nu l}}\right)\,,
\end{equation}
where $m_{\nu~l}$ denotes the lightest neutrino. Similarly the $d=6$ case
corresponds to the flat direction $udd$~\cite{Enqvist:1999mv,McDonald:1996cu}.
   

\subsubsection{Baryogenesis below the electroweak scale}

At temperatures below the electroweak scale, the sphaleron transitions are rather inactive. If the universe
reheats below the electroweak scale then it is a challenge to generate the required baryon asymmetry.
The leptogenesis based scenarios are hard to implement below the electroweak scale. However within
SM there exists a possibility of realizing cold electroweak baryogenesis as discussed in 
\cite{GarciaBellido:1999sv,Cornwall:2000eu,Cornwall:2001hq,Krauss:1999ng,Tranberg:2003gi,Tranberg:2006ip,Tranberg:2009de,Enqvist:2010fd}. SUSY further opens a door to realize baryon asymmetry at temperatures
even close to the BBN via R-parity violating interactions~\cite{Cline:1990bw,Scherrer:1991yu,Kitano:2008tk,Kohri:2009ka}. Here we will discuss both the scenarios.

\vspace{-0.5cm}
\paragraph{\bf Cold electroweak baryogenesis:}
There are mechanisms to obtain cold electroweak baryogenesis 
where it is assumed that the SM {\it d.o.f} are not in thermal equilibrium.
Moreover in a cold environment  the CP-violation in SM is much larger than at  the
electroweak temperatures, and most of the baryon asymmetry is produced at the initial 
quench when the Higgs field is rolling down the potential. Baryon production essentially stops after the 
first few oscillations, after which the coherent Higgs field will start decaying, thereby reheating the universe. However,
the hurdle is to obtain this fast quench without the presence of strong first order phase transition.

There are couple of possibilities of realizing cold initial condition and out of equilibrium condition. There could be a very 
low scale of inflation which might not be responsible for generating the seed perturbations, or the universe could be 
simply trapped in a vacuum where the SM {\it d.o.f} are not even excited.  The out-of-equilibrium condition 
can be obtained during the coherent oscillations of the scalar fields. In order to realize this idea, we would require 
a scalar field coupled to the SM Higgs, $\sigma^{2}H^{2}$.  During the coherent oscillations, it is possible to 
have the baryon number violating sphaleron transitions~\cite{GarciaBellido:1999sv,Tranberg:2003gi,Cornwall:2000eu}. 
This can happen since the Higgs oscillations can excite the electroweak gauge bosons from the time dependent vacuum fluctuations
with a very large occupation number, similar to case of preheating. These long wavelength fluctuations of the gauge 
fields are responsible for overcoming the sphaleron barriers which leads to the baryon number violation.
Furthermore, there could be extra sources of CP-violations during the oscillations as pointed out in \cite{Cornwall:2000eu,Cornwall:2001hq,Tranberg:2009de}.


\vspace{-0.5cm}
\paragraph{\bf R-parity violation and  baryogenesis:}
The current limits on some of the $R$-parity violating interactions are poorly understood.
Let us now consider a scenario where B and L are violated within 
MSSM, with a  superpotential:
\begin{equation}
W =\mu_i^{'} L_i H_u+\lambda_{ijk}L_i L_j e_k + \lambda^{'}_{ijk}L_i
Q_j d_k +\lambda^{''}_{ijk}u_i d_j d_k\,,\\
\label{rp-sup}
\end{equation}
where $L_i=(\nu_i, e_i)$, $Q_i=(u_i,d_i)$, $H_u=(h_u^+, h_u^0)^T$, 
$H_d=(h_d^0, h_d^-)^T$, {\it etc} are $SU(2)_L$ doublets and $u^c_i$, 
$d^c_i$ are $SU(2)_L$ singlet quarks. In Eq.(\ref{rp-sup}), the first three terms violate 
lepton number by one unit ($\Delta L=1$), while the last term violates baryon number 
by one unit ($\Delta B=1$). For the stability of proton we assume that $\lambda_{ijk}=
\lambda'_{ijk}=0$. This can be accomplished if there exists any conservation of lepton 
number, which then forces $\mu'_i$ to be zero. However, the 
electric dipole moment of neutron gives~\cite{Barbier:2004ez}
\begin{equation}
{\rm Im}\left( \lambda''_{312} \lambda''_{332} \right) < 0.03 \left(\frac{0.01}{V_{td}} 
\right) \left( \frac{\tilde{M} }{{\rm TeV}} \right)^2 
\end{equation}
and the non-observation of $n-\bar{n}$ oscillation gives an upper bound on 
$\lambda^{''}_{11k}$ to be ~\cite{Barbier:2004ez}
\begin{equation}
|\lambda''_{11k}| < \left( 10^{-6} - 10^{-5} \right) \frac{10^8 {\rm s} }{\tau_{\rm osc}} 
\left(\frac{\tilde{M}}{ \rm TeV} \right)^{5/2}
\end{equation} 
While $ \lambda''_{332}$ is hardly constrained and can be taken to be as large as
${\cal O}(1)$. Let us consider that a scalar field $\phi$ decays to MSSM {\it d.o.f} right
before BBN primarily into gauge bosons and gauginos via  $R$-parity violating couplings $\lambda''_{ijk}$.

Let us  assume that the gauginos are heavier than the quarks and squarks. As a result their decay 
to a pair of quark and squark through one loop quantum correction gives rise to a net CP 
violation. The magnitude of CP violation in the decay:  $\tilde{g}\rightarrow t \tilde{t}^c$ 
can be estimated as~\cite{Cline:1990bw}:
\begin{eqnarray}
\epsilon = \frac{\Gamma \left( \tilde{g} \rightarrow t \tilde{t}^c \right) - 
\Gamma \left( \tilde{g} \rightarrow \bar{t} \tilde{t} \right) }
{\Gamma_{\tilde{g}}^{\rm tot}} 
 \approx  \frac{\lambda''_{323}}{16 \pi} \frac{ {\rm Im} \left( A_{323}^* m_{\tilde{g}} \right)} 
{|m_{\tilde{g}}|^2}
\label{cp_violation}
\end{eqnarray}
where $A_{323}$ is the trilinear SUSY breaking term and we also assume a maximal 
CP violation. As a result the decay of gauginos produce more squarks (antisqarks) than 
antisquarks (squarks). The baryon number violating ($\Delta B=1$) decay, induced by $\lambda''_{323}$ 
of squarks (antisquarks) to quarks (antiquarks) then gives rise to a net baryon asymmetry. Note that 
the decay of squarks (anti-squarks) are much faster than any other processes that would erase the 
produced baryon asymmetry. Hence the B-asymmetry can simply be given by:
\begin{equation}
\eta_B\sim B_{\tilde{g}} \epsilon \frac{n_{\phi}}{s} \sim \frac{3}{4}B_{\tilde g}\epsilon 
\frac{T_{R}}{m_{\phi}}\,,
\label{B-asym}
\end{equation}
where $B_{\tilde{g}}\sim 0.5$ is the branching ratio of the decay of $\phi$ to $\tilde{g} \tilde{g}$,
and in the above equation $s$ is the entropy density resulted through the decay of $\phi$. 
For $T_R/m_{\phi}\sim 10^{-7}$ and 
$m_{\phi}\sim 10^{5}$~GeV. Therefore a reasonable CP violation of order $\epsilon\sim 0.01- 0.001$ 
could accommodate the desired baryon asymmetry of ${\cal O}(10^{-10})$ close to the temperature 
of $T\sim 10-1$~MeV~\cite{Cline:1990bw,Kohri:2009ka}.


\section{Dark matter}

There is a conclusive evidence that a considerable fraction of the current energy 
density is in the form of a non-baryonic dark matter. The dynamical motions of astronomical
objects such as rotation curves for spiral galaxies~\cite{Begeman:1991iy}, velocity dispersion of individual galaxies
in galaxy clusters, large x-ray temperatures of clusters~\cite{Flores:2005td}, bulk flows and the peculiar
motion of our own local group~\cite{Dressler:1986rv}, all implies the presence of a dark matter. The mass of 
galaxy clusters inferred by their gravitational lensing of background images is also consistent with the 
large dark-to-visible mass ratios~\cite{Bolton:2005nf}. Perhaps the most compelling
evidence, at a statistical significance of $8\sigma$ comes from the two colliding clusters of 
galaxies, known as the Bullet cluster~\cite{Clowe:2006eq,Markevitch:2003at}. It was found that the spatial offset of the center 
of the total mass from the center of the baryonic mass peaks cannot be explained with an alteration of gravitational force law. Furthermore, 
the large scale structure formation from the initial seed perturbations from inflation requires a significant non-baryonic dark 
matter component~\cite{Abazajian:2008wr}. In terms of the critical density, 
$\rho_c=3 H_0^2 M_{\rm P}^2/8\pi =1.88\times10^{-29}$g cm$^{-3}$ and with Hubble constant $H_0\equiv 100 h$
km\,sec$^{-1}$Mpc$^{-1}$, the dark matter density inferred from WMAP and large scale structure data 
is $\Omega_{{\rm DM}}\equiv \rho_{{\rm DM}}/\rho_c \sim 0.22$~\cite{Komatsu:2010fb}.

The dark matter is assumed to be a weakly interacting massive particle (WIMP), yet undiscovered. There are many well motivated 
particle physics candidates, e.g.~\cite{Jungman:1995df,Taoso:2007qk,Bertone:2004pz,Kusenko:2009up}, all of 
which arise from beyond the SM physics. The  dark matter is assumed to be stable on the scale of cosmological structure formation.
By virtue of new symmetries, for example R-parity conservation in SUSY allows the lightest SUSY particle (LSP) to be absolutely 
stable~\cite{Goldberg:1983nd,Ellis:1983ew}, or in the case of extra dimensions, the Kaluza-Klein (KK) parity leaves the lightest KK 
particle (LKP) stable~\cite{Servant:2002aq}.

In many cases some of these symmetries which protect the dark matter particle from decaying are broken by sufficiently suppressed 
higher-dimensional operators, such that the dark matter might as well have a finite life time comparable to the age of the universe. 
In the context of SUSY grand unification, operators with mass dimension $6$ are expected to make SUSY dark matter unstable, with a time-scale
\begin{equation}\label{eq:gut}
\tau\sim8\pi \left(\frac{M_{\rm GUT}^4}{m_{X}}\right)^5\sim 10^{27}{\rm sec}\left(\frac{\rm TeV}{m_X}\right)^5\left(\frac{M_{\rm GUT}}{2\times 10^{16}\ {\rm GeV}}\right)^4
\end{equation}
where $M_{\rm GUT}\sim 10^{16}$~GeV, and $m_{X}$ is the dark matter particle. The lower dimensional operators would yield much shorter 
time scale, as it would lead to dark matter decay long before the structure formation. Within SUSY one compelling candidate could be the 
gravitino with R-parity weakly broken in the hadronic sector, yielding the required baryon asymmetry also in the process~\cite{Kohri:2009ka}.

The widely accepted lore is that after radiation-matter equality, when the universe becomes matter
dominated, the density perturbations in the dark matter begin to grow, and drive the
oscillations of the photon-baryonic fluid around the dark matter
gravitational potential wells. Immediately after the epoch of recombination the baryons 
kinematically decouple from photons, which then free-stream through the universe; the baryons on
the other hand slowly fall into the potential wells created by the dark matter particles, eventually becoming 
light emitting galaxy, see for more details \cite{Kolb:1988aj,Peebles:1994xt,Dodelson:2003ft}. There are three broad categories 
of dark matter which have been central to our discussion.


\subsection{Types of dark matter}

\subsubsection{Hot Dark Matter}

If the dark matter particle is collisionless, then they can damp the 
fluctuations from higher to lower density regions above the free-streaming scale.
This  hot dark matter consists of particles which are relativistic at the time of structure
formation and therefore lead to large damping scales \cite{Bond:1983hb}.

The SM neutrinos are the simplest examples of hot dark matter. In the early universe they 
can be decoupled from a  relativistic bath at $T \sim 1$~MeV, leading to a
relic abundance today that depends on the sum of the flavor masses:
\begin{eqnarray}
\Omega_{\nu}h^2 = \frac{\sum_{i}m_{\nu_{i}}}{90\mbox{ eV}}.
\label{eqn:Omeganeutrino}
\end{eqnarray}
 Various observational constraints combining Ly-$\alpha$ forest,
CMB, SuperNovae and Galaxy Clusters data leads to \cite{Seljak:2006qw,Fogli:2008ig}: $\sum m_{\nu } < 0.17 \mbox{ eV} \mbox{  (95 \% CL)}$.
Similar limits can be applied to any generic hot dark matter candidate,
such as axions \cite{Hannestad:2010yi}
or to hot sterile neutrinos \cite{Dodelson:2005tp,Kusenko:2009up}. The free-streaming length for neutrinos is
\cite{Kolb:1988aj}:
\begin{equation}
\lambda_{FS} \sim 20 \left( \frac{30 \mbox{ eV}}{m_{\nu}} \right) \mbox{ Mpc}.
\end{equation}
For instance, the universe dominated by the eV neutrinos would lead to
suppressed structures at $600$~Mpc scale, roughly the size of supercluster.
Furthermore, hot dark matter would predict a top-down hierarchy in the
formation of structures, with small structures forming by
fragmentation of larger ones, while observations show
that larger galaxies have formed from the mergers of the initially small galaxies.


\subsubsection{Cold Dark Matter}

The standard theory of structure formation requires cold dark matter (CDM),  whose free-streaming length is
such that only fluctuations roughly below the Earth mass scale are 
suppressed~\cite{Hofmann:2001bi,Green:2005fa,Loeb:2005pm,Bertschinger:2006nq,Green:2003un}. 
The CDM candidates are heavy and non-relativistic at the time of their freeze-out from thermal plasma. The current paradigm 
of $\Lambda$CDM is falsifiable whose predictive power can be used to probe the structures at various cosmological scales,
such as the abundance of clusters at $z\leq 1$ and the galaxy-galaxy correlation functions have proven it a successful
and widely accepted cosmological model of large scale structure formation.

The N-body simulations based on $\Lambda$CDM provide a strong hint of a universal dark matter profile, with the 
same shape for all masses, and initial power spectrum. The halo density can be parametrized by:
\begin{equation}
\rho(r)=\frac{\rho_0}{(r/R_{s})^{\gamma}\left[1+(r/R_{s})^{\alpha}\right]^{(\beta-\gamma)/\alpha}}\,,
\end{equation}
where $\rho_{0}$ and the radius $R_{s}$ vary from halo to halo. the parameters $\alpha,~\beta$ and $\gamma$  vary slightly from one profile to other.
The four most popular ones are:
\begin{itemize}

\item Navarro, Frenk and White (NFW) profile~\cite{Navarro:1996gj}, where $\alpha =1, ~\beta=3, \gamma=1$, and $R_{s}=20$~Kpc. 

\item Moore profile~\cite{Moore:1999gc}, where $\alpha=1.5,~\beta=3,~\gamma=1.5$, and $R_{s}=28$~Kpc.

\item Kra profile~\cite{Kravtsov:1997dp}, where $\alpha=2,~\beta=3,~\gamma=0.4$, and $R_{s}=10$~Kpc.

\item Modified Isothermal profile~\cite{Bergstrom:1997fj}, where $\alpha=2,~\beta=3,~\gamma=0$, and $R_{s}=3.5$~Kpc.

\end{itemize}

Amongst all the four profiles, the scales where deviations are most pronounced (the inner few kiloparsecs) are also the most compromised by numerical uncertainties.  The power-law index value, $\gamma$, in the inner most regions is part of the numerical uncertainties and still under debate, as all four simulations provide different numbers. The simulations hint towards  a cuspy profile, as the density in the inner regions becomes large, while from the rotation curves of low surface brightness (LSB) galaxies point towards uniform dark matter density profile with constant density cores \cite{Gentile:2004tb}. In our own galaxy the situation is even more murky, as the observations of the velocity dispersion of stars  near the core suggests a supermassive black hole at the center of our Galaxy, with a mass $M_{SMBH}\approx 2.6\times 10^{6} M_{\odot}$~\cite{Ghez:1998ph}. Many galaxies have been found to host supermassive blackholes of $10^{6}-10^{8}M_{\odot}$. It has been argued that if supermassive blackhole exists at the galactic center, the accretion of dark matter by the blackhole would enhance the dark matter density~\cite{Peebles:1994xt}. To alleviate some of these problems, dark matter with a strong elastic scattering cross section~\cite{Spergel:1999mh,Dave:2000ar}, or large annihilation cross sections \cite{Kaplinghat:2000vt} have been proposed. 

There are further discrepancies between observations and numerical simulations. The number 
of satellite halos as predicted by simulations exceeds the number of observed Dwarf
galaxies in a typical galaxy like Milky-Way~\cite{Moore:1999nt,Klypin:1999uc}. However
recent hydrodynamical simulations with $\Lambda$CDM, including the supernovae induced outflows
suggest a fall in the dark-matter density to less than half of what it would otherwise be within the central Kpc.


\subsubsection{Warm Dark Matter}

Besides hot and cold dark matter, the early universe can also provide warm dark matter (WDM) candidates whose
velocity dispersion lies between that of hot and CDM. The presence of WDM reduces 
the power at small scales due to larger free-streaming length compared to that of a CDM~ \cite{Bode:2000gq,SommerLarsen:1999jx}.

The origin of WDM can be found within sterile states. For instance, the see-saw mechanism for the active neutrino masses from the SM singlet 
states~\cite{Minkowski:1977sc,Yanagida,Gell-Mann,Mohapatra:1980yp} would
naturally generate masses to the active $m(\nu_{1,2,3})\sim y^{2}\langle H\rangle^{2}/M_{N}$, 
and sterile neutrinos $m(\nu_{a})\sim M_{N}$ ($a>3$) in Eq.~(\ref{Linteract}), if we take $i, j=1,\cdots n+3$.
The typical mixing angles in this case are: $\theta_{a i}\sim y^{2}_{ai}\langle H\rangle^{2}/M_{N}^{2}$. In order to explain
the neutrino masses  from atmospheric and solar neutrino data, $n=2$ is sufficient, however for pulsar 
kicks~\cite{Kusenko:1996sr,Kusenko:1998bk,Kusenko:2006rh}, supernovae 
explosion~\cite{Fryer:2005sz,Hidaka:2006sg,Hidaka:2007se}, as well as sterile neutrino as a dark matter 
candidate~\cite{Dodelson:1993je,Shi:1998km,Dolgov:2000ew,Asaka:2005an,Abazajian:2001nj,Petraki:2007gq},
we require at least $n=3$, so in total $6$ sterile Majorana states, for a review on all these effects, see~\cite{Kusenko:2009up}.
The presence of such extra sterile neutrinos is also supported by $\bar \nu_{\mu}\rightarrow \bar\nu_{e}$ oscillations observed at LSND~\cite{Aguilar:2001ty}, 
and the recent results by MiniBoone~\cite{AguilarArevalo:2010wv}.

A sterile neutrino with a KeV mass can be an ideal WDM candidate
which can be produced in the early universe by oscillation/conversion of thermal active neutrinos,
with a momentum distribution significantly suppressed from a thermal spectrum \cite{Dodelson:1993je,Abazajian:2001nj}.
A typical free-streaming scale is given by, see \cite{Abazajian:2006yn}
\begin{equation}
\lambda_{FS} \approx 840 \mbox{ Kpc}\mbox{ h}^{-1} \left( \frac{1\mbox{ KeV}}{m_{s}}\right) \left( \frac{<p/T> }{3.15}       \right)\,,
\end{equation}
where $m_s$ is the mass of the sterile flavor eigenstate, $0.9 \geq \langle p/T \rangle/3.15 \geq 1$ is the 
mean momentum over temperature of the neutrino distribution and  ranges from $1$ (for a thermal) to $\sim 0.9$ (for a non-thermal) 
distribution. Very stringent bounds on the mass of WDM particles have been obtained by different groups. Typically,
the bounds range from $m_s \geq 10-20 \mbox{ KeV} \mbox{ (95 \% CL)  } (m_{WDM} \geq 2-4 \mbox{ KeV})$, see~\cite{Kusenko:2009up}.
It is quite plausible to imagine a mixed dark matter scenario, where more than one species contributed to the total
dark matter abundance. If there is a fraction of  sterile neutrinos or WDM, then the above bounds can even be relaxed.


\subsection{WIMP production}

\subsubsection{Thermal relics}\label{Th-relic}

At early times it is assumed that the dark matter particle, denoted by $X$ is in chemical and kinetic equilibrium, i.e. in 
local thermodynamic equilibrium. The dark matter will be in equilibrium as long as reactions can keep $X$ in chemical equilibrium and
the reaction rate can proceed rapidly enough as compared to the expansion rate of the universe, $H(t)$. When the reaction rate becomes
smaller than the expansion rate, then the particle $X$ can no longer be in its equilibrium, and thereafter its abundance with respect to 
the entropy density becomes constant.  When this occurs the dark matter particle is said to be ``frozen out.'' 

The equilibrium abundance of $X$ relative to the entropy density depends upon the ratio of the
mass of the particle to the temperature.  Let us define the variable $Y\equiv
n_X/s$, where $n_X$ is the number density of $X$ with mass $m_X$,
and $s =2\pi^{2}g_{*} T^3/45$  is the entropy density, where $g_{*}$ counts the number of relativistic {\it d.o.f}.  The equilibrium value of
$Y$, $Y_{EQ}\propto \exp(-x)$ for $x=m_{X}/T\gg 1$, while
$Y_{EQ}\sim$ constant for $x\ll 1$.

The precise value of $Y_{EQ}$ can be computed exactly by solving the Boltzmann equation~\cite{Kolb:1988aj}:
\begin{equation}
\label{Bz}
\dot n_X+3Hn_X=-\langle \sigma v\rangle (n_X^2-(n_X^{eq})^2)\,,
\end{equation}
where dot denotes time derivative, $\sigma $ is the total annihilation cross section, $v$ is the velocity, bracket denotes 
thermally averaged quantities, and $n^{eq}$ is the number density of $X$ in thermal equilibrium:
\begin{equation}
n^{eq}=g\left({mT}/{2\pi}\right)^{3/2}e^{-m_{X}/T} ,
\label{Max-Boltz}
\end{equation}
where $T$ is the temperature. In terms of $Y=n_{X}/s$ and $x=m_{X}/T$, and using the conservation of entropy per comoving volume
($sa^3=$~constant), we rewrite Eq.~(\ref{Bz}) as:
\begin{equation}
\label{Max-Boltz1}
\frac{dY}{dx}=-\frac{\langle\sigma v\rangle s}{H x} \left(Y^2-({Y^{eq}})^2\right) .
\end{equation}
In the case of heavy $X$, the cross section can be expanded 
with respect to the velocity in powers of $v^2$, 
$\langle \sigma v\rangle =a+b\langle v^2\rangle +{\cal O}(\langle v^4\rangle)+...\approx a+6b/x$, where
$x=m_X/T$ and $a, b$ are expressed in ${\rm GeV}^{-2}$. Typically $a\neq 0$ for s-wave annihilation, and
$a=0$ for p-wave annihilation. We can rewrite Eq.~(\ref{Max-Boltz1}) in terms of a new variable: 
$\Delta=Y-Y^{eq}$,
\begin{equation}\label{New-Boltz}
\Delta^{\prime}=-Y^{eq\prime}-f(x)\Delta(2Y^{eq}+\Delta)\,,
\end{equation}
where prime denotes $d/dx$, and
\begin{equation}
f(x)=\frac{\pi g_{\ast}}{45}m_{X}M_{\rm P}(a+6b/x)x^{-2}\,.
\end{equation}
One can find a simple analytic solution for Eq.~(\ref{New-Boltz}) for two extreme regimes
\begin{eqnarray}
\Delta =-\frac{Y^{eq\prime}}{f(x)(2Y^{eq}+\Delta)}\,,~x\ll \frac{m_{X}}{T_{f}},~\Delta'\ll Y^{eq\prime}\\
\Delta^{-2}\Delta^{\prime} =-f(x)\,,~~x\gg \frac{m_{X}}{T_{f}}\,,~\Delta'\gg Y^{eq\prime}\,.
\end{eqnarray}
Integrating the last equation for $(x_{f},\infty)$, and using $\Delta(x_{f})\gg\Delta_{\infty}$, we find
\begin{equation}
\Delta_{\infty}^{-1}\approx Y_{\infty}^{-1}=\sqrt{\frac{\pi g_*}{45}}M_{\rm P} \left(\frac{m_{X}}{x_f}\right)\left(a+\frac{3b}{x_f}\right). 
\end{equation}
In terms of the present energy density, $\rho_{X}=m_X n_X=m_X s_0 Y_{\infty}$, where 
$s_0=2889.2$ cm$^{-3}$ is the present entropy density, the relic abundance of dark matter particle 
in terns of the critical energy density is given by: 
\begin{equation}
\Omega_Xh^2\approx \frac{1.07\times 10^{9}}{M_{\rm P}}\frac{x_f}{\sqrt{g_{\ast}}(a+3b/x_f)}~{\rm GeV}^{-1}\,,
\end{equation}
where the freeze-out temperature is defined by solving this equation $\Delta(x_{f})=cY^{eq}(x_{f})$
iteratively for early and late time solutions, for $c\sim {\cal O}(1)$
\begin{equation}
x_{f}=\ln\left[c(c+2)\sqrt{45}\frac{g_{ast}}{2\pi^{5/2}}\frac{m_{X}M_{\rm P}(a+6b/x_{f})}{g_{\ast}^{1/2}\sqrt{x_f}}
\right]\,.
\end{equation}
An approximate order of magnitude 
estimation of the abundance can be written as:
\begin{equation}
\label{anni-abund}
\Omega_Xh^2\approx\frac{3\times 10^{-27}~\mathrm{cm}^3~\mathrm{s}^{-1}}{\langle \sigma v \rangle}\sim 
\frac{0.1~{\rm pbarn}}{\langle \sigma v \rangle}\,.
\end{equation}
For a WIMP interacting with a heavy gauge boson, would naturally yield an upper bound
on $m_{X}$. From the above Eq.~(\ref{anni-abund}), on dimensional grounds, 
$\Omega\sim 1/\langle \sigma v\rangle\sim 1/m_{X}^{2}$ for $\langle \sigma v\rangle\simeq \alpha^{2}(m_{X}/M^{2})^{2}$, where
$M$ is the mass of the new gauge boson. For $m_{X}\sim M\sim 1$~TeV, abundance of the dark matter becomes of 
order of unity, $\Omega_{X}h^{2 }\sim {\cal O}(1)$, for $\alpha\sim 0.1$. Actually, the unitarity bound limits the dark matter mass to be 
below $m_{X}\leq 300$~TeV~\cite{Griest:1989wd}. For a realistic scenario $\alpha\sim 0.01$, the unitarity bound would yield $m_{X}\leq 3$~TeV.

Note that the above  non-relativistic expansion of $\langle \sigma v\rangle\approx a+6b/x$ may not hold universally.
When a mass of second particle becomes nearly degenerate with the dark matter particle $X$ as in the case 
of coannihilation~\cite{Binetruy:1983jf,Griest:1990kh}, or the cross section is strongly varying function of the center of mass energy as in 
the case of a resonant annihilation~\cite{Griest:1990kh}. In the latter case,  $\sigma$, gets a boost by resonant annihilation when 
$m_{X}\approx m_{A}/2$, where $X$ annihilates with an exchange of particle, $A$, with a mass $m_{A}$.


\subsubsection{Coannihilating WIMPs}\label{coanni}

If  there are $N$ particles, $X_i$ ($i=1,\ldots,N$) with the lightest one, $X_{1}$, which have  nearly {\it degenerated }
masses $m_i$, such that $m_1 \leq m_2 \leq \cdots \leq m_{N-1} \leq m_N$, and internal {\it d.o.f}
(statistical weights) $g_i$. The next to lightest dark matter particle will be $N_{2}$. In this case  the above 
calculation of relic density,  Eq.~(\ref{Bz}), gets modified~\cite{Binetruy:1983jf, Griest:1990kh,Servant:2002aq}.
\begin{equation}
  \frac{dn}{dt} = -3Hn - \sum_{i,j=1}^N \langle \sigma_{ij} v_{ij} \rangle 
  \left( n_{i}n_{j} - n_{i}^{\rm{eq}}n_{j}^{\rm{eq}} \right),
\end{equation}
where $n= \sum_{i=1}^N n_{i}$ is the number density of the relic particle, since all other particles decay much 
before the long-lived $X_{1}$. The total annihilation rate for $X_i - X_j$ into a SM particle is given by:
\begin{eqnarray}
  \sigma_{ij}  & = & \sum_X \sigma (X_i X_j \rightarrow X_{SM})\,,~~~{\rm and}\\
  v_{ij} &=& \sqrt{(p_{i} \cdot p_{j})^2-m_{i}^2 m_{j}^2}/{E_{i} E_{j}}\,,
\end{eqnarray}
is the relative particle velocity, with $p_{i}$ and $E_{i}$ are the four-momentum and energy of 
particle $i$. One requires to define a thermal averaged $\langle\sigma_{ij}v_{ij}\rangle$, which  is defined by:
\begin{equation}
  \langle \sigma_{ij}v_{ij} \rangle = \frac{\int d^3{\bf
      p}_{i}d^3{\bf p}_{j} 
  f_{i}f_{j}\sigma_{ij}v_{ij}}
  {\int d^3{\bf p}_{i}d^3{\bf p}_{j}f_{i}f_{j}},
\end{equation}
where $f_{i}$ are distribution functions in the Maxwell-Boltzmann approximation.

Typically, when the scattering rate of  particles off SM particles in a
thermal background is much faster than their annihilation rate, then in the
above Eq.~(\ref{Bz}), $\langle \sigma v \rangle$ is replaced by:
\begin{equation} \label{Boltz-3}
  \langle \sigma_{\rm{eff}} v \rangle = \sum_{ij} ^{N}
   \sigma_{ij}\frac{g_{i}g_{j}}{g_{eff}^{2}}(1+\Delta_{i})^{3/2}(1+\Delta_{j})^{3/2}e^{-x(\Delta_{i}+\Delta_{j})}\,.
\end{equation}
where $\Delta_{i}=(m_{i}-m_{1})/m_{1}$,
and  $g_{eff}=\sum_{i}^{N}g_{i}(1+\Delta_{i})^{3/2}\exp(-x\Delta_{i})$. 
In the case of co-annihilation, the freeze-out temperature is determined by
\begin{equation}
x_{f}=\ln\left[c(c+2)\sqrt{45}\frac{g_{eff}}{2\pi^{5/2}}\frac{m_{X}M_{\rm P}(a_{eff}+6b_{eff}/x_{f})}{g_{\ast}^{1/2}\sqrt{x_f}}
\right]\,.
\end{equation}
where $a_{eff}$ and $b_{eff}$ are the coefficients of the Taylor expansion of $\sigma_{eff}$. The relic
abundance for $N_{1}$ is now given by
\begin{eqnarray}
&&\Omega_{N_{1}}h^{2}\approx \frac{3\times 10^{-27}~\mathrm{cm}^3~\mathrm{s}^{-1}}
{g_{*}^{1/2}x_{f}^{-1}(I_{a}+3I_{b}/x_{f})}\,,~~{\rm where}\\
I_{a}&=&x_{f}\int_{x_{f}}^{\infty}a_{eff}x^{-2}dx\,,~~I_{b}=2x_{f}^{2}\int_{x_{f}}^{\infty}b_{eff}x^{-3}dx\,.\nonumber
\end{eqnarray}
In any realistic framework  there are many particles which interact with  
the dark matter candidate $X$. They all eventually decay into $X$, and at the time of 
freeze-out the density of all heavy particles is exponentially suppressed except 
when there is mass degeneracy occurs between heavy particles and the $X$.
The details of coannihilation has been studied extensively within SUSY~\cite{Edsjo:1997bg}, and
publicly available numerical codes include coannihilations with all SUSY particles~\cite{Gondolo:2004sc}.


\subsubsection{Non-thermal relics}

The dark matter particle, $X$, can also be created in an out of equilibrium condition, i.e.
$X$ must not have been equilibrium when it froze out. A sufficient
condition for non-equilibrium is that the annihilation rate (per particle) must be 
smaller than the expansion rate of the universe: $n_X\langle \sigma v\rangle <H$.  

Let us assume that $X$ were non-relativistic at the time of production and they were 
never in local thermodynamical equilibrium. The largest dark matter density will thus be determined
by the largest freeze out temperature, which can be attainable in the universe. Assuming this 
to be the reheat temperature, $T_{R}$ and the universe follows a radiation domination,  
then the ratios of energy densities will be given by~\cite{Kolb:1998ki}:
\begin{equation} 
\frac{\rho_X(t_0)}{\rho_r(t_0)}
=\frac{\rho_X(t_{R})}{\rho_{r} (t_{R})}\:\left(\frac{T_{R}}{T_0}\right),
\label{TRT0}
\end{equation}
where $T_{0}$ is the present temperature and $t_{0}$ corresponds to the present time, 
$\rho_\gamma$ is the energy density in radiation, and $\rho_X=m_{X}n_{X}$
denotes the energy density in the dark matter with the number density $n_{X}$.  
If we further assume that $X$ particles were created at time
$t=t_{*} < t_{R}$, sometime during the coherent oscillations of the inflaton and 
before the completion of reheating, then both the $X$ particle energy density and the
inflaton energy density would redshift approximately at the same rate
until reheating is completed.  Therefore, 
\begin{equation} 
\frac{\rho_X(t_{R})}{ \rho_r(t_{R})} \approx 
\frac{\rho_X(t_{*})}{3M_{\rm P}^2 H^2(t_{*}) },
\end{equation} 
assuming that the inflaton energy density dominated the universe.
Since, $\Omega_X = \rho_X(t_0)/\rho_c(t_0)$, where $\rho_c(t_0)=3 H_0^2M_{\rm P}^2$ and
$H_0=100\: h$ km sec$^{-1}$ Mpc$^{-1}$, then using Eq.~(\ref{TRT0}), one obtains~\cite{Kolb:1998ki}:
\begin{eqnarray} 
\Omega_X h^2 &\approx& \Omega_r h^2\:
\left(\frac{T_{R}}{T_0}\right)\: 
 \left(\frac{M_X}{M_{\rm P}}\right)\:
\frac{n_X(t_{*})}{3M_{\rm P} H^2(t_{*})}\,.\\
&\sim &10^{17}\left(\frac{T_{R}}{10^9\mbox
{GeV}}\right)\frac{\rho_X(t_*)}{\rho_{inf}(t_*)}\,,
\label{TRT01}
\end{eqnarray}
where $\Omega_r h^2 \approx 4.31 \times 10^{-5}$ is the fraction of
critical energy density in radiation today, and $T_{0}\sim 2.3\times 10^{-13}$~GeV.
The above expression tells us that a non-thermal creation would require a very 
small fraction of the inflaton energy density to be transferred to the dark matter particle 
$X$, otherwise the universe would be dominated by the dark matter particles. 

For a singlet hidden sector dark matter, it is really a challenge not to {\it overproduce} them directly 
from the decay of the inflaton. If the inflaton sector belongs to the hidden sector, then it is natural to have
inflaton couplings to such hidden sector dark matter field. There are three possible ways to obtain a 
small fraction of $\rho_{X}(t_*)/\rho_{inf}(t_{*})$ in order to match the current observations.


\subsubsection*{(a)~Gravitational production}

 The dark matter can be created from the transition of the equation of state of the universe
 from inflation to matter domination or radiation domination, due to non-adiabatic evolution of the 
 vacuum~\cite{Kolb:1998ki,Chung:1998zb,Chung:2001cb,Chung:1999ve}. The underlying mechanism is similar to the metric fluctuations
 which seed the structure formation, except now the excitations can create massive particles.
 The gravitational production of dark matter is universal, and it can occur even if the dark 
 matter coupling to the inflaton is vanishingly small. 
 
 Let us consider a simple action for $X$ field with a 
 metric $ds^2= dt^2- a^2(t) d{\bf x}^2 = a^2(\eta)
\left[d\eta^2 - d{\bf x}^2\right] $, where $\eta$ is a conformal time.
\begin{equation}
S=\int \frac{dt}{2} \int d^3\!x\, a^3\left( \dot{X}^2 - \frac{(\nabla
X)^2}{a^2} - m_X^2 X^2 - \xi R X^2 \right)\,,
\end{equation}
where $R$ is the Ricci scalar.  Let us expand the $X$ field in terms of 
creation and annihilation operators which obey: $[a_{k_1}, a_{k_2}^\dagger] = \delta^{(3)}({\bf k}_1 -{\bf k}_2)$, and
\begin{equation}
X=\int \frac{d^3\!k}{(2 \pi)^{3/2} a(\eta)} \left[a_k u_k(\eta) e^{i
{\bf{k \cdot x}}} + a_k^\dagger u_k^*(\eta) e^{-i {\bf{k \cdot x}}}\right],
\end{equation}
where  the mode functions obey the identity $u_k u_k^{'*} - u_k'u_k^* = i$, and
prime denotes derivative w.r.t. $\eta$. The mode equation is given by:
\begin{equation}
u_k''(\eta) +[k^2 + m_X^2 a^2 + (6 \xi - 1) a''/a ] u_k(\eta) = 0,
\label{eq:modeequation}
\end{equation}
The parameter $\xi= 1/6,$ for conformal and $\xi=0$ for minimal
coupling.  Here we will consider the conformal coupling for simplicity.
The number density of $X$ particles can be estimated by a Bogoliubov
transformation: 
\begin{equation}
u_k^{\eta_1}(\eta)= \alpha_k u_k^{\eta_0}(\eta) +
\beta_k u_k^{* \eta_0}(\eta)\,,
\end{equation}
where $\eta_0=-\infty$, and  $\eta_1=+\infty$.
The energy density of produced particles is given by~\cite{Chung:1998zb}:
\begin{equation}
 \rho_X(\eta_1) = m_X
H_{inf}^3 (\tilde{a}(\eta_1))^{-3 }\int_0^{\infty}
\frac{d\tilde{k}}{2
\pi^2} \tilde{k}^2 |\beta_{\tilde{k}}|^2, 
\end{equation} 
where the number operator is defined at $\eta_1$. Assuming that the transition from
inflation-radiation or matter domination is smooth, the largest 
energy density can be obtained if $m_X/H_{inf} \sim 1$. If $0.04 \leq m_X/H_{inf}\leq 2$.
If $H_{inf} \sim m_\phi \sim 10^{13}$GeV and $m_\phi$ is the mass of the inflaton, then $X$ particles
produced gravitationally can match  the density today of the order of the critical density provided they are long lived.
Such super heavy massive dark matter particle $X$ is known as Wimpzillas!


\subsubsection*{(b)~Direct decay of the inflaton}

The dark matter can also be created from direct inflaton decay if $m_X < m_\phi/2$, with a rate
$\Gamma_X \sim h^{2}_{X} m_\phi/{8 \pi}$, where $h_{X}$ is the interaction strength. The total inflaton decay rate
is given by $\Gamma_d \sim \sqrt{1/3}~T^{2}_{R}/M_{\rm P}$, while
the inflaton number density at the time of decay is given by
$n_\phi \sim T^{4}_{\rm R}/m_\phi$. This constrains the overall
coupling to
\begin{equation}
h^{2}_{X} \leq 32 \pi \sqrt{1 \over 3} {T_{R} \over M_{\rm P}}
{10^{-9} \over m_X}\,,
\end{equation}
where $m_X$ is in units of GeV. This is required due to the fact that
the produced $X$ must not overclose the universe which,
for $\Omega_X \leq 0.22$ and $H_0 = 70$ km ${\rm sec}^{-1}
{\rm Mpc}^{-1}$, reads
\begin{equation}
{n_X / n_\gamma} \leq 4 \times 10^{-9}  m_X^{-1},
\end{equation}
when $m_X$ is expressed in units of GeV~\cite{Allahverdi:2001ux}. It is evident from the overclosure bound that $h_X$ needs to be very small.


\subsubsection*{(c)~Creation during reheating}

If the process of reheating is slow and not instantaneous, 
then it is possible to create WIMP from the ambient plasma which is in the
process of acquiring thermalization via scatterings. The Boltzmann equations for
inflaton energy density, $\rho_\phi$, radiation energy density, $\rho_r$, and  dark matter 
energy density $\rho_{X}$ are given by~\cite{Kolb:1988aj,Chung:1998rq}:
\begin{eqnarray}
\label{boltzm}
& &\dot{\rho}_\phi + 3H\rho_\phi +\Gamma_\phi\rho_\phi = 0
	\nonumber \\
& & \dot{\rho}_R + 4H\rho_R - \Gamma_\phi\rho_\phi
   - 	\frac{\langle\sigma|v|\rangle}{m_X}
	\left[ \rho_X^2 - \left( \rho_X^{eq} \right)^2 \right] =0
\nonumber \\
& & \dot{\rho}_X + 3H\rho_X 
    + \frac{\langle\sigma|v|\rangle}{m_X}
	\left[ \rho_X^2 - \left( \rho_X^{eq} \right)^2 \right] = 0  \ ,
\end{eqnarray}
where dot denotes time derivative, and  thermal averaged cross section is given 
by: $\langle\sigma|v| \rangle$.  The equilibrium energy
density for the $X$ particles, $\rho_X^{eq}$, is determined by the
radiation temperature, $T=(30\rho_R/\pi^2g_*)^{1/4}$. Following ~\cite{Chung:1998rq,Kolb:1988aj},
it is useful to introduce two dimensionless constants, $\alpha_\phi$
and $\alpha_X$, defined in terms of $\Gamma_\phi=\alpha_\phi m_\phi$,
$\langle \sigma|v| \rangle=\alpha_X m_X^{-2}$, and
$ \Phi \equiv \rho_\phi m_\phi^{-1} a^3 \ ; \quad
R    \equiv \rho_r a^4 \ ; \quad
X    \equiv \rho_X m_X^{-1} a^3 $.
With these parameters the Boltzmann equations are:
\begin{eqnarray}
\label{system}
\Phi' & = & - c_1 \ \frac{x}{\sqrt{\Phi x + R}}   \ \Phi \nonumber \\
R'    & = &   c_1 \ \frac{x^2}{\sqrt{\Phi x + R}} \ \Phi \
            + c_2 \ \frac{x^{-1}}{ \sqrt{\Phi x +R}} \
	           	         \left( X^2 - X_{eq}^2 \right) \nonumber \\
X'    & = & - c_3 \ \frac{x^{-2}}{\sqrt{\Phi x +R}} \
		\left( X^2 - X_{eq}^2 \right) \ .
\end{eqnarray}
where $x=am_{\phi}$, prime denotes $d/dx$, and 
the constants $c_1$, $c_2$, and $c_3$ are given by
$$c_1 = \sqrt{\frac{3}{8\pi}} \frac{M_{\rm P}}{m_\phi}\alpha_\phi\,,~~
c_2 = c_1\frac{m_\phi}{m_X}\frac{\alpha_X}{\alpha_\phi} \,,~~ 
c_3 = c_2 \frac{m_\phi}{m_X}. $$
The equilibrium value of $X$ is given in terms of the temperature $T$ and Eq.~(\ref{Max-Boltz}):
$X_{eq} =n_{X}m_{X}= ({m_X^3}/{m_\phi^3})({1}/{2\pi})^{3/2}x^3 ({T}/{M_X})^{3/2}\exp(-M_X/T)$.

It is straightforward to solve the system of equations in Eq.\
(\ref{system}) with initial conditions at $x=x_I$, $R(x_I)=X(x_I)=0$,
and $\Phi(x_I)=\Phi_I =(3/8\pi)(M_{\rm P}/m_{\phi}^{2}) (H_{I}^{2}/m_{\phi}^{2})x_{I}^{3}$.  
At early time solution for $R$ can be easily obtained:
\begin{equation}
\label{st}
R \simeq 0.4 c_1
     \left( x^{5/2} -  x_I^{5/2} \right) \Phi_I^{1/2}
			 \qquad (H \gg \Gamma_\phi) \ .
\end{equation}
By maximizing the above equation, which is obtained at $x/x_I = (8/3)^{2/5} = 1.48$, 
the largest temperature of the ambient plasma can be even larger than the reheat temperature~\cite{Kolb:1988aj}
\begin{equation}
\label{max}
{T_{MAX}}/{T_{R}} = 0.77 \left({9}/{5\pi^3g_*}\right)^{1/8}
		\left({H_{inf} M_{\rm P}}/{T_{R}^2}\right)^{1/4} \ .
\end{equation}
From Eq.\ (\ref{st}) when $x/x_I>1$, $T$ scales as $a^{-3/8}$, which implies that entropy
is created in the early-time regime.  
For the choices of $m_\phi$, $\alpha_\phi$, $g_*$, and $\alpha_X$, and  $\Omega_Xh^2 = 0.22$~\cite{Chung:1998rq}:
\begin{equation}
\Omega_X h^2 = \alpha_{X}^{2} \left(\frac{200}{g_{*}}\right)^{3/2}
\left(\frac{2000 T_R} {m_X} \right)^{7}
\end{equation}


\subsubsection*{(d)~Non-perturbative creation of dark matter}

The dark matter can be created non-perturbatively during the coherent oscillations of the inflaton. 

\vspace{-0.5cm}
\paragraph{\bf Superheavy dark matter during preheating:}
If the dark matter couples to the inflaton directly, then it is more efficient to excite them from the 
coherent oscillations of the inflaton during preheating. One of the most interesting applications 
of preheating is the copious production of particles which have a mass greater than
the inflaton mass $m_{\phi}$.  Such processes are impossible in perturbation 
theory and in the theory of narrow parametric resonance. 

Following Eq.~(\ref{intLag}), let us suppose that the dark matter $X$ is coupled to the inflaton
with an interaction term: $(1/2)g^{2}X^{2}\phi^{2}$.
During the broad resonance regime, as we have discussed in section \ref{PPR},
superheavy $X$-particles with mass $m_{X} \gg m_{\phi}$ can be produced. The momentum dependent frequency
 $\omega_k(t)$ violates the adiabatic condition of time dependent vacuum, see Eq.~(\ref{adiab}), when
\begin{equation}\label{adiabAAA}
{k^2 + m^2_{X}} \lesssim (g^2\phi m_\phi
\hat\phi)^{2/3} - g^2\hat\phi^2 \ ,
\end{equation}
where $\hat\phi$ is the amplitude of the inflaton oscillations.
The maximal range of momenta for
which particle production occurs corresponds to $\phi(t) = \phi_*$, where
$\phi_* \approx {\textstyle {1 / 2}} \sqrt {(m_\phi\hat\phi)/ g}$. The
maximal value of momentum for particles produced at that epoch can be
estimated by $k^2_{\rm max} + m^2_{\chi} = {(g m_\phi \hat\phi) / 2}$.  The resonance becomes efficient for
$ g m_\phi \hat\phi \gtrsim 4 m^2_{X} $.
Thus, the inflaton oscillations may lead to a copious production of
superheavy particles with $m_{X} \gg m_{\phi}$ if the amplitude of the field $\phi$ is
large enough, $g\hat\phi \gtrsim 4m^2_{X}/m_{\phi}$~\cite{Kofman:1997yn}.

\vspace{-0.5cm}
\paragraph{\bf Dark matter from the fragmentation of a scalar condensate:}
Let us assume that coherent oscillations of a scalar condensate
in a potential $U(\phi)$ has a frequency which
is large compared to the expansion rate of the universe. The equation of state is 
obtained by averaging, $p/\rho=(\vert\dot\phi\vert^2/\rho)-1$, over one oscillation
cycle $T$. The result is: $p=(\gamma -1)\rho$, 
where $ \gamma = (2/T)\int_0^T \left( 1-{U(\phi)}/{\rho} \right)dt$ \cite{Turner:1983he}.
For the case  $U\sim m^2\phi^2$, one finds $\gamma =1$, so that one
effectively obtains the usual case of pressureless, non-relativistic
cold matter. When the motion of the condensate is not simply oscillatory,
such as in the case of a rotating trajectory with a phase, one can generalize
the above calculation by integrating over the orbit of the condensate. Let us consider the potential
\begin{equation}
U(\phi)=\frac 12 m_{\phi}^2|\phi|^2\left({\phi^2/\mu^2}\right)^x\,,
\end{equation}
one finds that $\gamma = (1+x)/( 1+x/2)~,~~p=x/( 2+x )$.
There arises a negative pressure  whenever $x<0$. This is a sign of an instability of the 
scalar field under arbitrarily small perturbations. The
quantum fluctuations in the condensate grow according to when effective mass of the scalar 
field is much larger than the expansion
\begin{equation}
\ddot{\delta}_{\bf k} = -K{\bf k}^2\delta_{\bf k}~.
\end{equation}
If $K=2x<0$, quantum fluctuations of the condensate at the
scale, $\lambda= 2\pi/\vert{\bf k}\vert$, will grow exponentially in time:
\begin{equation}
\delta\phi_{\bf k}(t)=\delta\phi(0){\rm exp}\left(-K{\bf k}^2 t\right)~.
\end{equation}
In reality the onset of non-linearity sets the scale at which the spatial
coherence of the field can no longer be maintained and the condensate
fragments. The initial fluctuations in the condensate owes to the inflationary perturbations.
If the condensate carries a global charge, due to charge conservation the energy-to-charge ratio
changes as the the condensate fragments. This is what happens in the case of MSSM. The AD condensate 
which was responsible for generating the baryon asymmetry at the first instance could fragment~\cite{Kasuya:1999wu,Kasuya:2000wx,Kusenko:2009cv}. The ground state 
of these fragmented lumps is a non-topological soliton with a fixed charge, called the
$Q$-ball \cite{Coleman:1985ki,Lee:1991ax,Kusenko:1997zq}. In gauge mediated SUSY breaking scenarios
the $Q$-balls can absolutely stable and can be a candidate for CDM~\cite{Kusenko:1997vp,Kusenko:1997si,Kusenko:2009iz}. The $Q$-balls can also be formed
from the fragmentation of the inflaton~\cite{Enqvist:2002si,Enqvist:2002rj}.
The slow surface evaporation of a $Q$-ball will also create SUSY LSP, which we will discuss below.


\subsection{Candidates}

In this subsection we will discuss some of the dark matter candidates which are well motivated, and the challenges they face.

\subsubsection{Primordial blackholes}

The primordial blackholes (PBH) can be created in the early universe~\cite{Hawking:1971ei,Carr:1974nx}, and they can survive the age of the universe with a typical lifetime of an evaporating blackhole which is given by:
\begin{equation}
\frac{\tau}{10^{17}~{\rm sec}}\approx \left(\frac{M}{10^{15}~{\rm grams}}\right)^{3}\,,
\end{equation}
If the initial mass $M\approx10^{15}$~g, the blackhole will be evaporating now, for heavier 
blackholes the Hawking evaporation is negligible, and they can be a CDM candidate. When $M\sim 10^{9}$~g, the blackholes would
decay at the time of BBN. The PBHs are formed from the collapse of 
order one perturbations, $\delta \equiv \delta \rho/\rho\sim {\cal O}(1)$, inside the Hubble patch.  The 
detailed numerical simulation suggest $\delta_{c}\sim 0.7$~\cite{Niemeyer:1997mt,Niemeyer:1999ak,Jedamzik:1999am}. In a
radiation dominated epoch the blackhole mass is bounded by the Hubble mass
\begin{equation}
M_{H}\approx 10^{18}~g\left(\frac{10^{7}~{\rm GeV}}{T}\right)^{2}\,,
\end{equation}
where $T$ is temperature of the thermal bath during radiation. In spite of novelty in 
this idea, the detailed calculations suggest that it is hard to form primordial 
blackholes just from the collapse of sub-Hubble over densed regions - one 
requires more power on small scales $n<1.25-1.30$ in ${\cal P}_{\zeta}(k)=Ak^{n-1}$
\cite{Carr:1993aq,Green:1997sz,Drees:2011hb}, while the CMB data points towards $n\sim 0.96$.
It was shown in a hydrodynamical simulation  \cite{Jedamzik:1999am} that primordial blackholes
can also be produced in a first order phase transition, and during preheating~\cite{Green:2000he}.

The abundance of PBH contains many uncertainties, as the details of the initial gravitational collapse and the 
initial number density of $n_{_{PBH}}$ depends on many physical circumstances. These uncertainties can however 
be encoded in terms of the ratio determined by the initial time, $t_{i}$; $\rho_{_{PBH}}(t_{i})/\rho(t_{i})=Mn_{_{PBH}}(t_{i})/\rho(t_{i})=
4Mn_{_{PBH}}/3T_{i}s(T_{i})$, by assuming $\rho=3sT/4$,
\begin{equation}
\beta'(M)\approx 7.98\times 10^{-29}\left(\frac{M}{M_{\odot}}\right)^{3/2}\left(\frac{n_{_{PBH}}(t_{0})}{1~{\rm Gpc}^{-3} }\right)\,.
\end{equation} 
where $t_{0}$ corresponds to present time. In terms of this fraction $\beta'$, the PBH abundance is given by \cite{Carr:1975qj,Green:1999xm}
\begin{equation}
\Omega_{_{PBH}}h^{2}\approx 0.5 \left(\frac{\beta'(M)}{1.15\times10^{-8}}\right)\left(\frac{M}{M_{\odot}}\right)^{-1/2}\,.
\end{equation}
 The value of $\beta'(M)$ can be constrained from $\Omega_{_{PBH}}\leq \Omega_{CDM}$, which yields
 $\beta'(M)< 2.04\times 10^{-18}(\Omega_{CDM}/0.25)(M/10^{15}~{\rm g})^{1/2}$, for mass $M\geq10^{15}$~g.
A tighter constrain on $\beta'$ arises from a range of astrophysical observations, such as BBN, CMB anisotropy, 
and $\gamma$-ray backgrounds for $M\leq 10^{15}$~g, the bound weakens for larger mass blackholes~\cite{Carr:2009jm}. 


\subsubsection{Axions}

The axions were introduced to solve the strong CP problem~\cite{Peccei:1977hh,Peccei:1977ur} 
which requires a new global chiral symmetry $U(1)_{PQ}$ that is broken
spontaneously at the Peccei-Quinn (PQ) scale $f_a$ (for reviews, see~\cite{Kim:1986ax,Turner:1989vc,Raffelt:1990yz,Sikivie:2006ni}). 
The corresponding pseudo-Nambu-Goldstone boson is the axion
$a$~\cite{Weinberg:1977ma,Wilczek:1977pj}, which couples to the gluons
\begin{equation}
        {\cal L} 
        = \frac{a}{f_a/N}\,
        \frac{g_s^2}{32\pi^2}\,G_{\mu\nu}^a\widetilde{G}^{a\mu\nu}
        \ ,
\label{a-gluon-gluon}
\end{equation}
where $N$ is the color anomaly of the PQ symmetry depends  on the interactions. 
This interaction term compensates the vacuum contribution in the QCD Lagrangian
${\cal L}_{\Theta}=\Theta (g_{s}^{2}/32\pi^{2})G^{a}_{\mu\nu}\widetilde G^{a \mu\nu}$,
in a way that $\Theta \rightarrow \bar\Theta+{\rm Arg~(det}M)<10^{-9}$~\cite{Nakamura:2010zzi}, in order to match the 
electric dipole moment of the neutron. The dynamical solution yields when
 $\langle a\rangle =-\bar\Theta f_{a}/N$, at which the effective potential for the axion has its minimum.

The axion can interact via heavy quark while all other SM fields do not carry any PQ charge,
in which case $N=1$~\cite{Kim:1979if,Shifman:1979if}.  The axion can directly couple to the SM, and at the lowest order 
it will induce non-renormalizable coupling with the gluons, where $N=6$~\cite{Dine:1981rt,Zhitnitsky:1980tq}.

The axion searches, various astrophysical and cosmological observations suggest 
that the PQ scale~\cite{Sikivie:1999sy,Raffelt:2006cw,Nakamura:2010zzi} must be large,
\begin{equation}
f_a/N \gtrsim 6 \times 10^8~{\rm GeV} \,,
\label{Eq:faMin}
\end{equation}
and  the axion mass must be very small, $m_a \leq 0.01$~eV. The cosmological constraints on $f_{a}>2\times 10^{7}$~GeV
($m_{a}\leq  0.3$~eV) arises from BBN, accounting for the current bound on the relativistic species, 
$N^{eff}=3.1^{+1.4}_{-1.2}$~\cite{Iocco:2008va}. Another interesting bound arises from isocurvature perturbations from CMB. At best 
one can allow less than $10\%$ of the total perturbations to arise from sources other than the inflaton fluctuations--the axions being 
massless during inflation can account for such fluctuations, which limits $f_{a}\geq 10^{12}-10^{13}$~GeV, however, it depends on 
the scale of inflation~\cite{Beltran:2006sq,Steffen:2008qp}.

The axion life time depends on the axion-photon interaction, which  gives a long life time compared to
the age of the universe, i.e., $ \tau_a = \Gamma_{a\to\gamma\gamma}^{-1}=64\pi/(g_{a\gamma\gamma}^2 m_a^3)\sim
10^{40}(f_{a}N^{-1}/10^{10}~{\rm GeV})^{5}$~s for $m_{u}/m_{d}\sim 0.56$~\cite{Kolb:1988aj}.  

Axion is massless for $T\geq 1~{\rm GeV} \geq \Lambda_{QCD}$ and it acquires mass
only through instanton effects for $T\leq \Lambda_{QCD}$. For  $f_a/N\leq 3\times 10^{7}$~GeV
(corresponding to $m_a\geq 0.2$~eV), the axion is a thermal relic that decouples after the
quark--hadron transition, $T_{f}\leq150$~MeV. The axion is kept in thermal equilibrium 
with $\pi\pi \leftrightarrow \pi a $. The relic  thermal abundance  is given by 
$ \Omega_a h^2 = 0.077\left({10}/{g_{*}(T_{f})}\right) \left({m_a}/{10~{\rm eV}}\right) $,
where $g_{*}$ denotes the number of effectively massless degrees of
freedom.

In an opposite limit, when  $f_a/N$ is very large, the axions are never in thermal
equilibrium, and in particular when $T_{R} < f_{a}$ the PQ symmetry is never restored.
The main production mechanism is due to the coherent oscillations of the axion due to the 
initial misalignment angle $\Theta_i$ of the axion. At $ T\sim \Lambda_{QCD}$, the axion 
obtains a {\it temperature dependent} effective mass and 
oscillate coherently around its minimum when $m_a(T_{osc})\simeq H(T)$.  These oscillations of the axion 
condensate behaves as cold dark
matter~\cite{Preskill:1982cy,Abbott:1982af,Dine:1982ah}
with a relic density that is governed by the initial misalignment
angle $-\pi<\Theta_i\leq\pi$~\cite{Beltran:2006sq,Sikivie:2006ni}:
\begin{equation}
        \Omega_a h^2 \sim 0.15\,\xi\, 
        f(\Theta_i^2)\,
        \Theta_i^2
        \left(\frac{f_a/N}{10^{12} {\rm GeV}}\right)^{ 7/6}
\label{OAM}
\end{equation}
with $\xi={\cal O}(1)$ parametrizing theoretical uncertainties related
to details of the quark--hadron transition and $f(\Theta_i^2)$
accounting for anharmonicity in $\Theta_i$ -- $f(\Theta_i^2)\to
1$ for $\Theta_i^2\to 0$. For $10^{10}\,{\rm GeV}\leq f_a/N \leq 10^{13}$~GeV, this
``misalignment mechanism'' can provide the correct dark matter abundance.

\nopagebreak
\subsection{SUSY WIMP}

The most general gauge invariant and renormalizable superpotential
would also include baryon number $B$ or lepton number $L$ violating terms, with 
each violating by one unit: $W_{\Delta L=1}=\frac{1}{2}\lambda^{ijk}L_{i}L_{j}e_{k}+
\lambda^{\prime ijk}L_{i}Q_{j}d_{k}+\mu^{\prime i}L_{i}H_{\mu}$ and
$W_{\Delta B=1}=\frac{1}{2}\lambda^{\prime \prime ijk}u_{i}d_{j}d_{k}$,
where $i=1,2,3$ represents the family indices. The chiral supermultiplets
carry baryon number assignments $B=+1/3$ for $Q_{i}$, $B=-1/3$ for
$u_{i},  d_{i}$, and $B=0$ for all others. The total lepton number
assignments are $L=+1$ for $L_{i}$, $L=-1$ for $e_{i}$, and $L=0$ for
all the others. Unless $\lambda^{\prime}$ and $\lambda^{\prime\prime}$ terms are very
much suppressed, one would obtain rapid proton decay which violates
both $B$ and $L$ by one unit.

There exists a discrete $Z_2$ symmetry, which can forbid baryon and lepton number violating terms,
known as $R$-parity \cite{Fayet:1979yb}. For each particle:
\begin{equation}
P_{R}=(-1)^{3(B-L)+2s}\,
\end{equation}
with $P_{R}=+1$ for the SM particles and the Higgs bosons, while $P_{R}=-1$
for all the sleptons, squarks, gauginos, and Higgsinos.
Here $s$ is spin of the particle. Besides forbidding $B$ and $L$ violation from the 
renormalizable interactions, $R$-parity has  interesting phenomenological and cosmological consequences.
The lightest sparticle with $P_{R}=-1$, the LSP, must be absolutely stable.
If electrically neutral, the LSP is a natural
candidate for dark matter \cite{Ellis:1983ew,Dimopoulos:1988jw}. The advantage here 
is that their cross sections are governed by the SM gauge group -- and therefore the dark matter paradigm is embedded 
within a visible sector. However, there are some exceptions which we will discuss first. 


\subsubsection{Gravitino}

The gravitino is a spin-$3/2$ supersymmetric partner of the graviton,
which is coupled to all the sectors universally, e.g. visible and hidden sectors, with a 
Planck suppressed interaction. In this respect gravitino is truly a singlet dark matter candidate, if it happens to be the 
LSP. The gravitino mass can vary ($m_{3/2}\sim {\cal O}(100)~{\rm GeV}-{\cal O}(1)$~KeV) 
depending on the details of the SUSY breaking schemes,
for instance, in gauge and gravity-mediated
SUSY breaking scenarios~\cite{Dine:1994vc,Dine:1995ag,Giudice:1998bp}.
Indeed, without considering the SUSY breaking mechanisms and the SUSY
breaking scale, we can treat the gravitino mass as a free parameter.

\vspace{0.5cm}
{\bf Production:} Gravitinos with both the helicities can be produced from a thermal
bath, they are never in thermal equilibrium. They are produced 
mainly through the scatterings--within MSSM there are many scattering 
channels which include fermion, sfermion, gauge and gaugino quanta all of which have a cross-section
$\propto 1/M^2_{\rm P}$. The thermal abundance is given by (up to a logarithmic correction)
for $g_{\ast} =228.75$ as in the case of MSSM:
$Y_{\pm 3/2} \simeq (T_{R}/ 10^{10}~{\rm GeV})\times 10^{-12}$~\cite{Ellis:1984eq}, and 
$Y_{\pm 1/2} \simeq  [1 + M^2_{\widetilde g}(T_{R})/ 12 m^2_{3/2} ]
({T_{R} /10^{10}~{\rm GeV}})\times 10^{-12}$~\cite{Bolz:2000fu},
where $M_{\widetilde g}$ is the gluino mass. 

\begin{figure}[t]
  \centerline{\includegraphics[width=0.30\textwidth]{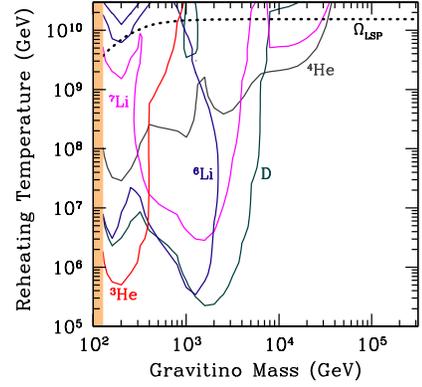}}
  \caption{The bound on reheat temperature $T_{R}$ with respect to an unstable gravitino mass $m_{3/2}$,
  where neutralino is the LSP with  a mass $117$~GeV
  ( indicated by the shaded light-orange region in which $m_{3/2}\leq m_{\bar \chi^{0}_{1}}$
    for $(m_{1/2},~m_{0})=(300,~141)$~GeV,
    $A_0=0$, $\tan\beta=30$. The thermal relic density is given by:
    $\Omega_ {\bar \chi^{0}_{1}}h^2=0.111$. Above the dotted line
    labeled as $\Omega_{\rm LSP}$, the ${\bar \chi^{0}_{1}} $ density from
    decays of thermally produced gravitinos exceeds
    $\Omega_ {\bar \chi^{0}_{1}} h^2=0.118$, see~\cite{Kawasaki:2008qe}.}
  \label{GravitinoConstraints}
\end{figure}

Note that for
$M_{\widetilde g} \leq m_{3/2}$ both the helicity states have
essentially the same abundance, while for $M_{\widetilde g} \gg
m_{3/2}$ production of helicity $\pm 1/2$ states is enhanced due to
their Goldstino nature. The net abundance  $\Omega_{3/2} h^2$ can be approximated by 
the convenient expression for the universal gaugino masses $M_{1,2,3}=m_{1/2}$ at 
$M_{GUT}$ and $m_{3/2}\ll M_{1, 2, 3}$~\cite{Pradler:2007is}
\begin{equation}
  \Omega_{3/2} h^2 \simeq 0.32 \Big( \frac{10~{\rm GeV}}{m_{3/2}} \Big) 
  \Big( \frac{m_{1/2}}{1~{\rm TeV}} \Big)^2 \Big(
    \frac{T_{R}}{10^{8} {\rm GeV}} \Big)
  \label{eq:omega-3/2}
\end{equation}
Thermally produced gravitinos have a negligible free-streaming velocity today. However, the gravitinos created from 
decays can be warm or hot dark matter. 

Besides thermal production, gravitino can be produced non-thermally from the decay of the
NLSP (next-to-lightest SUSY particle). Obviously different NLSP's give slightly different abundances.
For sneutrino, see Refs.~\cite{Feng:2004mt,Arina:2007tm,Ellis:2008as}, for stop NLSP, 
see~\cite{Kang:2006yd,DiazCruz:2007fc,Berger:2008ti}.  A
simple approximation yields~\cite{Steffen:2006hw,Pospelov:2008ta}
\begin{equation}
        Y_{\widetilde l_{1}}^{\rm dec}\equiv \frac{n_{\widetilde l_{1}}}{s}
        \simeq
        0.7 \times 10^{-12}
        \left(\frac{m_{\widetilde l_{1}}}{1~{\rm TeV}}\right)\ ,
\label{Yslepton}
\end{equation}
where the total slepton number density is given by assuming an equal number density
of positively and negatively charged slepton's. The NLSP's can
have a long lifetime $\tau_{\widetilde l_{1}}$. For a slepton NLSP, one finds in the limit $m_{l}\to 0$,
\begin{equation}
  \tau_{\widetilde l_{1}} 
  \simeq 
  \Gamma^{-1}
  = 
  \frac{48 \pi m_{3/2}^2 M_{\rm P}^2}{m_{\widetilde l_{1}}^5}
  \left(1-\frac{m_{3/2}^2}{m_{\widetilde l}^2}\right)^{-4},
\label{StauLifetime}
\end{equation}
which holds not only for a charged slepton NLSP but also for the
sneutrino NLSP. Similar expressions for the lifetimes for the 
neutralino NLSP \cite{Feng:2004mt} and the stop NLSP can be found in\cite{DiazCruz:2007fc}.
If the NLSP decays into the gravitino LSP after
BBN, the SM particles emitted in addition to the gravitino
can affect the abundances of the primordial light elements. 
Also the presence of a long-lived negatively charged particle, champ, i.e.
$\widetilde l_{1}^{-}$, can lead to bound states that catalyze BBN reactions~\cite{Dimopoulos:1989hk,DeRujula:1989fe}. The new acceptable limits 
on champs: $100(q_{X}/e)^{2}\leq m_{X}\leq 10^{8}(q_{X}/e)^{2}$~TeV, virtually ruled out any low scale SUSY champs~\cite{Chuzhoy:2008zy}.
It was suggested in Ref.~\cite{Pospelov:2006sc} that bound-state formation of champ with
$He^{4}$ can lead to a substantial production of primordial $Li^{6}$, which puts the constraint 
$\tau_{\widetilde l_{1}} \lesssim 5\times 10^3$~sec.
~\cite{Pospelov:2006sc,Hamaguchi:2007mp,Bird:2007ge,Takayama:2007du,Pradler:2007is,Pospelov:2008ta}.

\vspace{0.5cm}
{\bf Uncertainties:} The main uncertainties on gravitino abundance arise from the {\it hidden sectors}. If the inflaton
sector is embedded within a hidden sector then there are many more sources of gravitino 
production--the inflaton could decay directly into gravitino during reheating or preheating~\cite{Maroto:1999ch,Giudice:1999am,Nilles:2001my,Kallosh:1999jj,Kallosh:2000ve,Nilles:2001vc,Frey:2005jk}, the inflaton couplings to other hidden sectors can similarly excite gravitinos giving rise to large
uncertainties in their total abundance. 
\begin{equation}\label{non-therm-gr}
\Omega_{3/2}h^{2}=\Omega_{3/2}^{MSSM}h^{2}+\Omega_{3/2}^{Inflaton}h^{2}+\Omega_{3/2}^{Hidden}h^{2}\,,
\end{equation}
All these contributions can easily overproduce gravitinos, i.e. $\Omega_{3/2}h^{2}\sim 1$, especially the last two 
sectors are largely unconstrained by particle physics. These uncertainties can be minimized
if the last phase of inflation occurs within MSSM, as discussed in Sect.~\ref{GGIOM}, then the only predominant source of 
gravitino production arises from the decay of the MSSM inflaton and from the MSSM thermal bath.

Another solution has been put forward -- for high scale and  hidden sector models of inflation -- since
the flat directions of MSSM can be displaced from their minimum during inflation, the flat direction VEV at early times would 
generate time dependent masses to the MSSM fields which are coupled to the flat direction. As a  result, the inflaton might 
not even decay into all the MSSM {\it d.o.f} due to kinematical blocking~\cite{Allahverdi:2005fq,Allahverdi:2005mz,Allahverdi:2007zz}.
Furthermore, the flat direction VEV also generates masses to gauge bosons and gauginos which participate in scatterings, therefore
delaying the actual thermalization process and lowering the reheating temperature, i.e. $T_{R}$. Both these effects address the 
thermal gravitino overproduction problem without any need of extra assumptions~\cite{Allahverdi:2005fq}.


\vspace{0.5cm}
{\bf Unstable gravitino:} An unstable gravitino decays to particle-sparticle pairs, and its
decay rate is given by $\Gamma_{3/2} \simeq m^{3}_{3/2}/4 M^2_{\rm P}$.
If $m_{3/2} < 50$ TeV, the gravitinos decay during or after BBN, which can ruin its successful
predictions for the primordial abundance of light
elements~\cite{Cyburt:2002uv,Kawasaki:2004yh}. If the gravitinos decay radiatively, the most
stringent bound, $\left(n_{3/2}/s\right) \leq 10^{-14}-10^{-12}$,
arises for $m_{3/2} \simeq 100~{\rm GeV}-1$ TeV~\cite{Cyburt:2002uv}.
On the other hand, much stronger bounds are derived if the gravitinos
mainly decay through the hadronic modes. In particular, for a hadronic
branching ratio $\simeq 1$, and in the same mass range,
$\left(n_{3/2}/s\right) \leq 10^{-16}-10^{-15}$ will be
required~\cite{Kawasaki:2004yh,Kawasaki:2008qe}. This puts constraint on reheat temperature of the universe, i.e. $T_{R}$, 
at which these unstable gravitinos are produced, see Fig.~\ref{GravitinoConstraints}.
An intriguing possibility arises if R-parity is broken. The gravitino LSP with $m_{3/2}\sim 1$~GeV can still
be a long-lived cold dark matter, while evading the bounds on $T_{R}$ from Fig.~\ref{GravitinoConstraints}. 
The gravitino in this case cannot decay into hadrons which is kinematically 
suppressed, and the three-body decay life time is typically larger than the age of the universe~\cite{Kohri:2009ka}.
For a GeV mass gravitino, the present day free-streaming velocity is $\leq 10^{-9}$~km/s, which corresponds to 
that of a cold dark matter.


\vspace{0.5cm}
{\bf Detection:} The direct detection of gravitino will be impossible at the LHC, their production will be extremely suppressed. 
If the NLSP is long lived (quasi-stable) as in the case of stau then they would
penetrate the collider detector in a way similar to muons~\cite{Drees:1990yw,Nisati:1997gb,Feng:1997zr}.  If the 
produced staus are slow, then from  the associated highly ionizing tracks and
time of flight measurements one can determine their mass~\cite{Ellis:2006vu}. This might 
give some indirect handle on gravitinos.


\subsubsection{Axino }

\begin{figure}[t]
\includegraphics[width=.30\textwidth]{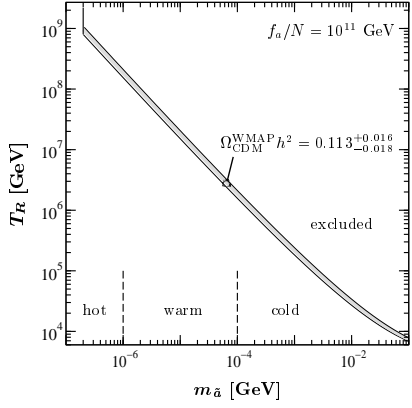}
\caption{The plot shows the reheating temperature $T_{R}$ with respect to the axion mass for $f_a/N = 10^{11}$~GeV. The gray 
band indicates $\Omega_{\widetilde a}h^2= 0.113^{+0.016}_{-0.018}$.  Thermally produced axinos can be classified as hot, warm, and cold dark matter~\cite{Covi:2001nw} as indicated in the plot. The plot is taken from~\cite{Brandenburg:2004du}.}
\label{Fig:adml}
\end{figure}

The axino, $\widetilde a$, is a superpartner of the axion, is another example of a gauge singlet dark matter 
candidate~\cite{Nilles:1981py,Kim:1983dt,Kim:1983ia}. It interacts extremely weakly since its couplings are
suppressed by the PQ scale $f_a\gtrsim 10^8$~GeV~\cite{Sikivie:2006ni,Raffelt:2006rj,Raffelt:2006cw},
and its mass can range from eV to GeV.  In the hadronic axion model~\cite{Kim:1979if,Shifman:1979if} in a SUSY setting, 
the axino couples to MSSM field indirectly via loops of heavy (s)quarks. Typically $\widetilde a$ decouples early at a 
temperature $T_{f}\geq 10^{9}$~GeV, below this temperature, they are mainly created from scatterings of MSSM fields in an 
kinetic equilibrium. The thermal abundance is given by~\cite{Brandenburg:2004du,Covi:2001nw,Choi:2007rh}
\begin{eqnarray}
        \Omega_{\widetilde a}h^2
        &\simeq&
        5.5g_{s}^6(T_R) \ln\left(\frac{1.108}{g_{s}(T_R)} \right) 
        \left(\frac{10^{11}~{\rm GeV}}{f_a/N}\right)^{2}
\nonumber\\
        &&
        \times
        \left(\frac{m_{\widetilde a}}{0.1~{\rm GeV}}\right)
        \left(\frac{T_R}{10^4~{\rm GeV}}\right)
\label{Eq:AxinoDensityTP}
\end{eqnarray}
with the axion-model-dependent color anomaly $N$ of the PQ symmetry breaking scale $f_{a}$.
Thermally produced axinos can be hot, warm, and cold dark matter~\cite{Covi:2001nw} as shown 
in Fig.~\ref{Fig:adml}.

Non-thermal production of $\widetilde a$ has many uncertainties. The $\widetilde a $ can be created from the decay of the NLSP, 
direct decay from the inflaton or moduli, or any other hidden sector. The expression for the final abundance is similar to 
that of Eq.~(\ref{non-therm-gr}), where $3/2 \longrightarrow \widetilde{a}$. In this regard $\widetilde a$ faces similar challenges 
as that of a gravitino dark matter.

The $\widetilde a$ LSP is inaccessible in direct/indirect
dark matter searches if R-parity is conserved. Their direct 
production at collider is strongly suppressed.  Nevertheless,
quasi-stable $\widetilde \tau$'s could appear in collider detectors (and
neutrino telescopes~\cite{Albuquerque:2006am,Ahlers:2006pf}) as a possible signature of
the $\widetilde a$ LSP. 


\subsubsection{Neutralino}

In the MSSM the binos $\widetilde B$ (superpartner of B), winos $\widetilde W$ (superpartner of $W$)
and Higgsinos ($\widetilde H_{u}^{0}$ and $\widetilde H_{d}^{0}$) mix into $4$ Majorana fermion eigenstates, called
neutralinos with $4$ mass eigenstates:
$\widetilde \chi_{1}^{0},~\widetilde \chi_{2}^{0},~\widetilde \chi_{3}^{0},~\widetilde \chi^{0}_{4}$, ordered with increasing mass.
The LSP is thus denoted by $\widetilde \chi_{1}^{0}=N_{11}\widetilde B+N_{12}\widetilde W+N_{13}\widetilde H_{u}^{0}+N_{14}\widetilde H_{d}^{0}$.
The gaugino fraction, $f_{G}=N_{11}^{2}+N_{12}^{2}$, and Higgsino fraction, $f_{H}=N_{13}^{2}+N_{14}^{2}$,  are determined by the mixing 
matrix, $N$, which diagonalizes the neutralino mass matrix~\cite{Kane:1993td,Jungman:1995df}. 
 
 
 \begin{figure}[t]
\includegraphics[width=4.6cm]{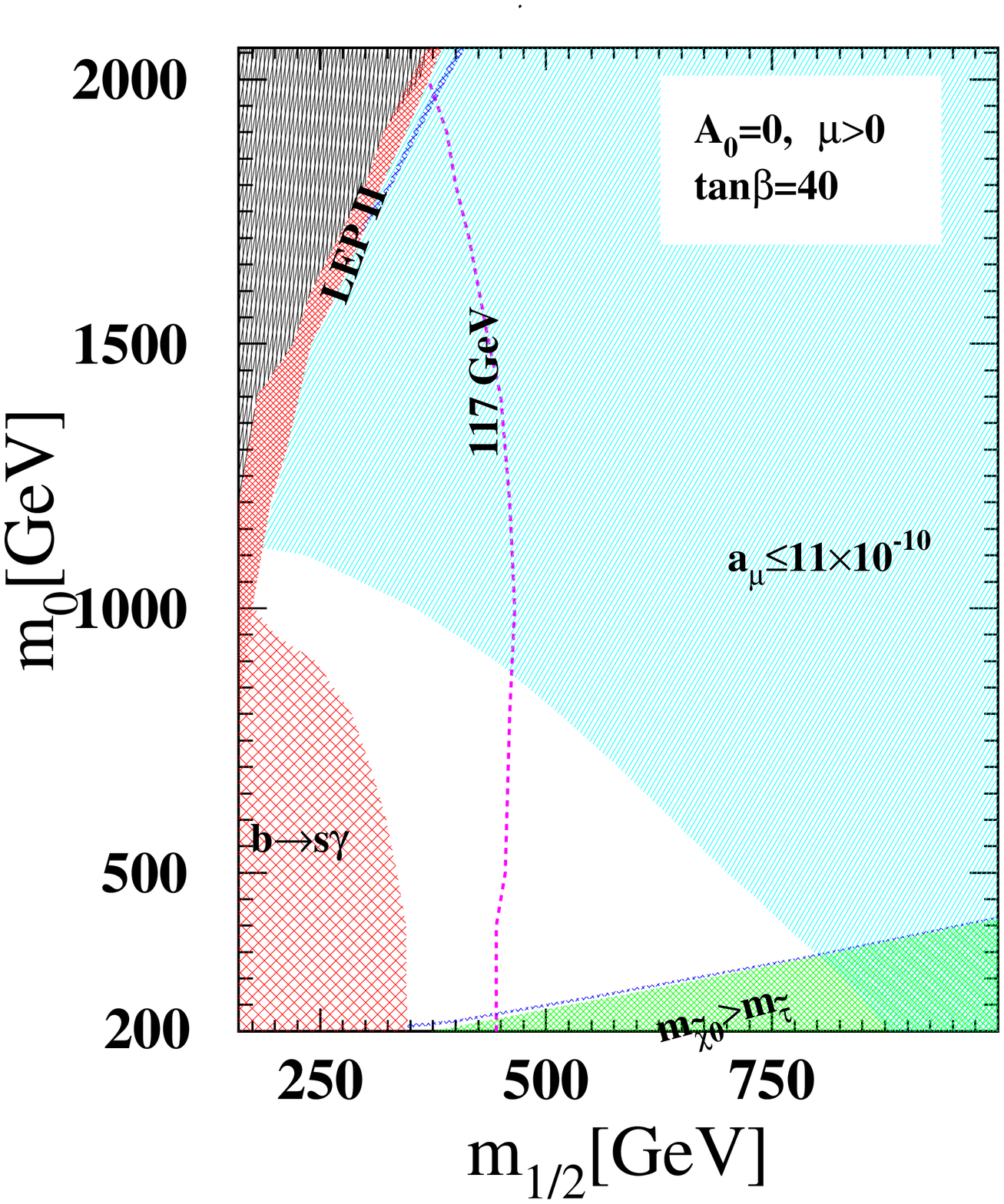}~
\includegraphics[width=4.6cm]{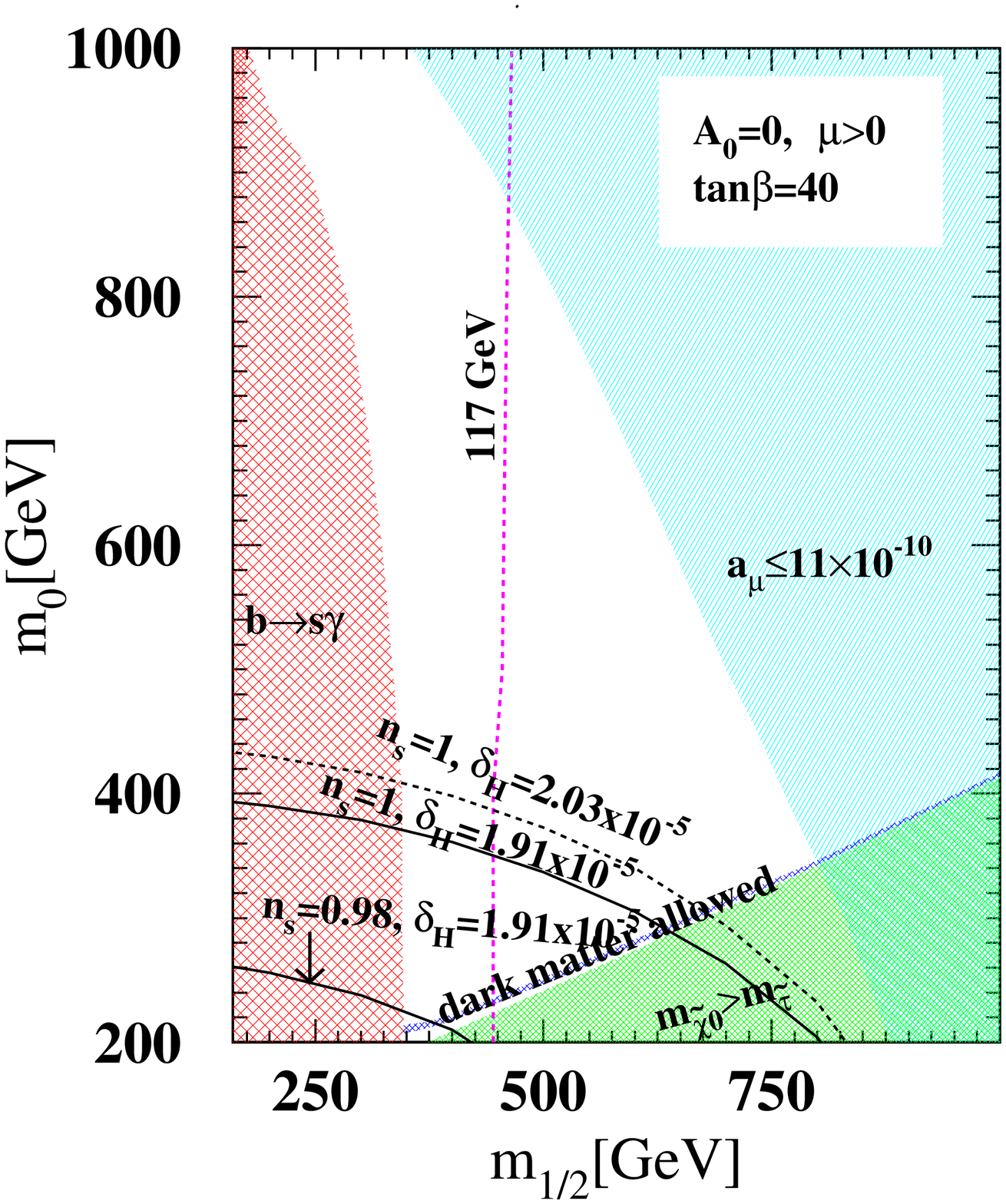}
\caption{On the left hand panel the co-annihilation band is shown for ($m_0-m_{1/2}$) 
plane for $\tan\beta=40$. The constraints are shown in Figure, see~\cite{Nath:2010zj}. 
On the right hand panel the overlapping contours between MSSM inflation and neutralino dark 
matter for different values of ($n_s.~\delta_H$) within $95\%$c.l.
are shown for the same parameter region~\cite{Allahverdi:2007vy}.
\label{Neut-fig}}
\label{40flat}
\end{figure}

 The $\widetilde \chi_{1}^{0}$'s were in thermal equilibrium for primordial
temperatures of $T>T_{f}\sim\ m_{\widetilde\chi_{1}^{0}}/20$. At $T_{f}$,
the annihilation rate of the (by then) non-relativistic
$\widetilde \chi_{1}^{0}$'s becomes smaller than the Hubble rate so that they
decouple from the thermal plasma, see Sect.~\ref{Th-relic}.  
 
It is easy to work with  a limited set of parameters, the mSUGRA model is a 
simple model which contains only five parameters:
\begin{equation} m_0,~\ m_{1/2},~\ A_0,~\
\tan\beta~\ {\rm and}~\ sign(\mu ).
\end{equation}
$m_0$ is the universal scalar soft breaking parameter, $m_{1/2}$ is the universal gaugino mass,
 $A_0$ is the universal cubic soft breaking mass, measures at $M_{\rm GUT}$,
 and $\tan\beta = \langle \widetilde {H}^{0}_u \rangle / \langle \widetilde{H}^{0}_d
\rangle$ at the electroweak scale. 

The model parameters are already constrained by different
experimental results.  (i)~the light Higgs mass bound of $M_{h^0} > 114$~GeV
from LEP~\cite{Barate:2003sz}, (ii)~the $b\rightarrow s \gamma$ branching
ratio bound of $1.8\times10^{-4} < {\cal B}(B \rightarrow X_s
\gamma) < 4.5\times10^{-4}$ (we assume here a relatively broad
range, since there are theoretical errors in extracting the
branching ratio from the data)~\cite{Alam:1994aw}, (iii)~the 2$\sigma$
bound on the dark matter relic density: $0.095 < \Omega_{\rm CDM}
h^2 <0.129$~\cite{Komatsu:2008hk}, (iv)~the bound on the lightest chargino
mass of $M_{\tilde \chi^{\pm}_{1}} >$ 104~GeV from LEP \cite{Nakamura:2010zzi} and (v) the
muon magnetic moment anomaly $a_\mu$,  where one gets a 3.3$\sigma$
deviation from the SM from the experimental
result~\cite{Bennett:2004pv}.

The allowed mSUGRA parameter space, at present, has mostly three
distinct regions: (i)~the stau-neutralino
($\tilde\tau_1~-~\tilde\chi^1_0$), coannihilation region where
$\tilde\chi^1_0$ is the lightest SUSY particle (LSP), (ii)~the
$\tilde\chi^1_0$ having a dominant Higgsino component (focus point)
and (iii)~the scalar Higgs ($A^0$, $H^0$) annihilation funnel
(2$M_{\tilde\chi^1_0}\simeq M_{A^0,H^0}$). These three regions have
been selected out by the CDM constraint. There stills exists a bulk
region where none of these above properties is observed, but this
region is now very small due to the existence of other experimental
bounds. The allowed parameter space for the neutralino dark matter (blue region) for $\tan(\beta)=40$ is shown 
in  the left panel of Fig.~\ref{Neut-fig}.


\vspace{0.5cm}
{\bf Detection:} In general the observable signals for SUSY at LHC are: $n$ leptons+ $m$ jets +$\met$ (missing transverse energy), where 
either $n$ or $m$ could be $0$. The existence of missing energy in the signal will tell us the possibility of dark matter candidate.
There are SM backgrounds, e.g. $W$ and $Z$ bosons decaying to neutrinos providing $\met$. The clean signal for 
SUSY would be jets + $\met$, without isolated leptons. One of the key analysis for mSUGRA is to measure the $M_{eff}$ which is the sum of the
transverse momenta of the four leading jets and the missing transverse energy: $M_{eff} = p_{T,1} + p_{T,2} + p_{T,3} + p_{T,4} + \met$.
One has to further measure the masses (squarks, sleptons), $A_{0} $ and $\tan\beta$, and the mixing matrices which lead to the calculation of the relic density. One particularly favored parameter space is the coannihilation  region where the stau
and the neutralino masses are close for smaller values of $m_0$. The mass difference, $\Delta m$, governs the relic abundance due to the Boltzmann suppression factor $e^{-\Delta M/T}$, see section~\ref{coanni}. Therefore measuring $\Delta M$ directly gives handle on measuring the relic abundance at the LHC, see for a detailed 
discussion in~\cite{Nath:2010zj}. 


\vspace{0.5cm}
{\bf MSSM inflation and dark matter:} After considering all these bounds it was found that there
exists an interesting overlap between the constraints from MSSM inflation
and the $\widetilde\chi_{1}^{0}$ abundance, see the right hand panel of Fig.~\ref{40flat} for $\tan\beta=40$.
The constraints on the parameter space arising from the inflation
appearing to be consistent with the constraints arising from the
dark matter content of the universe and other experimental results.

It is also interesting to note that the allowed region for $u_{1}d_{2}d_{3}$ as an MSSM inflaton with a mass
$m_{\phi}$, required by the CMB observations for $\lambda=1$, see Fig.~\ref{nsdel0} in Sect.~\ref{GIIOM}, lies in the
stau-neutralino coannihilation region which requires smaller values
of the SUSY particle masses~\cite{Allahverdi:2007vy}. Similar analysis were performed in \cite{Balazs:2004ae,Balazs:2004bu}, 
where the authors studied the overlap between MSSM parameters for the electroweak baryogenesis and the $\widetilde\chi_{1}^{0}$
dark matter abundance.


\subsubsection{Sneutrino}

The lightest right handed  (RH) sneutrino $\widetilde N$ can be a good dark matter candidate
when the SM gauge group is augmented to $SU(3)_{C}\times SU(2)_{L}\times U(1)_{Y}\times U(1)_{B-L}$,
with a superpotential~\cite{Allahverdi:2007wt,Allahverdi:2009sb}
\begin{equation}
W=W_{MSSM}+W_{B-L}+yNH_{u}L\,.
\end{equation}
The model introduces new gauge boson $Z'$, two Higgs fields $H_{1}',~H_{2}'$, and their superpartners, the Yukawa 
coupling is denoted by $y$. The spontaneous breaking of $U(1)_{B-L}$ will generate Majorana neutrino masses, or if 
$y\approx 10^{-12}$, then Dirac neutrino masses, or a mixture of both Dirac and Majorana~\cite{Allahverdi:2010us}.
If the right handed sneutrino is the LSP then it provides another compelling candidate which is well motivated from particle 
theory and can be embedded with least unknown uncertainties~\cite{Allahverdi:2007wt,Arina:2007tm,Lee:2007mt}.

Scatterings via the new $U(1)$ gauge interactions bring the RH
sneutrino into thermal equilibrium.
In order to calculate the relic abundance of the RH sneutrino, we
need to know the masses of the additional gauge boson $Z^{\prime}$
and its SUSY partner ${\tilde Z}^{\prime}$, the new Higgsino masses,
Higgs VEVs which break the new $U(1)_{B-L}$ gauge symmetry, the RH
sneutrino mass, the new gauge coupling, and the charge assignments
for the additional $U(1)$.   The primary diagrams responsible to provide the right amount of relic
density are mediated by ${\tilde Z}^{\prime}$ in the $t$-channel

\begin{figure}[t]
\includegraphics[width=6cm]{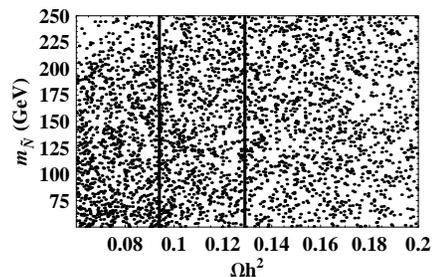}
\caption{$\Omega h^2$ vs $m_{\tilde N}$. The solid lines from left
to right are for $\Omega h^2 =$ 0.094 and 0.129 respectively. The
$Z^{\prime}$-ino mass is equal to the Bino mass since the new $U(1)$
gauge coupling is the same as the hypercharge gauge coupling. The plot is taken from~\cite{Allahverdi:2007wt}.
}\label{sneutrinoominf}
\end{figure}

By assuming that the new gauge symmetry is
broken around  TeV in Fig.~\ref{sneutrinoominf}, we show the relic density values for smaller masses\
of sneutrino where the lighter stop mass is $ \leq 1$ TeV. The
smaller stop mass will be easily accessible at the LHC. By
varying new gaugino and Higgsino masses and the ratio of the VEVs of
new Higgs fields, the
WMAP~\cite{Komatsu:2008hk} allowed values of the relic density, i.e.,
$0.094-0.129$ is satisfied for many points. In the case of a larger
sneutrino mass in this model, the correct dark matter abundance can
be obtained by annihilation via $Z^{\prime}$ pole~\cite{Lee:2007mt,Allahverdi:2007wt}.

{\bf Detection:}
Since the dark matter candidate, the RH sneutrino, interacts with
quarks via the $Z^{\prime}$ boson, it is possible to see it via the
direct detection experiments. The detection cross sections are not
small as the interaction diagram involves $Z^{\prime}$ in the
$t$-channel. The typical cross section is about 2$\times 10^{-8}$ pb
for a $Z^{\prime}$ mass around 2 TeV. It is possible to probe this
model in the upcoming dark matter detection experiments.
The signal for this scenario at the LHC will contain standard
jets plus missing energy and jets plus leptons plus missing energy.
The jets and the leptons will be produced from the cascade decays of
squarks and gluinos into the final state containing the sneutrino.


\subsubsection{Stable and evaporating Q-ball, LSP dark matter}

The AD condensate fragments to form Q-balls, for reviews see~\cite{Enqvist:2003gh,Dine:2003ax}, and a finite size Q-ball has a minimum of 
energy and it is stable with respect to decay into free quanta if $U(\phi)/\phi^{2}=$~min, for $\phi_{0}>0$. 

For a sufficiently large $Q$, the energy of a soliton is then given by~\cite{Coleman:1985ki,Kusenko:1997zq,Lee:1991ax}:
$E=|\nu Q|< m_{\phi}|Q|$,
which ensures its stability against decay into plane wave solutions
with $\varphi \simeq \varphi_{0}$ inside and $\varphi \simeq 0$ outside
the soliton, where $\phi(t, x)=e^{i\omega t}\varphi(x)$. The value of $\nu$ was computed in~\cite{Kusenko:1997ad,Kusenko:1997zq}.
 Note that the global $U(1)$ symmetry is thus broken inside
the soliton by the VEV, however, remains unbroken outside the soliton.
The  most crucial piece is the presence of a global $U(1)$
charge of the $Q$-ball which actually prevents it from decaying and
makes the soliton stable. 

In the context of gauge mediated SUSY breaking the AD potential
takes the form $U(\varphi)\approx m^4_{\phi}\log\left(1+{|\varphi|^2}/
{m_{\phi}^2}\right)$~\cite{Kusenko:1997si,Dvali:1997qv},
where $m_{\phi} \sim {\cal O}({\rm TeV})$, represents the SUSY
breaking scale.  The profile of the $Q$-ball is given by $\varphi(r)\sim \exp(-m_{\phi}r)$, where
the energy, radius and the VEV of a $Q$-ball goes as~\cite{Kusenko:1997si,Enqvist:2003gh}
\begin{equation}
\label{thick1}
E\approx \frac{4\sqrt{2}}{3}\pi m_{\phi}Q^{3/4}, R \approx \frac{Q^{1/4}}{\sqrt{2}m_{\phi}}, \varphi_{0} \approx \frac{m_{\phi}}{\sqrt{2\pi}}Q^{1/4}
\end{equation}
This allows for the existence of some entirely stable Q-balls with a large
baryon number $Q\sim B$ (B-balls).  Indeed, if the mass of a B-ball is $M_{B} \sim
({\rm 1~TeV}) \times B^{3/4}$, then the energy per baryon number
$(M_{B}/B)\sim ({\rm 1~TeV}) \times B^{-1/4}$ is less than 1~GeV for $B >
10^{12}$.  Such large B-balls cannot  dissociate into protons and neutrons
and are entirely stable -- thanks to the conservation of energy and the baryon
number.  If they were  produced in the early universe, they would exist at
present as a form of dark matter~\cite{Kusenko:1997si}. There are astrophysical and terrestrial lilmits~\cite{Kusenko:1997vp,Kusenko:2009iz}, 
and direct searches for the Q-balls, which places a lower limit on  $Q>10^{22}$~\cite{Arafune:2000yv}.

In the gravity mediated case the $B$-balls are not stable, but they evaporate via surface evaporation.
In this process AD condensate can generate the required baryon asymmetry and also create dark mater.
The appropriate candidate will be the $udd$ flat direction, lifted by $n=6$ operator, which carries the baryon 
number and the right dark matter abundance, see Eq.~(\ref{assym2}).

When a $B$-ball decays, for each unit of $B$ produced,
corresponding to the decay of $3$ squarks to quarks, there will be at least
three units of $R$-parity produced, corresponding to at least $3$ 
LSPs (depending on the nature of the cascade produced by the squark decay
and the LSP mass, more LSP pairs could be produced). Let $N_{_{LSP}} \geq 3$
be the number of LSPs produced per baryon number and $f_{B}$ be the fraction
of the total $B$ asymmetry contained in  $B$-balls. Then the
baryon to dark matter ratio, $r_{B} = {\rho_{B}}/{\rho_{DM}}$,  and the dark matter abundance are given by~\cite{Enqvist:1997si,Enqvist:1998en},
\begin{equation}
r_{B} = \frac{m_{n}}{N_{_{LSP}}f_{B}m_{_{LSP}}},~~ \Omega_{LSP}\approx 3f_{B}\left(\frac{m_{_{LSP}}}{m_{n}}\right)
\end{equation}
where $m_{n}$ is the nucleon mass and $m_{_{LSP}}$ is the 
LSP mass. It is rather natural to have $r_{B} < 1$.

The LSPs produced in $B$-ball decays will collide with themselves
and with other weakly interacting particles in the background and
settle locally into a kinetic equilibrium. Thermal contact can be
maintained until $T_f\sim m_{_{LSP}}/20$, and
a rough freeze-out condition for LSPs (if they were initially in thermal
equilibrium) will be given by: $n_{_{LSP}}\langle \sigma_{\rm eff}v\rangle \approx H_f {m_{\widetilde\chi_{1}^{0}}/T_f}$,
where $\sigma_{\rm eff}$ is the LSP annihilation cross-section and
the subscript $f$ refers to the freeze-out values. The thermally
averaged cross section can be written as
$\langle\sigma_{\rm eff}v\rangle =a+bT/m_{\widetilde\chi_{1}^{0}}$, where $a$ and $b$
depend on the couplings and the masses of the light fermions \cite{Jungman:1995df}.

Assuming an efficient LSP production,
so that $f_{B} =1$, one finds for the LSP density for $b \approx H m_{\widetilde\chi_{1}^{0}}^{2} T_{f}^{-2} n_{f}^{-1}$,
where $n_f \approx (m_{\widetilde\chi_{1}^{0}} T_f)^{3/2}\exp[{-m_{\chi_{1}^{0}}/T_f}]$.
The LSPs produced in $B$-ball decays will spread out by a random
walk with a rate $\nu$ determined by the collision frequency
divided by a thermal velocity $v_{th} \approx (T/m_{\widetilde \chi_{1}^{0}} )^{1/2}$. Since
the decay is spherically symmetric, it is very likely that the LSPs have a Gaussian distribution.
In terms of the density parameter $\Omega_{\widetilde\chi_{1}^{0}}$, the neutralino abundance
can be written as \cite{Fujii:2002kr}
\begin{eqnarray}
 \Omega_{\widetilde\chi_{1}^{0}}
 \simeq
  0.5
   \frac{m_{\chi_{1}^{0}}}{100 {\rm GeV}}\cdot
    \frac{10^{-7}{\rm GeV}^2}{\langle{\sigma v}\rangle}\cdot
     \frac{100 {\rm MeV}}{T_d}
     \left(
      \frac{10}{g_*(T_d)}
      \right)^{1/2}
      \,.
      \label{Omega-ana}
\end{eqnarray}
where the decay temperature of the B-ball is given by
\begin{equation}
T_d\ll 21 \left({m_\chi \over 100~{\rm GeV}}\right)^{3/16}\left(
{10^{20}\over N_{\widetilde\chi_{1}^{0}}^{tot}}\right)^{1/8}
\left(\frac{100}{g(T)}\right)^{3/16} ~{\rm GeV}.
\end{equation}
Below this temperature the annihilations of $\widetilde\chi_{1}^{0}$ are negligible.
Similar analysis can be performed for other LSP candidates. For example, if the gravitino is an LSP,
the gravitino abundance from the Q-balls decay will be  given by~\cite{Shoemaker:2009kg}:
\begin{equation}
\Omega_{3/2}h^{2}\approx 0.11\left(\frac{m_{3/2}}{1~{\rm GeV}}\right)\left(\frac{N_{g}f_{B}}{3}\right)
\left(\frac{\Omega_{b}h^{2}}{0.02}\right)\,,
\end{equation}
where $N_{g}=3$ and $f_{b}\sim 1$. The above expression is valid for temperatures below $10^{7}$~GeV. Similar expression can be derived for 
the Q-balls decaying into axino dark matter~\cite{Roszkowski:2006kw}.

\subsection{Detection of WIMP}

The direct detection of WIMP is the cleanest way to seek the identity of the dark matter, their detection is possible
through elastic collision with the nuclei at terrestrial targets~\cite{Goodman:1984dc}. This method is especially promising for detecting SUSY WIMP candidates, such as neutralino or sneutrino. There are also ways of inferring WIMP in the sky by using the galaxy itself as a detector in the indirect dark matter searches via studying the gamma rays, high energy neutrinos, charged leptons, proton, anti-proton background from the decay or annihilation of the dark matter particles.

\subsubsection{Direct detection of WIMPs}

Important quantity is the recoil energy deposited by the WIMP interaction with the nucleus of mass $m_{N}$ in an elastic collision,
$E_r = m_{r}^{2}v^{2}(1- \cos\theta)/m_{N}$, where $m_{r}$ is the WIMP nucleus reduced mass, $\theta$ is 
the scattering angle in the dark matter-nucleus center-of-mass frame, and $v$ is the velocity 
relative to the detector, and it is of the order of the galactic rotation velocity $\sim 200$ km/s. Typical recoil energies 
are $E_{r}\sim {\cal O}(1-100)$~ keV. 

The differential rate for WIMP elastic scattering off nuclei is given by~\cite{Lewin:1995rx}
\begin{equation}
\frac {dR}{dE_R}=N_{T}
\frac{\rho_{0}}{m_{W}}
                    \int_{v_{\rm min}}^{v_{\rm max}} \,d \vec{v}\,f(\vec v)\,v
                     \,\frac{d\sigma}{d E_R}\,  
\label{eq1}
\end{equation}
where $N_T$ represents the number of the target nuclei,
$m_X$ is the dark matter mass and $\rho_0\sim 0.3~{\rm GeV}/{\rm cm}^{3}$ is
the local WIMP density in the galactic halo,
$\vec v$ and $f(\vec v)$ are the WIMP
velocity and velocity distribution function in the Earth frame, which we take it to be Maxwellian, 
 and ${d\sigma}/{d E_R}$ is the WIMP-nucleus differential cross section. The velocity $v_{min}=\sqrt{(m_N E_{r}/2m_{\rm r}^2)}$, and 
 $v_{max}$ is the escape velocity of the WIMP in the Earth frame, $v_{esc}=544^{+64}_{-46}$~km/s.

In fact, the Earth velocity with respect to the dark matter halo must be written as $v_e = v_0 \,(1.05 + 0.07 \cos \omega t)$ where $1.05\,v_0$ is the galactic velocity of the Sun and $\omega = 2\pi / 1\,$year. The $7\%$ modulation is due to the rotation of the Earth around the Sun~\cite{Drukier:1986tm,Freese:1987wu}. In the above expression, $f(v)$ must be replaced by $f(|\vec{v}-\vec{v_e}|)$. There also exists a forward-backward asymmetry in a directional signal as first pointed out in \cite{Spergel:1987kx,Copi:1999pw}.

For a given momentum transfer $q$, the differential cross section depends on the nuclear form factor
\begin{equation}
\frac{d\sigma}{dq^2} = \frac{\sigma_0}{4\,m_r^2\,v^2} F^2(q)\,,
\end{equation}
where $F(q)$ is a dimensionless  {\it form factor} such that $F(0)=1$, in which case $\sigma_0$ corresponds to the total cross-section.
It is possible to estimate the parameters $\sigma_0$ and $F(q)$,  for example in the case of neutralino WIMP, which is a Majorana fermion therefore it only has axial and scalar couplings~\cite{Jungman:1995df}. 

\vspace{-0.5cm}
\paragraph{\bf Spin-dependent cross-section:} The axial part of the neutralino-quark interaction is mediated via Z boson and squark exchange 
$\mathcal{L}_{q\chi} \sim (\bar{X} \gamma^{\mu} \gamma_5 X)\,(\bar{q}\gamma_{\mu} \gamma_5 q)$. At the level of nutralino-nucleon interaction by considering the nucleon matrix element $\langle n|\bar q\gamma^{\mu}\gamma^{5}|n\rangle $, the effective 
Lagrangian is given by: $\mathcal{L}_{n X}^{\rm eff} = 2 \sqrt{2} \,G_F \,a_{(n)}\, (\bar{X} \gamma^{\mu} \gamma_5 X)\,(\bar{n}\gamma_{\mu} \gamma_5 n)$, where $G_{F}$ is the Fermi constant and $a_{n}$ is a dimensionless parameter. For a nucleus of spin J, with $\langle S_p \rangle$ and $\langle S_n \rangle$ being the average spins ``carried" by protons and neutrons respectively, the cross-section at zero momentum transfer is given by~\cite{Jungman:1995df}
\begin{equation}
\frac{d\sigma}{dq^2} (q=0) = \frac{8}{\pi v^2} G_F^2 \,\Lambda^2\, J(J+1)\,.
\end{equation}
where $\Lambda = ({a_p \langle S_p \rangle + a_n \langle S_n \rangle})/{J}$. Additional corrections to the form factor is required to take into account of the  non-zero momentum transfer. There are many experiments which are sensitive to spin-dependent
cross section with pure proton couplings,  DAMA~\cite{Bernabei:2005hj}, PICASSO~\cite{Archambault:2009sm}, KIMS~\cite{Lee.:2007qn}. 
Recently  COUPP~\cite{Behnke:2010xt} has set the best constraint on spin-dependent cross section down to 
$7\times 10^{-38}~{\rm cm}^{2}$ for a WIMP mass $\sim 30$~GeV.

\begin{figure}[t!]
\centering
\includegraphics[width=0.8\columnwidth]{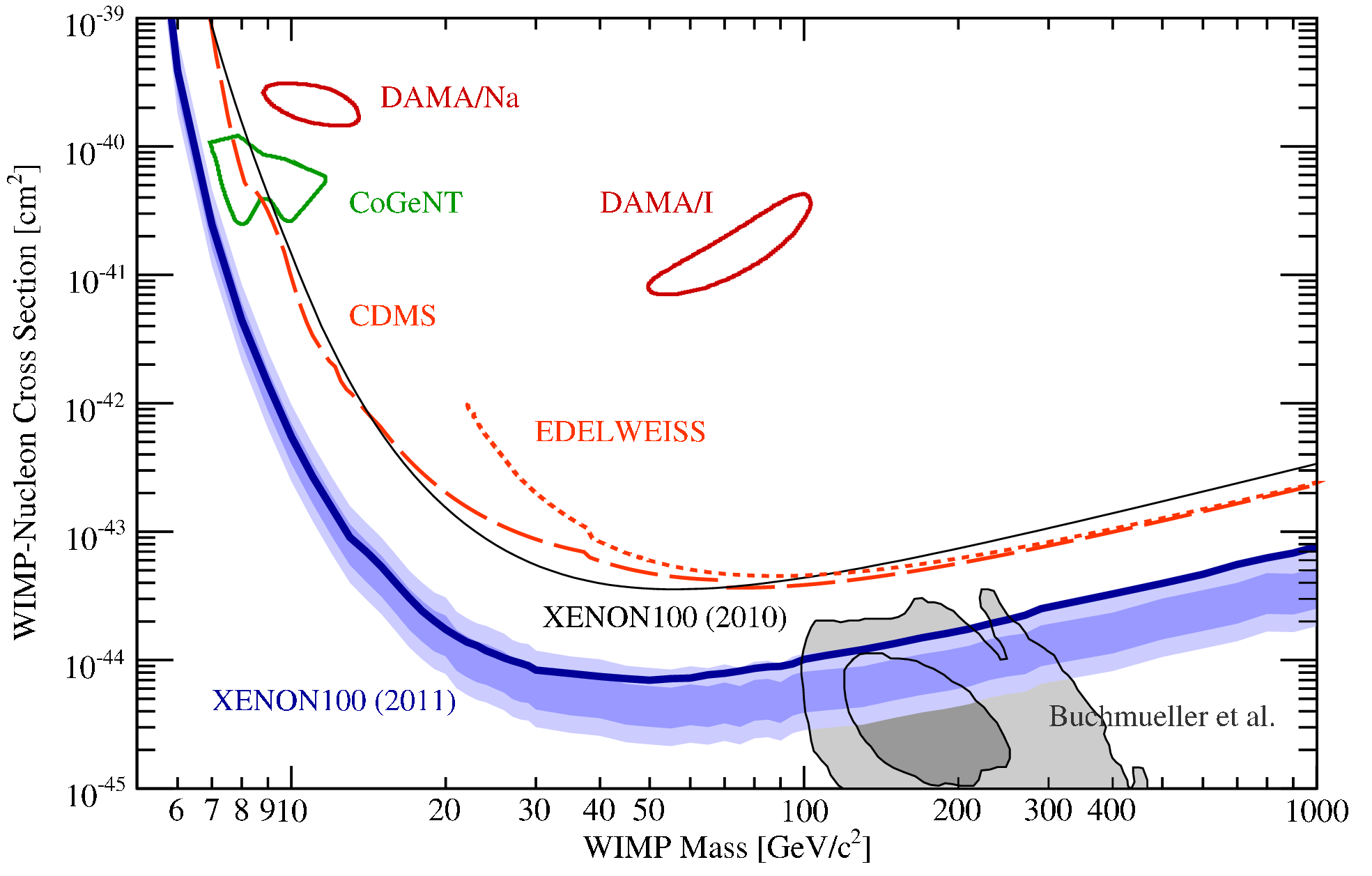}
\caption{Spin-independent elastic WIMP-nucleon cross-section $\sigma$ as function of WIMP mass $m_{\chi}$. The new XENON100 limit at 
$90\%$~CL is shown as the thick (blue) line together with the $1\sigma$ and $2\sigma$ sensitivity of this run (shaded blue band). From \cite{Aprile:2011hi}. }
\label{dm-limit}
\end{figure}

\vspace{-0.5cm}
\paragraph{\bf Spin-independent cross-section:} The scalar part of the neutralino-quark interaction is mediated via Higgses and squark exchanges: $\mathcal{L}_{qX} = f_q (\bar{q}q)(\bar{X}X)$. To express the neutralino-nucleon coupling one needs the nucleon matrix element $m_q \langle n | \bar{q} q | n \rangle \equiv m_{(n)}$,  the effective interaction has the form: $\mathcal{L}_{nX}^{\rm eff} = f_{(n)}\,(\bar{X}X)(\bar{n}n)$, where $f_{n}$ contains the information about hadron physics, and typically it is the same for proton or neutron, i.e., $f_{p}\sim f_{n}$. In the case of right handed sneutrino also there exists no spin-dependent part as it has no axial-vector interactions, and the cross-section is dominated by the spin-independent part. One can define a single spin-independent WIMP-nucleon cross section $\sigma_0\equiv \sigma^{\rm SI}$, independent of the spin of the nucleon~\cite{Jungman:1995df}. 
\begin{equation}
\frac{d\sigma}{dq^2} = [Z\,f_p + (A-Z)\,f_n]^2 \frac{F^2(q)}{\pi v^2}\,,
\end{equation}
In the spin-independant case, for low nuclear recoil energies $F(q)\sim 1$, there is a coherence effect which boosts the WIMP-nucleus cross section by a factor $A^2 m_r^2$. As a result this technique is better suited to detect the WIMP candidate with heavy nucleus targets. Recently 
the most stringent limits on the elastic spin-independent WIMP-nucleon cross-section has been given by number of experiments, such as 
CDMS-II~\cite{Ahmed:2009zw}, EDELWEISS-II~\cite{Armengaud:2011cy} and XENON100~\cite{Aprile:2011hi}. These limits are shown in Fig.~\ref{dm-limit}.
The shaded gray area also shows the expected region of CMSSM for the WIMP mass and the cross section are indicated at $68\%$ and 
$95\%$~CL~\cite{Buchmueller:2011aa}. The current results also covers the $90\%$~CL areas favored by CoGeNT (green)~\cite{Aalseth:2010vx} and 
DAMA (light red)~\cite{Savage:2008er}.

It should be noted that CoGeNT~\cite{Aalseth:2010vx}, and DAMA/NaI~\cite{Bernabei:2005hj} and 
DAMA/LIBRA~\cite{Bernabei:2008yi} collaborations have observed an annual modulation signal. The combined results of the latter group stands 
at greater than  $8\sigma$ statistical significance. The modulation signal phase matches well with the expected annual signal of WIMPS, and subsequent 
data \cite{Bernabei:2010mq} has increased the statistical significance of the modulation signal. However the annual modulation claim has not 
been verified by any other experiments, especially the null results from CDMS, XENON10, and XENON100 data. 
The CDMS data~\cite{Ahmed:2009zw} shows $2$ signal events with $0.6\pm 0.1$ events expected as background, but due to low statistics this result has not provided sufficient evidence for the dark matter.


\subsubsection{Indirect detection}

The indirect dark matter detection depends on the nature of the WIMP. If the WIMP belongs to the visible sector, or if 
it has some SM interactions, then their annihilation or decay would yield to the known particle physics spectrum, for a recent review on indirect 
detection, see \cite{Porter:2011nv}. However 
the spectrum would depend on where they are produced, their energy deposition, and what are their final
states, e.g. $\gamma,~e^{\pm}, \cdots $ etc. The signal could be a hard spectrum with a monochromatic line if WIMPS annihilate directly into 
photons~\cite{Bergstrom:1997fj},
or a continuum spectrum if they annihilate into a pair of intermediate particles ($(q=u,d,c,s,t,b),~\bar q,~Z,~g,~W^{\pm},l^{\pm}$). The former 
process is generically suppressed compared to the latter. Most of these latter particles, i.e. $W, ~Z, ~g$ decay 
into  $p,~\bar{p}$, $\pi^{0}$, and a tiny fraction of deuterium or anti-deuterium $D$/$\bar{D}$. The $\pi^0$s decay to gamma rays, while 
the $\pi^\pm$ decays produce $e^\pm$. If the final states of decay or annihilations are  $e^\pm$s or $\mu^\pm$s, they
dominantly produce a hard $e^\pm$ spectrum, with the $\mu^\pm$ decays into $\nu_\mu$ and $\nu_e$.  If the final states have
$\tau^\pm$, they produce a softer $e^\pm$ spectrum and a strong neutrino signal.  The $\tau^\pm$ can also decay hadronically to 
pions and thus can also produce a strong $\gamma$-ray signal.  

The source spectrum is generically given by 
\begin{equation}
\label{eqn:PPfactor}
\Phi_s(E)=\frac{1}{4 \pi} \frac{\langle \sigma v \rangle}{2 m^2 _X}
\sum\limits_f {dN_f\over dE}B_{f,s}\, ,
\end{equation}
where $f$ denotes the annihilating final states, each with branching fraction $B_{f,s}$ with $E$ being the energy of  secondary particles.
The production rate per annihilation of species $f$ is given by $dN_f/dE$. For a decaying dark matter
$\langle \sigma v \rangle/2 m^2 _X$ can be replaced by  $\Gamma/ M_X$,
where $\Gamma$ is the decay rate.

The flux of such final states would depend on the annihilation or decay rate, which 
in turn would depend on the dark matter density $\propto \rho_{X}^{2}$. Therefore, the natural sources to look at in the sky
are the nearby galactic centers -- where there are large astrophysical uncertainties, dwarf galaxies -- which have small 
astrophysical background, and galactic centers -- where the dark matter density is very large but distant sources would
yield a local tiny flux. 

Gamma rays and neutrinos are perhaps the cleanest signals if they are produced as a result of WIMP annihilations or decays
as they are undeflected by magnetic fields and effectively indicating the direction to their source. The flux is given by the integral of the
WIMP density-squared along the line of sight from the observer to the source, multiplied by the production
spectrum $\phi_\gamma(E,\psi)=  J(\psi)\times\Phi_\gamma(E)$
\begin{equation}
\Phi_{\gamma}(E,\psi)=\frac{dN_{f}}{dE}\frac{\langle \sigma v\rangle}{4\pi m^{2}_{X}} B_{f, s}\int_{\rm l. o. s}ds\rho^{2}(r(s,\psi))\,,
\end{equation}
where $``f''$ denotes the final states and the coordinate $s$ runs along the line of sight, $E$ is the gamma-ray energy, and
$\psi$ is the elongation angle with respect to the center of the source. The astrophysics related term is hidden in~\cite{Bergstrom:1997fj}
\begin{equation}
\label{eqn:astrofactor}
J(\psi)=\frac{1}{8.5~{\rm Kpc}}\left(\frac{1}{0.3 ~{\rm GeV/cm}^3}\right)^{2}\int_{\rm l.o.s.}\rho^2(\ell)\,d\ell
\end{equation}
where the integration is in the direction $\psi$ along the line $\ell$. The above expression is also valid for neutrinos if the source is not
far away from us.

On the other hand the charged particles do not have the directional sensitivity,  they lose it in their course of path in a  
random motion due to the interstellar magnetic field. The motion of $e^{\pm}$ are deflected by the interstellar radiation
field by synchroton radiation in presence of magnetic field. If produced at energies $\geq 100$~GeV and if their sources are 
within few kiloparsecs, then they can reach the solar system.  Furthermore, the cosmic rays from the primary and secondary 
products can also generate $\gamma$ rays during their course through the ISM, all these effects can be captured numerically in 
the publically available code GALPROP \cite{Strong:1998pw,Strong:1998fr,Strong:2004de}, for a review see~\cite{Strong:2007nh}.

The main challenge is to disentangle the WIMP signals from astrophysical backgrounds, the powerful discriminator is the spectral tilt 
in the power spectrum. It is quite possible to have a significant fraction of WIMP annihilation or decay into monoenergetic photons, 
giving rise a distinctive line in the gamma-ray spectrum~\cite{Bergstrom:1997fj}, but the likely signal would be to have a relatively 
hard continuum spectrum with a bump or edge near the WIMP mass that is above the astrophysical background. 

\begin{figure}
\includegraphics[width=2.6in]{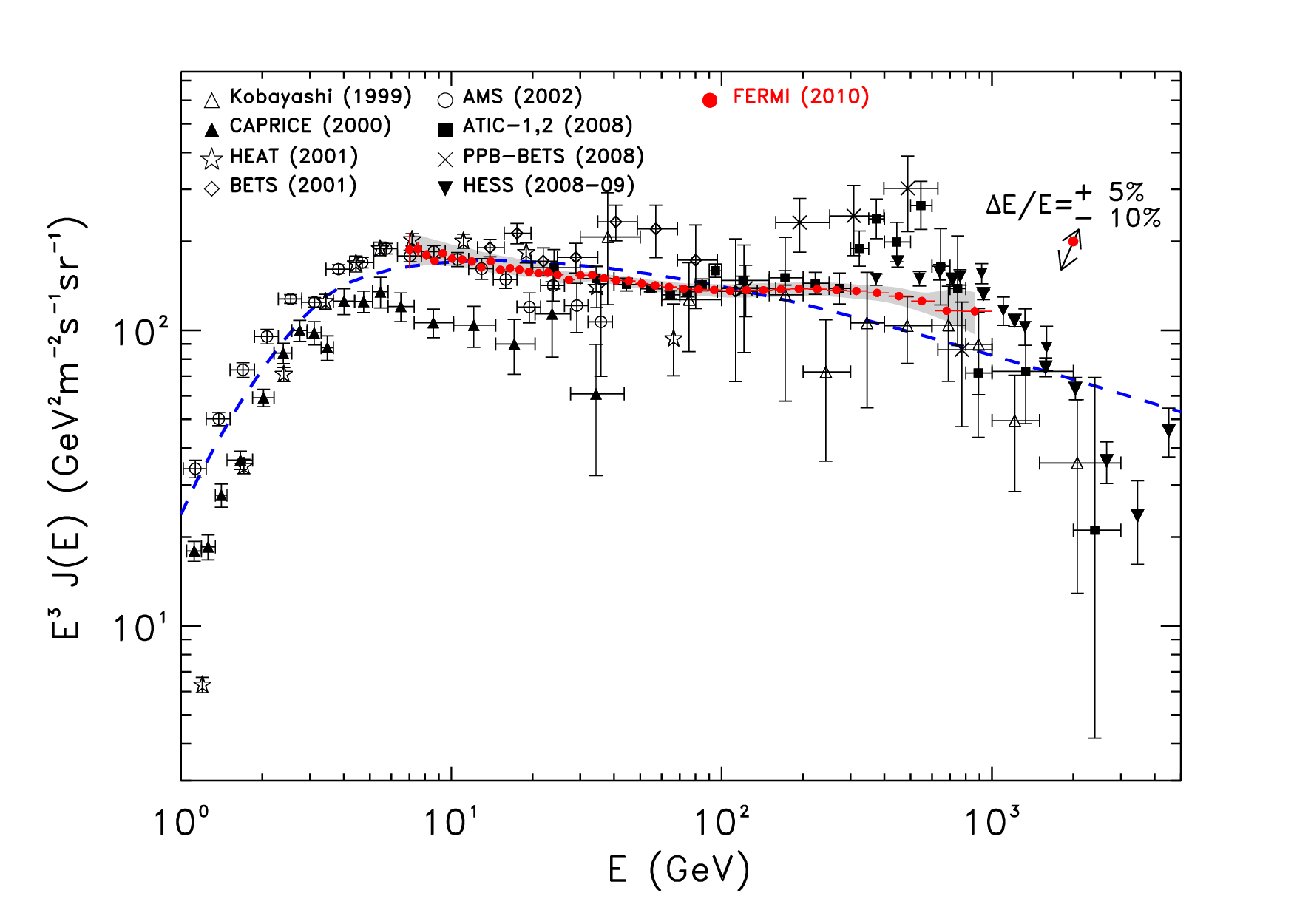}
\caption{
Combined electron and positron spectrum as measured by Fermi-LAT
for one year of observations, together with other measurements~\cite{Abdo:2009zk,Ackermann:2010ij}.
The systematic errors for the  measurement are shown by the grey band. The systematic uncertainty associated with the absolute energy scale
is shown by the non-vertical arrow. The dashed line shows the background from secondary $e^{\pm}$ in cosmic rays from GALPROP.}
\label{Fermi-lat:electron}
\end{figure}

In recent years there have been many new experiments which have propelled the indirect search for dark matter research vigorously. 
For instance,  the ATIC data -- which shows a significant bump in the electron flux around $300-800$~GeV~\cite{Chang:2008zz}, where 
conventional astrophysical sources would have predicted a decaying power law spectrum, and the 
PAMELA -- which shows a positron fraction which has a rising slope above $10$~GeV~\cite{Adriani:2008zr,Adriani:2010rc}, but the anti-proton 
flux and the fraction matches well with the expectations of the astrophysical origins. 
The Fermi-LAT collaboration which has also produced an electron spectrum from $7$~GeV to $1$~TeV~\cite{Abdo:2009zk, Ackermann:2010ij}
does not confirm the rising slope of the ATIC, see Fig.~\ref{Fermi-lat:electron}, rather the data matches well with the HESS at higher end of the 
spectrum~\cite{Aharonian:2008aa,Aharonian:2009ah}.


The rise in the $e^{+}$ fraction in the PAMELA data may have its roots in the astrophysical objects such as nearby pulsars, 
Monogem at a distance of $d=290$pc, and Geminga at a distance of $d=160$pc. Both are nearby objects to Earth which can 
contribute significantly to the $e^{\pm}$ flux, and can match both the PAMELA and Fermi-LAT data sets, see~\cite{Grasso:2009ma}. 
A dark matter interpretation of these data sets requires an ad-hoc assumption on the leptohilic nature of the WIMP with an annihilation 
cross-section $\langle \sigma v\rangle \sim 10^{-24}~{\rm cm^{2}}$, more than $2-3$ orders of magnitude enhancement from thermally
produced WIMPS at the TeV scale, se Eq.~(\ref{anni-abund}). Such an enhancement in the cross-section is possible at low energies 
and with low velocities of dark matter provided that the range of the interaction responsible for annihilation is much shorter than the
long-range attractive potential, known as a Sommerfeld enhancement~\cite{ArkaniHamed:2008qn}, 
see for a detailed derivation~\cite{Slatyer:2009vg} in the mass range of $500-900$~GeV~\cite{Grasso:2009ma}.  It is also possible to 
conceive a scenario of non-thermal dark matter production in order to 
evade the strict bound on cross-section, or decaying dark matter scenarios, i.e.~\cite{Fairbairn:2008fb,Arvanitaki:2009yb}. 

Finally, one would also expect a gamma  ray signal from the galactic center, the typical signature for a WIMP annihilation will be 
a line at the WIMP mass, due to the 2$\gamma$s, or $\gamma Z$ production channels. The Fermi-LAT collaboration has 
released a diffused galactic and extra galactic $\gamma$-ray background \cite{Abdo:2010nz}, however
no lines were observed yet~\cite{Abdo:2010nc}.

The high energy neutrinos are being another frontier for the indirect dark matter searches as the construction of new experiment IceCube
with large volume  is underway, in which case even Earth could be used as a detector to study the nature of dark matter, e.g.
~\cite{Albuquerque:2003mi}, or neutrino emission from the passage of Q-balls~\cite{Kusenko:2009iz}.  Searching for
WIMPs in more than one type of experiments;  direct, indirect, and the LHC will be necessary in order to 
stamp the origin of such elusive particles.

 
\section{Acknowledgements}
 The author would like to thank A. Kusenko for many useful discussions and collaborating at the initial stages of this review. 
 He would also like to thank R. Allahverdi, R. Brandenberger, C. Boehm, A. Chatterjee, P. Dayal, B. Dutta, K. Enqvist, A. Ferrantelli, J. Garcia-Bellido, S. Hotchkiss, K. Kazunori, 
 D. Lyth, S. Nadathur, K. Petraki, N. Sahu, Q. Shafi, and P. Stephens for helpful discussions.


\bibliography{KM-rev}


\end{document}